\documentclass[a4paper,11pt]{article}
\pdfoutput=1

\usepackage{jheppub} 

\usepackage[T1]{fontenc} 

\usepackage{amssymb}

\title{NLO QCD corrections to full off-shell production of
  $\boldsymbol{t\bar{t}Z}$ including leptonic decays}

\author[a]{Giuseppe Bevilacqua,}
\author[b]{Heribertus Bayu Hartanto,}
\author[c]{Manfred  Kraus,}
\author[d]{Jasmina Nasufi}
\author[d]{and Malgorzata  Worek}

\affiliation[a]{ELKH-DE Particle Physics Research Group, University of
  Debrecen, H-4010 Debrecen, \\P.O. Box 105, Hungary}
\affiliation[b]{Cavendish Laboratory, University of Cambridge, 
Cambridge CB3 0HE, United Kingdom} 
\affiliation[c]{Physics Department, Florida State University,
  Tallahassee, FL 32306-4350, USA}
\affiliation[d]{ Institute for Theoretical Particle Physics
and Cosmology, RWTH Aachen University, \\D-52056 Aachen, Germany}

\emailAdd{giuseppe.bevilacqua@science.unideb.hu}
 \emailAdd{hbhartanto@hep.phy.cam.ac.uk}
 \emailAdd{mkraus@hep.fsu.edu}
 \emailAdd{jasmina.nasufi@rwth-aachen.de}
 \emailAdd{worek@physik.rwth-aachen.de}

 \abstract{
Motivated by ongoing new physics searches in the top-quark sector at
the Large Hadron Collider we report on the calculation of NLO QCD
corrections to the Standard Model  $pp\to t\bar{t}Z+X$ process in the
tetra-lepton decay channel. This calculation is based on the matrix
elements for the $e^+ \nu_e\, \mu^-\bar{\nu}_\mu\, b\bar{b}
\,\tau^+\tau^-$ final state and includes all resonant and non-resonant
Feynman diagrams,  interferences and off-shell effects of the top quark
as well as the $W$ and $Z$ gauge bosons. Also incorporated are 
photon-induced contributions. As it is customary for such studies,
we show theoretical predictions for both fixed and dynamical
factorisation and renormalisation scale choices and different PDF sets.
Furthermore,  we study the main theoretical uncertainties that are
associated with  neglected higher-order terms in the perturbative
expansion and with  the parameterisation of the PDF sets. In order to
investigate the  size of off-shell effects and higher-order corrections
in top-quark  decays, we perform a second computation for this process,
which is  based on the narrow-width-approximation where the top quarks,
$W$  and $Z$ gauge  bosons are kept on-shell. Results are given for the 
integrated and differential fiducial cross sections for the LHC Run II
center-of-mass energy of $13$ TeV.}

\dedicated{\rm TTK-22-02,  P3H-22-002, CAVENDISH-HEP-22/03}


\begin{document} 
\maketitle
\flushbottom

% =============================================
%
\section{Introduction}
\label{sec:introduction}
%
% =============================================

The frontier center-of-mass energy and the ever-increasing luminosity 
of proton collisions at the Large Hadron Collider
(LHC) enable precise studies of Standard Model 
(SM) processes with small production rates, such as the associated 
production of a $t\bar{t}$ pair and a $Z$ boson. Among various possible 
decay channels for this process the final state containing four charged
leptons (electrons or muons) is the cleanest from the experimental point 
of view and the most interesting for theoretical studies. This channel 
is often referred to as the  tetra-lepton channel and abbreviated 
as $4\ell$.  Indeed, the measurements of the charged multi-lepton final
states are particularly precise at the LHC due to the excellent lepton
identification and selection in the ATLAS and CMS detectors. In fact,
both ATLAS and CMS collaborations have already used events with four
charged leptons and measured the inclusive cross section for
$t\bar{t}Z$ production \cite{CMS:2017ugv,ATLAS:2019fwo}. In addition,
the differential cross section was measured as a function of a few
observables \cite{CMS:2019too,ATLAS:2021fzm}. Processing the full Run II 
data $({\cal L} =139$ fb${}^{-1})$, fairly good agreement between data and
theoretical predictions at the differential level has been found. 
Interestingly, observables directly connected to the $Z$ boson show poorer
agreement. The upcoming LHC Run 3 and
the future high-luminosity LHC upgrade would definitely require more
accurate modelling of production and decays for the $t\bar{t}Z$
process.  A good theoretical knowledge of the $t\bar{t}Z$ process is
needed not only for an accurate prediction of fiducial cross section and
better understanding of the various properties of the top quark, but it is
also a fundamental prerequisite for the correct interpretation of possible
new physics signals that may arise in this channel.  The $t\bar{t}Z$
production process provides direct access to the coupling of
the top quark to the neutral electroweak gauge bosons. It gives important
insights into the top quark, that are complementary to $t\bar{t}$ and
single top quark production as well as to the $t\to W b$ decay. Any
deviations of the coupling strength of the top quark to the $Z$ boson
from its SM value might imply the existence of new physics effects. In
particular, the $t\bar{t}Z$ coupling might be affected by a tree-level
mixing with additional $Z^\prime$ gauge bosons and vector-like
leptons, see e.g.  \cite{Greiner:2014qna,Cox:2015afa,Kumar:2015tna,
Kim:2016plm,Fox:2018ldq,Alvarez:2020ffi,Bissmann:2020lge}. Such
deviations can be probed for example in the context of effective field
theory \cite{Baur:2004uw,Baur:2005wi,Berger:2009hi,Rontsch:2014cca,
Rontsch:2015una,Buckley:2015lku,BessidskaiaBylund:2016jvp,
Schulze:2016qas,Hartland:2019bjb,Maltoni:2019aot,Durieux:2019rbz,
Brivio:2019ius,Afik:2021xmi}. Additionally, $t\bar{t}Z$ production is
among the most important background processes for several new physics
scenarios. They comprise final states with multiple charged leptons,
missing transverse momentum and $b$-jets and are vigorously searched
for at the LHC \cite{ATLAS:2016dlg,CMS:2016mku,ATLAS:2017tmw,
CMS:2017tec,ATLAS:2019fag,CMS:2020cpy}. 
Finally, the $t\bar{t}Z$ process plays a very prominent
role in studies of important SM processes. Here, good examples 
comprise $t\bar{t}$ production in association with a Higgs boson
\cite{ATLAS:2017ztq,ATLAS-CONF-2019-045, CMS:2020mpn} and single
top quark production in association with a $Z$ boson
\cite{ATLAS:2020bhu,CMS:2021ugv}. Therefore, providing a reliable and
accurate description of $pp \to t\bar{t}Z$ production 
to advance our understanding of the process and 
to facilitate comparisons with constantly increasing and
improving LHC data,  is more timely than ever. To improve our
understanding of $t \bar {t} Z$  various effects must be considered
and  taken into account  already at the matrix element level,
omitting, as much as possible, approximations when
incorporating them. One might expect to at least examine where the
approximations used may fail. Furthermore, it might not be sufficient to
incorporate various effects, such as spin-correlated decays, at the lowest 
order in the perturbative expansion. Indeed,
higher-order corrections not only induce  substantial normalisation and
shape differences but also strongly reduce systematical 
uncertainties associated with theoretical predictions.  So it is
necessary to include next-to-leading order (NLO) QCD and electroweak
corrections wherever possible to claim high-precision predictions for
$ pp \to t \bar {t} Z$.

For the inclusive $t\bar{t}Z$ production, with stable top quarks and 
an on-shell $Z$ gauge boson, NLO QCD corrections have 
been around for over ten years
\cite{Lazopoulos:2008de}. They have been afterwards recomputed in Refs.
\cite{Kardos:2011na,Maltoni:2015ena}.  Furthermore,  results with NLO
electroweak (EW) corrections have been provided in the literature 
\cite{Frixione:2015zaa}. Besides NLO QCD and EW corrections, a
further step towards a more precise modelling of the 
$t\bar{t}Z$ production process has been achieved by including soft
gluon resummation effects at next-to-next-to-leading logarithmic
accuracy \cite{Broggio:2017kzi,Kulesza:2018tqz,Broggio:2019ewu,
Kulesza:2020nfh}. The calculations mentioned above provide important
information about the impact of higher-order corrections to the  total 
$t\bar{t}Z$ production rate. However, since decays of  unstable top 
quarks and $Z$ bosons are not taken into account, they are neither 
capable to ensure a reliable description of the fiducial phase space 
regions nor can they give us a glimpse into the top quark radiation 
pattern. Thus, for more 
realistic studies the decays of unstable particles must  be taken 
into account. For the latter various modelling approaches are available 
in the literature. Firstly, NLO QCD theoretical predictions for stable top
quarks and a $Z$ boson have been matched with parton shower (PS) Monte
Carlo (MC) programs \cite{Garzelli:2011is,Garzelli:2012bn}. In this 
case top quark and $Z$ gauge boson decays have been treated in the
parton shower approximation omitting spin correlations even at leading
order. Very recently, improved NLO + PS predictions for $t\bar{t}Z$
have been provided \cite{Ghezzi:2021rpc}. Specifically, the $pp\to
t\bar{t}\ell^+\ell^-$ process with $\ell$ denoting either $e^\pm$ or
$\mu^\mp$ have been matched to PS programs including both resonant and
non-resonant $Z$- as well as photon-induced contributions. 
Furthermore, decays of the top quarks with full
tree-level spin correlations in the narrow-width-approximation (NWA)
have been taken into account according to the method proposed in
Ref. \cite{Frixione:2007zp}. A different approach has been
considered in Ref. \cite{Rontsch:2014cca}, where both top quark and 
$Z$ gauge boson are treated in NWA, but NLO QCD corrections in both 
production and decay stages are taken into account. In this way, 
contributions which are parametrically
suppressed by ${\cal O}(\Gamma/m)$, arising from off-shell top quarks
or $W/Z$ bosons, have been neglected. Also the contribution from the
 $t\to WbZ$ decay has been neglected due to the tiny
available phase space and the size of the $t \to WbZ$ branching
ratio. The latter is of the order of ${\cal B}_{t\to WbZ} \approx 2
\times 10^{-6}$ for the top quark mass range $m_t \in
(170-180)$ GeV \cite{Altarelli:2000nt}. Finally, state-of-the-art NLO
QCD predictions have been provided for the $t\bar{t}Z$ process in the
di-lepton top quark decay channel \cite{Bevilacqua:2019cvp}. More
precisely, NLO QCD corrections to the $pp\to e^+ \nu_e\,
\mu^- \nu_\mu \, b\bar{b} \,\nu_\tau \bar{\nu}_\tau+X$ final state have
been calculated. All double-, single- and non-resonant Feynman diagrams,
interferences, and off-shell effects of the top quarks have been
properly incorporated at NLO QCD. Non-resonant and
off-shell effects due to the finite $W$- and $Z$-boson width have been
included as well. Furthermore, full off-shell $t\bar{t}Z$
predictions at NLO in QCD have been compared to those computed in the
NWA approach \cite{Hermann:2021xvs}. Similar studies for the
production of $t\bar{t} Z$ with the $Z$ boson decaying into charged
leptons, however, are still missing.

The purpose of this article is to mitigate this situation and to
calculate for the first time NLO QCD corrections to the $e^+ \nu_e \,
\mu^- \bar{\nu}_\mu \, b\bar{b}\, \tau^+ \tau^-$ final state for the
LHC Run II center-of-mass system energy of $\sqrt{s}=13$ TeV.
 We only simulate decays of the weak bosons to different
lepton generations, however, these interference effects are at the
per-mille level for inclusive cuts. Thus, the complete $pp\to \ell^+
\nu_\ell \, \ell^- \bar{\nu}_\ell \, b\bar{b}\, \ell^+ \ell^- +X$ cross
section (with $\ell =e^\pm, \mu^\pm$) can be obtained by multiplying
the results from this paper with a lepton-flavour factor of $8$.  Our
calculations comprise all quantum effects at the matrix
element level. We scrutinise the size of higher-order corrections and
theoretical uncertainties in such a complex environment. We
additionally address the choice of a judicious renormalisation and
factorisation scale setting.  Afterwards, the size of off-shell
effects of the top quarks and gauge bosons is examined at the
integrated and differential level. The latter study is carried out
with the help of a second computation that we perform for this process
based on the NWA approach. Specifically, we compare results from the
full off-shell calculation against predictions based on the full NWA
as well as on the NWA with leading order (LO) top-quark decays
(abbreviated as ${\rm NWA}_{\rm LOdec}$). By employing ${\rm NWA}_{\rm
LOdec}$ results, we are able to estimate the size of the NLO QCD
corrections to top quark decays. It should be clear that, in the full
off-shell picture, the contribution of the $Z \to \tau^+ \tau^-$ decay
must be complemented by its photon-induced counterpart, $\gamma^* \to
\tau^+ \tau^-$, in order to preserve gauge invariance. Therefore, like
Ref. \cite{Ghezzi:2021rpc}, we are in the condition of examining more
realistically the impact of photon contributions and in particular of
the $Z/\gamma^*$ interference on integrated and differential fiducial
cross sections. Finally, the effect of applying an additional cut on
the invariant mass of the $\tau$ lepton pair is studied. In
particular, we investigate the requirement that the invariant mass of
the $\tau$ leptons is set in a specific mass window around the $Z$
boson mass.

The paper is organised as follows. In Section \ref{details} we briefly
summarise the framework of our calculation and discuss technical
aspects of the computation. We outline the theoretical setup for LO
and NLO QCD results in Section \ref{setup}. Results for the integrated
fiducial cross sections are presented in Section
\ref{integrated}. They are provided for the LHC center-of-mass system
energy of 13 TeV and for a fixed and dynamical renormalisation and
factorisation scale choice. In Section \ref{differential} predictions
for a few differential fiducial cross sections are
given. Additionally, theoretical uncertainties associated with the
neglected higher order terms in the perturbative expansion and
different parameterisations of the parton distribution functions are
discussed in Sections \ref{integrated} and  \ref{differential}.
Theoretical predictions for $t\bar{t}Z$ in the NWA are provided in
Section \ref{comparison}. Also there the size of off-shell effects is
examined together with the impact of  higher-order corrections to 
top-quark decays. Finally, we summarise the results and outline our
conclusions  in Section \ref{summary}.

% =============================================
%
\section{Details of the calculation}
\label{details}
%
% =============================================
%

\begin{figure}[t]
  \begin{center}
    \includegraphics[width=\textwidth]{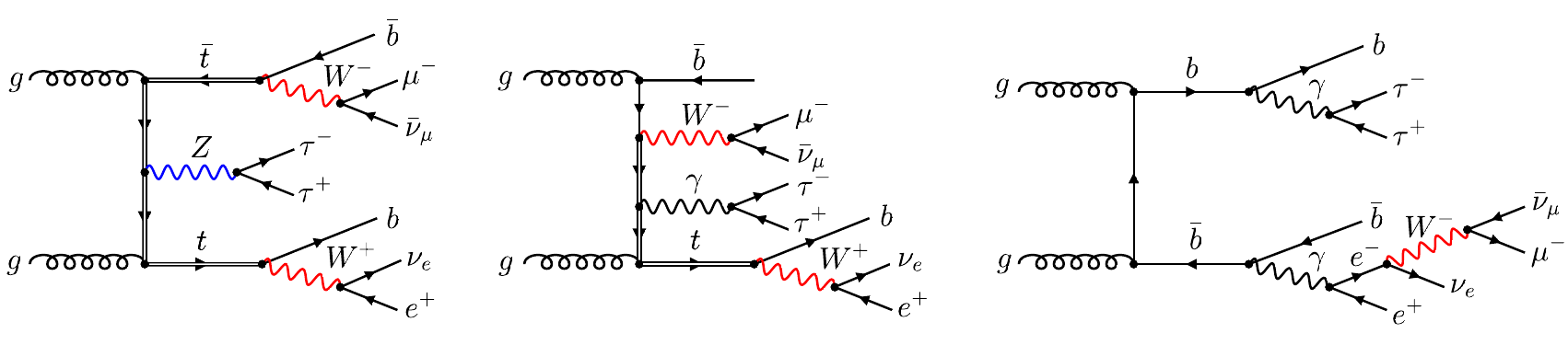}
\end{center}
\caption{\label{fig:fd-LO} \it Representative Feynman diagrams with
double- (left), single- (middle) and no top-quark
resonances (right) contributing to the
$gg\to e^+\nu_e\, \mu^-\bar{\nu}_\mu \, b\bar{b}\,\tau^+\tau^-$
partonic subprocess at leading order. The middle and right diagram  
comprise no $Z$ resonances whereas the third one involves only a
single $W$ boson. They contribute to $Z$ and $W$  off-shell
effects. }
\end{figure}

\begin{figure}[t]
  \begin{center}
    \includegraphics[width=\textwidth]{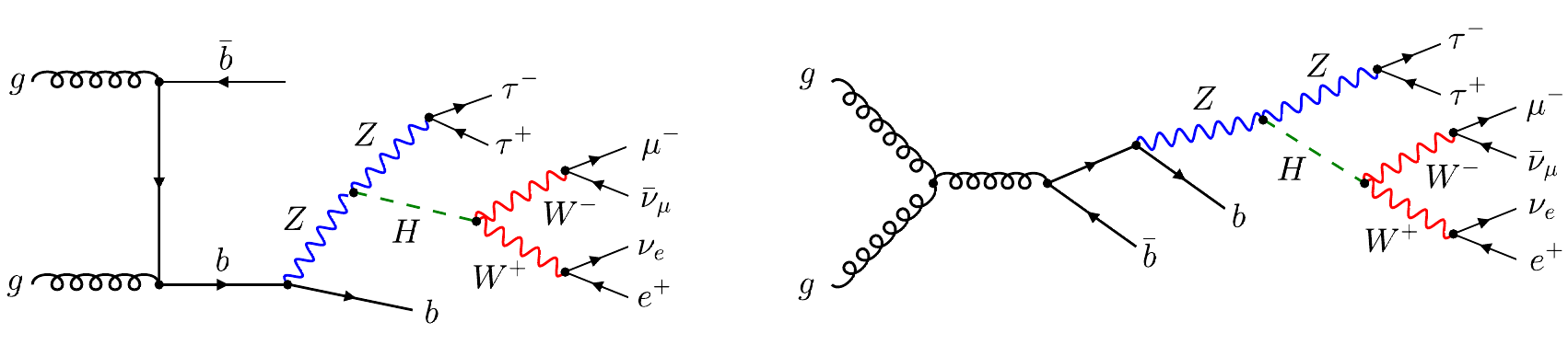}
\end{center}
\caption{\label{fig:fd-Higgs} \it Representative Feynman diagrams with
  the Higgs-boson-exchange contribution  that appear in the
$gg\to e^+\nu_e\, \mu^-\bar{\nu}_\mu \, b\bar{b}\,\tau^+\tau^-$
partonic subprocess at leading order even though the $b$ quarks and
$\tau^\pm$ leptons are treated as massless particles.}
 \end{figure}

In $pp$ collisions at the LHC the $e^+\nu_e\, \mu^-\bar{\nu}_\mu \,
b\bar{b}\,\tau^+\tau^-$ final state is produced via the scattering of
two gluons or a $q\bar{q}$ pair, where $q$ stands for up- or down-type
quarks.  The contributions to the tree-level squared amplitude at
${\cal O}(\alpha_s^2\alpha^6)$ can be subdivided into three classes:
diagrams containing two top-quark propagators that can become resonant,
diagrams containing only one top-quark resonance and finally diagrams
without any top-quark resonance. For the $Z$ and $W$ gauge bosons,
resonant and non-resonant contributions are present.  In the former case
also photon-induced contributions and the $Z/\gamma^*$ interference 
effects are included. Examples of Feynman diagrams are depicted in 
Figure \ref{fig:fd-LO}. The \textsc{FeynGame} program
\cite{Harlander:2020cyh} is employed to draw all Feynman diagrams in
this article. In total, there are $1836$ LO diagrams for the $gg\to
e^+\nu_e\, \mu^-\bar{\nu}_\mu \, b\bar{b}\,\tau^+\tau^-$ partonic
reaction and $980$ diagrams for each $q\bar{q}\to e^+\nu_e\,
\mu^-\bar{\nu}_\mu \, b\bar{b}\,\tau^+\tau^-$ subprocess. Even though
we do not employ Feynman diagrams in our calculations we present their
numbers as a measure of the complexity of the calculation. Let us note 
that, even in the case where $b$ quarks and $\tau$ leptons are treated as
massless particles, a number of Higgs-boson exchange diagrams contribute 
to the amplitude.  Specifically, there are $1844$ diagrams for the $gg$
initiated subprocess that need to be considered if
the Higgs boson is included and $984$ for each $q\bar{q}$
channel. Examples of the corresponding Feynman diagrams for the
$gg\to e^+\nu_e\, \mu^-\bar{\nu}_\mu \, b\bar{b}\,\tau^+\tau^-$
partonic reaction are depicted in Figure \ref{fig:fd-Higgs}. The Higgs
boson  contribution, however, is well below $0.1\%$ level as we have checked
by an explicit LO calculation using $m_H=125$ GeV and $\Gamma_H=4.07
\times 10^{-3}$ GeV. Thus, it is neglected throughout our
calculations. Furthermore, LO contributions induced by the bottom-quark 
parton density are at the permille level, i.e. they are of the order of 
$0.3\%$. Also these contributions are not taken into account
in our study.
%
%=============================================
%=============================================
%=============================================

The calculation of LO contributions is performed
automatically with the help of the \textsc{Helac-Dipoles} package
\cite{Czakon:2009ss}. The results are cross checked with the
\textsc{Helac-Phegas} program \cite{Cafarella:2007pc}. Both packages
are based on the well-known off-shell iterative algorithm for
scattering amplitudes
\cite{Draggiotis:1998gr,Kanaki:2000ey,Draggiotis:2002hm}. The
integration over the momentum fractions $x_1$, $x_2$ of the 
initial-state partons is optimised with the help of \textsc{Parni}
\cite{vanHameren:2007pt}. The phase space integration is executed with
the help of \textsc{Kaleu} \cite{vanHameren:2010gg} and cross checked
with \textsc{Phegas} \cite{Papadopoulos:2000tt}. The inclusion of the
decays of top quarks, $W^\pm$ and $Z$ gauge bosons is performed in the
complex mass scheme \cite{Denner:1999gp,Denner:2005fg}. Such a scheme 
fully respects gauge invariance and is straightforward to apply.

The virtual corrections consist of the 1-loop corrections to the LO
contributions. For our process one can classify the corrections into
self-energy, vertex, box-type, pentagon-type, hexagon-type and
heptagon-type topologies. In Table \ref{tab:one-loop-gg} the number of
one-loop Feynman diagrams, that corresponds to each type of correction
for the dominant $gg\to e^+\nu_e \, \mu^- \bar{\nu}_\mu \, b\bar{b} \,
\tau^+ \tau^-$ partonic subprocess as obtained with \textsc{Qgraf}
\cite{Nogueira:1991ex}, is given. In Table \ref{tab:one-loop-qq}
similar results are displayed for the $q\bar{q}\to e^+\nu_e \, \mu^-
\bar{\nu}_\mu \, b\bar{b} \, \tau^+ \tau^-$ partonic subprocess. To
evaluate the virtual corrections, \textsc{Helac-1Loop} 
\cite{vanHameren:2009dr} is used, which is based on the
\textsc{Helac-Phegas} program to calculate all tree-level like
ingredients and the OPP reduction method \cite{Ossola:2006us}. The
cut-constructible part of the virtual amplitudes as well as the
rational term $R_1$ of the amplitude is computed using the
\textsc{CutTools} code \cite{Ossola:2007ax,Ossola:2008xq}. The $R_2$
term, on the other hand, is obtained with the help of extra Feynman
rules \cite{Draggiotis:2009yb}. At the one-loop level the appearance
of a non-zero top-quark width in the propagator requires the
evaluation of scalar integrals with complex masses. For this task the
program \textsc{OneLOop} \cite{vanHameren:2010cp} is employed.  We
chose to renormalise the strong coupling in the $\overline{\rm MS}$
scheme with five active flavours and the top quark decoupled, while
the mass renormalisation is performed in the on-shell scheme. 
The correctness of the calculation is checked by a series of tests performed
both at the preparation stage and at runtime. We have checked that our
one-loop amplitudes agree with the results of
\textsc{MadGraph5}${}_{-}$\textsc{aMC@NLO}
\cite{Alwall:2014hca} for a few phase space points and for both $gg$ and 
$q\bar{q}$ subprocesses. Additionally, the cancellation of the infrared
$1/\epsilon^2$ and $1/\epsilon$ poles between virtual and real corrections, 
as provided by the ${\cal I}$-operator, has been checked numerically. We
also monitor the numerical stability by checking Ward identities at
every phase space point. The events which do not pass this runtime 
check are not discarded from the calculation of the finite part, but 
rather recalculated with higher precision. For the $q\bar{q}$ subprocess 
we use the so-called scale test \cite{Badger:2010nx}, which is based on
momentum rescaling.
%
%=============================================
\begin{table}[t]
\begin{center}
  \begin{tabular}{cc}
    \hline \hline
    &\vspace{-0.3cm}\\
\textsc{One-Loop}& \textsc{Number Of} \\
    \textsc{Correction} &\textsc{Feynman Diagrams} \\
    &\vspace{-0.3cm}\\
    \hline \hline
    &\vspace{-0.3cm}\\
 \textsc{Self-Energy} &  32040 \\
 \textsc{Vertex} &  39288  \\
 \textsc{Box-Type}                &   21452  \\
 \textsc{Pentagon-Type} & 8604    \\
   \textsc{Hexagon-Type}              & 1996   \\
    \textsc{Heptagon-Type}  &   180  \\
    & \vspace{-0.3cm} \\
    \hline \hline
 & \vspace{-0.3cm} \\
    \textsc{Total Number}  &  103560   \\
    &\vspace{-0.3cm}\\
    \hline \hline
 \end{tabular}
\end{center}
\caption{\label{tab:one-loop-gg} \it
  Number of one-loop Feynman diagrams for the dominant $gg\to
e^+\nu_e \, \mu^- \bar{\nu}_\mu \, b\bar{b} \, \tau^+ \tau^-$ partonic
subprocess at ${\cal O}(\alpha_s^3 \alpha^6)$ for the $pp\to e^+ \nu_e\,
\mu^- \bar{\nu}_\mu \, b\bar{b} \, \tau^+ \tau^- +X$ process. The
Higgs boson exchange contributions are not considered and the
Cabibbo-Kobayashi-Maskawa mixing matrix is kept diagonal.}
\end{table}
%=============================================
\begin{table}[t]
\begin{center}
  \begin{tabular}{cc}
        \hline \hline
    &\vspace{-0.3cm}\\
\textsc{One-Loop}& \textsc{Number Of} \\
    \textsc{Correction} &\textsc{Feynman Diagrams} \\
    &\vspace{-0.3cm}\\
    \hline \hline
    &\vspace{-0.3cm}\\
 \textsc{Self-Energy} &  18338   \\
 \textsc{Vertex} &  10272 \\
 \textsc{Box-Type}                &  5218  \\
 \textsc{Pentagon-Type} &  1900  \\
   \textsc{Hexagon-Type}              & 380   \\
    \textsc{Heptagon-Type}  &   32    \\
     & \vspace{-0.3cm} \\
    \hline \hline
    & \vspace{-0.3cm} \\
    \textsc{Total Number}  &  36140      \\
    &\vspace{-0.3cm}\\
    \hline \hline
 \end{tabular}
\end{center}
\caption{\label{tab:one-loop-qq} \it  As in Table
\ref{tab:one-loop-gg} but for the $q\bar{q}\to e^+\nu_e \, \mu^-
\bar{\nu}_\mu \, b\bar{b} \, \tau^+ \tau^-$ partonic subprocess. }
\end{table}
%=============================================
\begin{table}[t]
\begin{center}
  \begin{tabular}{cccc}
       \hline \hline
    &\vspace{-0.3cm}\\
    \textsc{Partonic}& \textsc{Number Of}
  &\textsc{Number Of}&  \textsc{Number Of}\\
\textsc{Subprocess} &\textsc{Feynman Diagrams} & \textsc{CS Dipoles}  
    & \textsc{NS Subtractions} \\
   &\vspace{-0.3cm}\\ 
    \hline\hline
    &\vspace{-0.3cm}\\
    $gg\to e^+ \nu_e \, \mu^- \bar{\nu}_\mu \, b\bar{b} \,
    \tau^+ \tau^- \,g$
&  12378  & 27  &  9 \\
    $gq\to e^+ \nu_e \,\mu^- \bar{\nu}_\mu \, b\bar{b} \,
    \tau^+ \tau^-\, q$ 
& 6416 & 15 & 5\\
    $g\bar{q}\to e^+ \nu_e \,\mu^- \bar{\nu}_\mu \, b\bar{b}
    \, \tau^+
  \tau^- \,\bar{q}$ 
& 6416 & 15 & 5 \\
    $q\bar{q}\to e^+ \nu_e \, \mu^- \bar{\nu}_\mu \, b\bar{b} \,
    \tau^+ \tau^-\,  g$
    &  6416 & 15 &  5\\
    &\vspace{-0.3cm}\\
    \hline \hline
 \end{tabular}
\end{center}
\caption{\label{tab:real-emission} \it  List of partonic
subprocesses contributing to the subtracted real emission at ${\cal
O}(\alpha_s^3 \alpha^6)$ for the $pp\to e^+ \nu_e \,\mu^-
\bar{\nu}_\mu \, b\bar{b} \, \tau^+ \tau^- +X$ process where $q = u, d, c,
s$. Also shown are the number of Feynman diagrams, as well as the
number of Catani-Seymour and Nagy-Soper subtraction terms that
correspond to these partonic subprocesses.}
 \end{table}
%=============================================
%
 
For the real corrections the generic subprocesses are listed in Table
\ref{tab:real-emission} where again $q$ stands for up- or down-type
quarks. All subprocesses include all possible contributions of the
order of ${\cal O}(\alpha_s^3\alpha^6)$. The complex mass scheme for
unstable top quarks and gauge bosons has been implemented in complete
analogy to the LO case. For the calculation of the real emission
contributions, the  \textsc{Helac-Dipoles} package is employed. It
implements the Catani-Seymour dipole formalism
\cite{Catani:1996vz,Catani:2002hc} for arbitrary helicity eigenstates
and colour configurations of the external partons
\cite{Czakon:2009ss}. Furthermore, it comprises the Nagy-Soper
subtraction scheme \cite{Bevilacqua:2013iha}, which makes use of
random polarisation and colour sampling of
the external partons. In Table \ref{tab:real-emission} we provide
additionally the number of Catani-Seymour dipoles and Nagy-Soper
subtraction terms. Even though we employ the Nagy-Soper subtraction scheme 
for the full off-shell calculation, we show numbers 
for both schemes to underline the difference between them.
Indeed, the difference between the number of Catani-Seymour dipoles 
and Nagy-Soper subtraction terms corresponds to the total number of 
possible spectators that are only relevant in the Catani-Seymour 
subtraction scheme. To check our calculation we  have explored the 
independence of the real emission results on the unphysical cutoff in 
the dipole subtraction phase space, see e.g.  
\cite{Nagy:1998bb,Nagy:2003tz,Bevilacqua:2009zn,Czakon:2015cla}.

The \textsc{Helac-1Loop} program and \textsc{Helac-Dipoles} are part
of the \textsc{Helac-NLO} framework
\cite{Bevilacqua:2011xh}. Theoretical predictions obtained with the
help of the \textsc{Helac-NLO} software are stored in the form of
modified Les Houches Event Files \cite{Alwall:2006yp} and ROOT Ntuples
\cite{Antcheva:2009zz}. Incorporating ideas outlined in
Ref. \cite{Bern:2013zja} we store each ``event'' with 
supplementary matrix element and PDF information. This allows us to
obtain results for different scale settings and PDF choices by
reweighting.  Furthermore, storing ``events'' has clear advantages when
different observables and/or binning are needed, and when different,
more exclusive, sets of selection cuts are to be used. Indeed, no
additional time-consuming code running is required in such cases.

% =============================================
%
 \section{Computational setup}
 \label{setup}
%
% =============================================

We consider the $pp\to e^+ \nu_e \,\mu^- \bar{\nu}_\mu \, b\bar{b} \,
\tau^+ \tau^- +X$ process at the LHC Run II center-of-mass energy of
$\sqrt{s}=13$ TeV. Specifically, we calculate $\alpha_s$ corrections
to the born-level process at ${\cal O}(\alpha_s^2\alpha^6)$. We only
simulate decays of the weak bosons to different lepton generations.
Interference effects related to same-flavor leptons, however, are at the
permille level for inclusive cuts. We have checked this by an explicit
leading-order calculation for the $pp\to e^+ \nu_e \, e^- \bar{\nu}_e \,
b\bar{b} \, e^+ e^-+X$, $pp\to e^+ \nu_e \,\mu^- \bar{\nu}_\mu \,
b\bar{b} \, e^+ e^- +X$ as well as $pp\to e^+ \nu_e \, e^-
\bar{\nu}_e\, b\bar{b} \, \tau^+ \tau^-+X$ process and found differences
up to $0.2\%$ only. In our calculations the Cabibbo-Kobayashi-Maskawa
mixing matrix is kept diagonal. The unstable particles, top quarks,
$W$ and $Z$ gauge bosons, are treated within the complex-mass scheme. 
The Higgs boson and initial state bottom quark contributions are neglected, 
as explained in the previous section. The Standard Model parameters are 
given  within the $G_\mu$ scheme 
\begin{equation}
\begin{array}{lll}
 G_{ \mu}=1.16638 \cdot 10^{-5} ~{\rm GeV}^{-2}\,, &\quad \quad \quad
                                                     \quad \quad
&   m_{t}=173 ~{\rm GeV} \,,
\vspace{0.2cm}\\
 m_{W}= 80.351972  ~{\rm GeV} \,, &
&\Gamma^{\rm NLO}_{W} =  2.0842989  ~{\rm GeV}\,, 
\vspace{0.2cm}\\
  m_{Z}=91.153481   ~{\rm GeV} \,, &
  &\Gamma^{\rm NLO}_{Z} = 2.4942664~{\rm GeV}\,.
\end{array}
\end{equation}
All other particles, including bottom quarks and $\tau$ leptons, are
considered massless. The top-quark  width is treated as a fixed
parameter throughout this work and its value corresponds to a fixed
scale $\mu_R = m_t$ that is most natural for top-quark decays. The
$\alpha_s(m_t)$ parameter used in the calculation of $\Gamma_t^{\rm
  NLO}$ is independent of $\alpha_s(\mu_0)$ that
goes into the matrix element calculations as well as PDFs, since the
latter describes the dynamics of the whole process. Computed  for
unstable $W$ bosons while  neglecting the bottom-quark mass the 
top-quark width  reads
\begin{equation}
\begin{array}{lll}
  \Gamma_{t}^{\rm LO} =  1.443303   ~{\rm GeV}\,, &
\quad \quad \quad \quad \quad \quad   \quad \quad \quad  
  &\Gamma_{t}^{\rm
     NLO} =  1.3444367445  ~{\rm GeV}\,.               
\end{array}
\end{equation}
For the NWA case with on-shell $W$ gauge boson, in the limit 
$\Gamma_W/m_W \to 0$, we obtain instead
\begin{equation}
\begin{array}{lll}
  \Gamma_{t, \,{\rm NWA}}^{\rm LO} =  1.466332    ~{\rm GeV}\,,
  &\quad \quad \quad \quad \quad \quad  \quad \quad
  &\Gamma_{t,\, {\rm NWA}}^{\rm
     NLO} =  1.365888   ~{\rm GeV}\,.              
\end{array}
\end{equation}
The LO and NLO top-quark widths are computed using formulas from
Refs. \cite{Jezabek:1988iv,Basso:2015gca}. All final-state $b$ and light 
quarks as well as
gluons with pseudorapidity $|\eta| < 5$ are recombined into jets with
the separation parameter $R = 0.4$ in the rapidity-azimuthal angle
plane via the IR-safe $anti$-$k_{T}$ jet clustering algorithm
\cite{Cacciari:2008gp}. Moreover, we impose additional cuts on the
transverse momenta and the rapidity of  $b$-jets
\begin{equation}
\begin{array}{lllll}
  p_{T,\,b} > 25 ~{\rm GeV} \,,
  &\quad \quad \quad  \quad &|y_b| < 2.5 \,,
  &\quad \quad \quad  \quad &  \Delta R_{bb} > 0.4\,,
\end{array}
\end{equation}
where $b$ stands for the two $b$-jets. The following selection
criteria are imposed to ensure that all charged leptons  are observed
inside the detector and are well separated from each other
\begin{equation}
\begin{array}{lllll}
  p_{T,\,\ell} > 20 ~{\rm GeV} \,,
  &\quad \quad \quad  \quad &|y_\ell| < 2.5 \,,
  &\quad \quad \quad  \quad &  \Delta R_{\ell\ell} > 0.4\,,
\end{array}
\end{equation}
where $\ell = e,\,\mu,\,\tau$. The minimum on the missing transverse 
momentum from undetected neutrinos is set to be
\begin{equation}
  p_T^{miss} > 40 ~{\rm GeV}\,.
\end{equation}
We set no restrictions on the kinematics of the extra light jet. We
require $2$ $b$-jets, $4$ charged leptons and missing transverse
momentum. The applied cuts are motivated by  recent
experimental analyses from ATLAS \cite{ATLAS:2021fzm} and CMS
\cite{CMS:2019too}. Following the recommendations of PDF4LHC
\cite{Butterworth:2015oua} we consistently use the NNPDF3.1
\cite{Ball:2017nwa} sets of parton distribution functions. The
two-loop (one-loop) running of $\alpha_s$ at NLO (LO) is provided by
the LHAPDF interface \cite{Buckley:2014ana}. Both the LO and NLO PDF
sets are obtained with $\alpha_s(m_Z) = 0.118$. The number of active
flavours is set to $N_F = 5$.  Unless stated otherwise, all results
are presented using the NNPDF3.1 PDF set. Nevertheless, to assess the
differences among various PDF sets we will also present our findings 
for the  integrated fiducial NLO cross sections as obtained with 
the following PDF sets: CT10 \cite{Lai:2010vv}, CT14 \cite{Dulat:2015mca},
CT18 \cite{Hou:2019efy}, MSTW2008 \cite{Martin:2009iq}, MMHT14
\cite{Harland-Lang:2014zoa}, MSHT20 \cite{Bailey:2020ooq}, NNPDF3.0
\cite{NNPDF:2014otw}, NNPDF4.0 \cite{Ball:2021leu} and ABMP16
\cite{Alekhin:2018pai}. For the central value of the renormalisation 
and factorisation scales we assume $\mu_R = \mu_F=\mu_0$. The scale
uncertainties, however, are estimated  by varying $\mu_R$ and $\mu_F$
independently in the following range
\begin{equation}
\label{variation}
  \left(
    \frac{\mu_R}{\mu_0},\frac{\mu_F}{\mu_0}\right) 
    = \left\{
(2,1),(0.5,1),(1,2),(1,1),(1,0.5),(2,2),(0.5,0.5)
 \right\}     \,.
\end{equation}
and choose the minimum and maximum of the resulting cross
sections. The central values of the factorization and
renormalisation scales have been set first to the following fixed
scale setting
\begin{equation}
\mu_0=m_t+\frac{1}{2}m_Z \,,
\end{equation}  
and subsequently the following dynamical scale choice is used 
\begin{equation}
  \mu_0=\frac{1}{3}H_T \,,
\end{equation} 
where $H_T$ is calculated on an event-by-event basis according to 
\begin{equation}
  H_T= p_{T,\,b_1} + p_{T,\,b_2} + p_{T,\,e^+} 
  +p_{T,\,\mu^-}+p_{T,\,\tau^+}
  +p_{T,\,\tau^-} + p_T^{miss}\,.  
\end{equation}
The extra jet, even if resolved, is not included in the definition of the
scale at NLO. Let us stress that the $H_T$ based scale setting 
is blind to the fact that in the $pp \to e^+ \nu_e \,\mu^- \bar{\nu}_\mu 
\, b\bar{b} \, \tau^+ \tau^- $ process Feynman diagrams 
with one or two top-quark resonances might appear. Nor does it assume 
anything about the $Z$ boson. In other words, information about the 
underlying  resonant structure of the event is not used at all.

% =============================================
%
\section{Integrated fiducial cross sections}
\label{integrated}
%
% =============================================

%=============================================
\begin{figure}[t]
  \begin{center}
\includegraphics[width=0.49\textwidth]{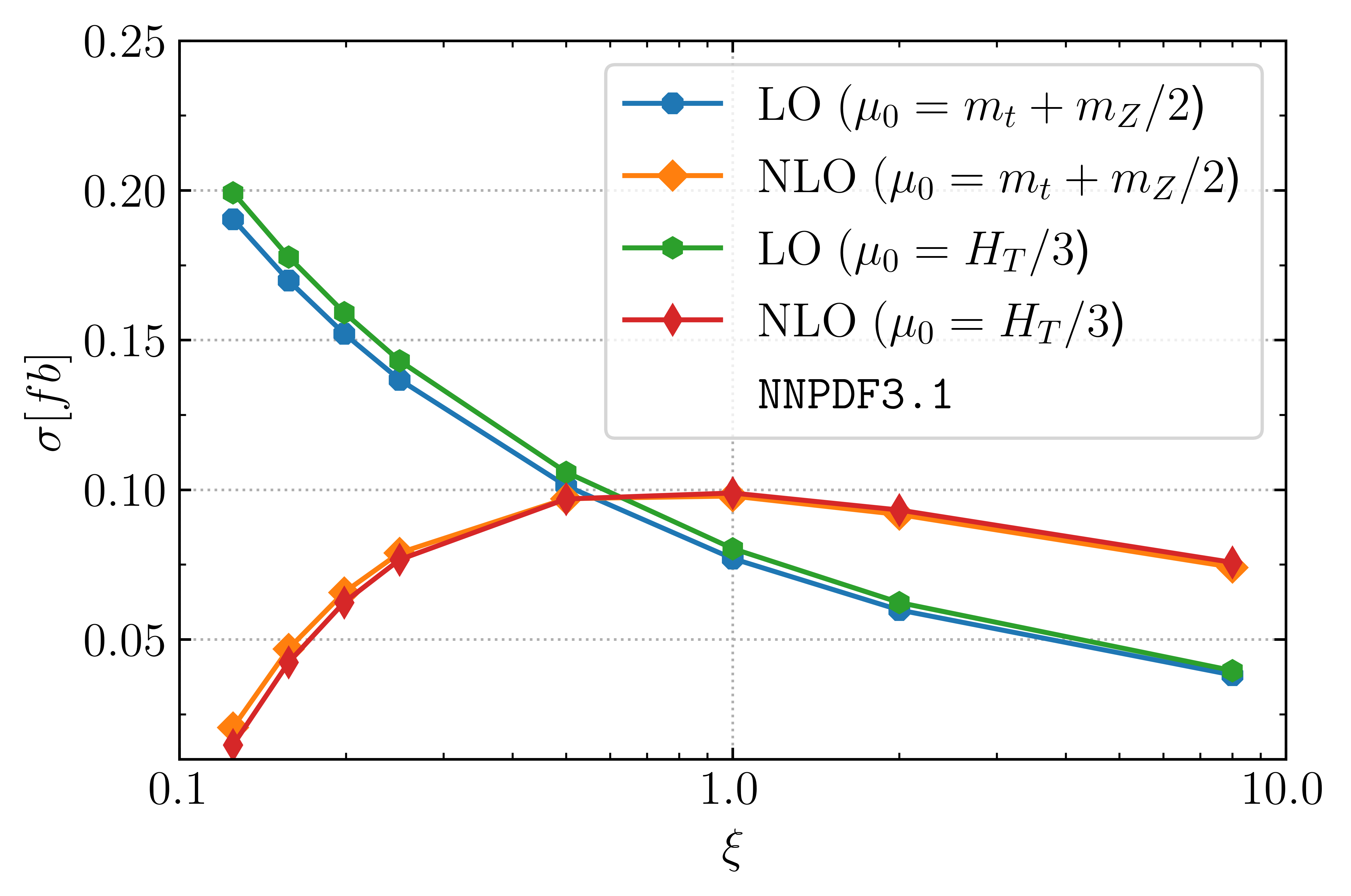}\\
\includegraphics[width=0.49\textwidth]{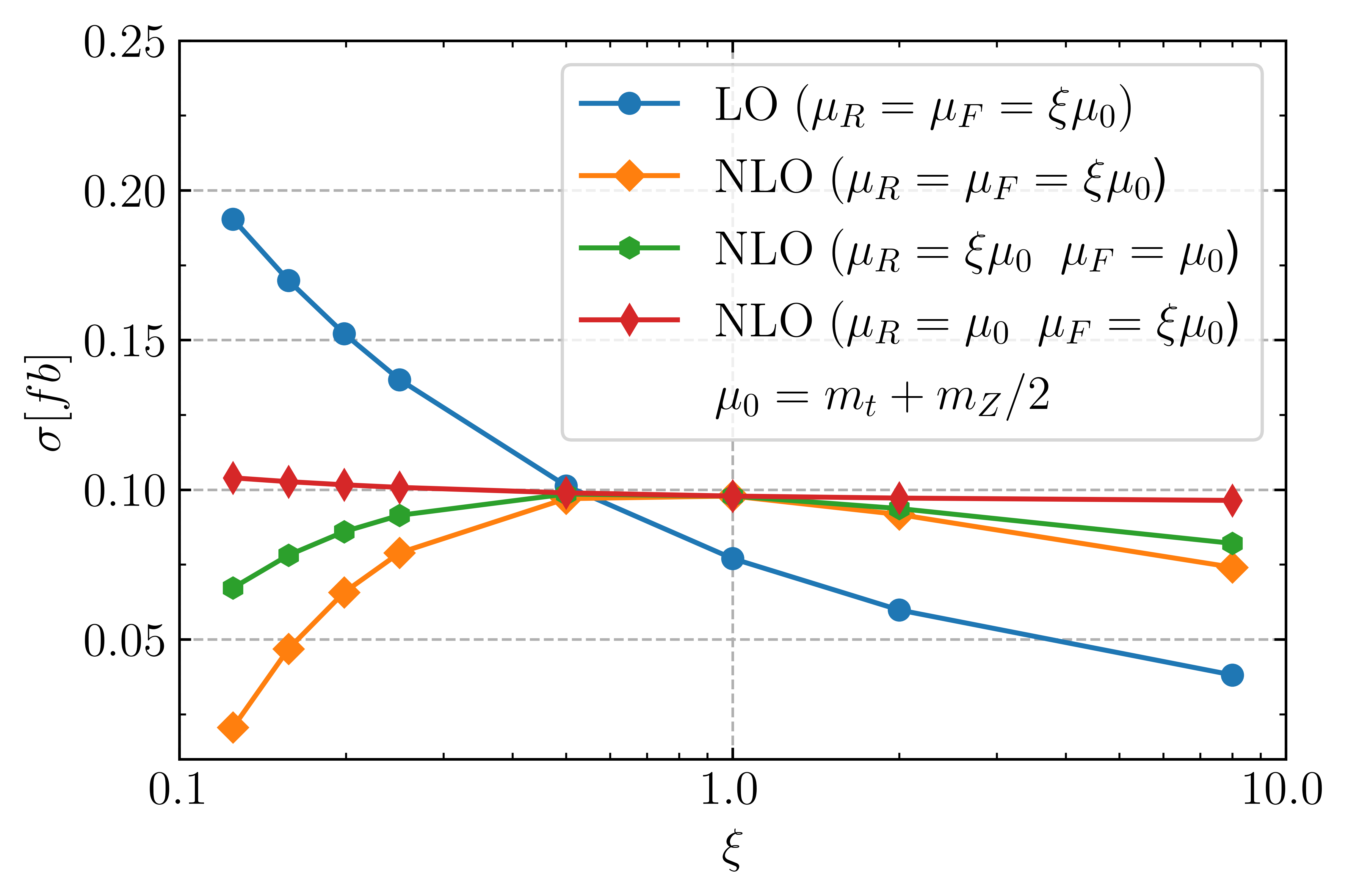}
\includegraphics[width=0.49\textwidth]{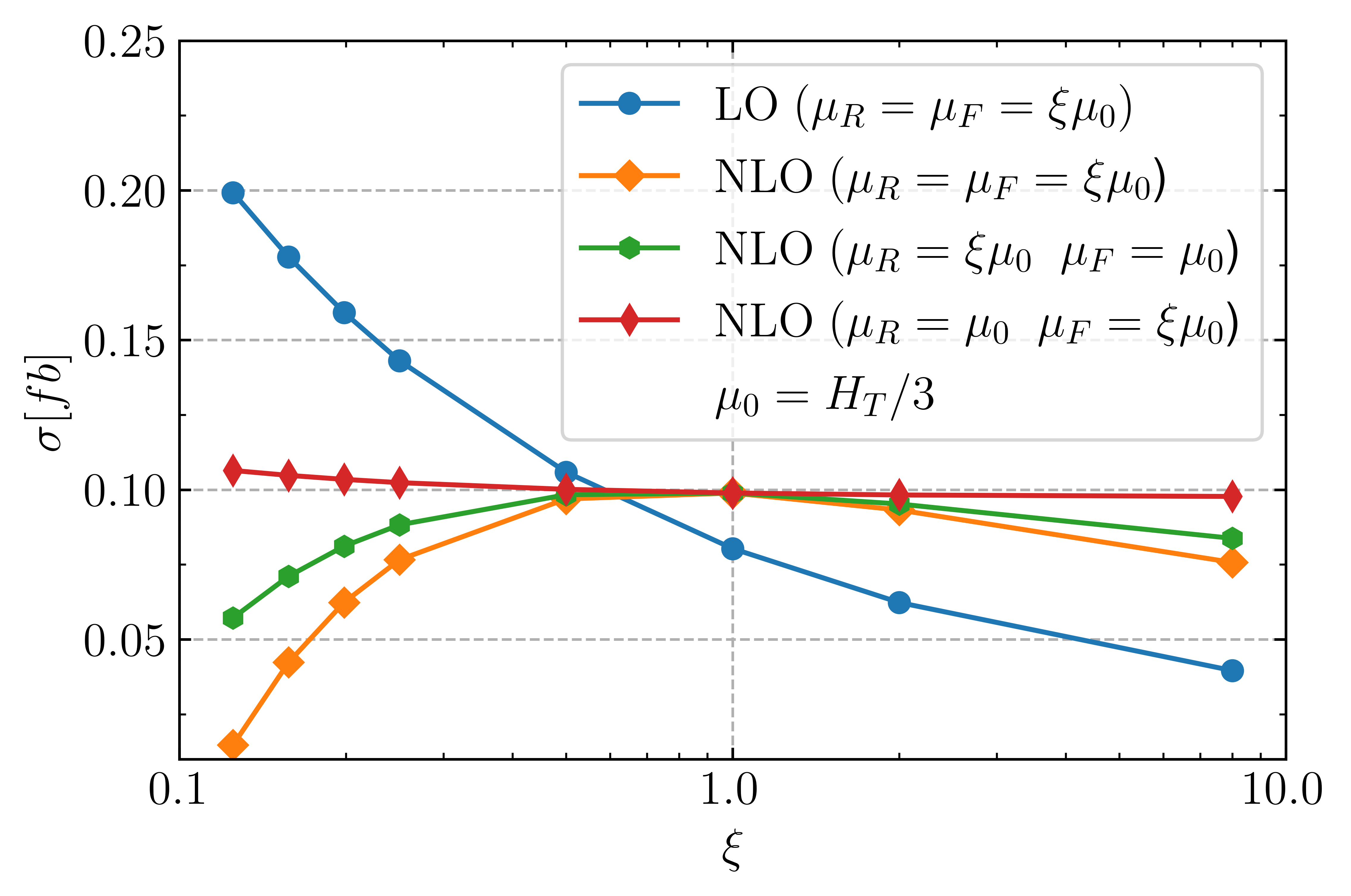}
\end{center}
\caption{\label{fig:scale} \it Scale dependence of the LO and NLO
integrated fiducial cross sections for $pp \to e^+\nu_e\, \mu^-
\bar{\nu}_\mu\,b\bar{b} \, \tau^+\tau^-+X$ at the LHC with $\sqrt{s} =
13$ TeV. In the upper plot renormalisation and factorisation scales 
are set to $\mu_R=\mu_F=\xi\mu_0$ where $\mu_0=m_t+m_Z/2$ and $\mu_0 =
H_T/3$. The LO and NLO NNPDF3.1 PDF sets are employed. In the two lower plots, the variation of $\mu_R$ with fixed $\mu_F$ and the reverse case 
are given for each case of $\mu_0$.}
\end{figure}
%=============================================
\begin{figure}[t]
  \begin{center}
    \includegraphics[width=0.7\textwidth]{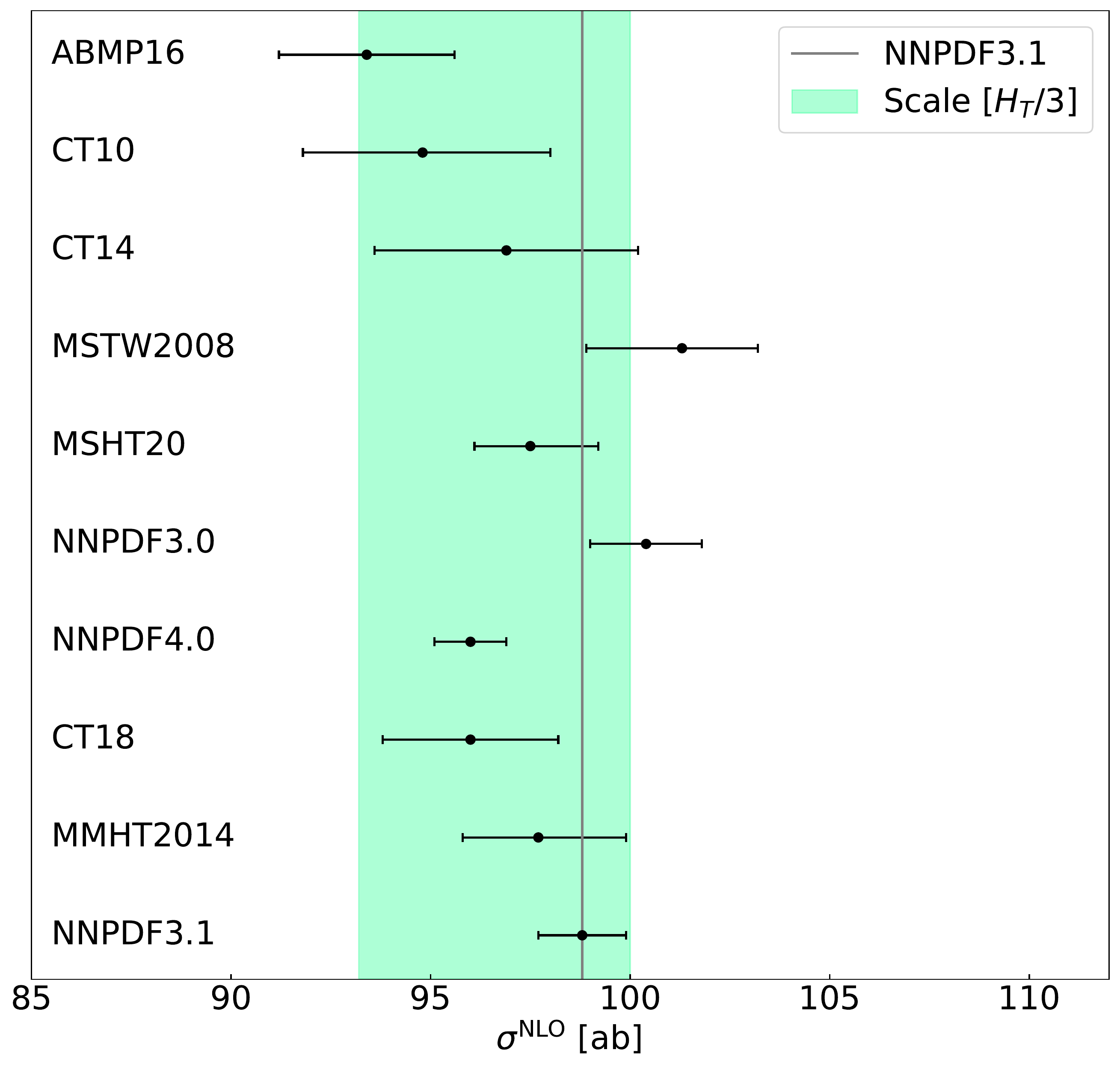}
\end{center}
\caption{\label{fig:pdf-all} \it
  NLO integrated fiducial cross sections for $pp \to e^+\nu_e\,
  \mu^-\bar{\nu}_\mu\, b\bar{b}\, \tau^+\tau^-+X$ at the LHC with
  $\sqrt{s}= 13$  TeV. Theoretical results are provided for
  $\mu_R=\mu_F= \mu_0=H_T/3$ as well as for various PDF sets. Scale
  uncertainties are reported for the NNPDF3.1 PDF set and are shown as
  a blue band.}
\end{figure}
%=============================================
\begin{figure}[t!]
  \begin{center}
    \includegraphics[width=0.7\textwidth]{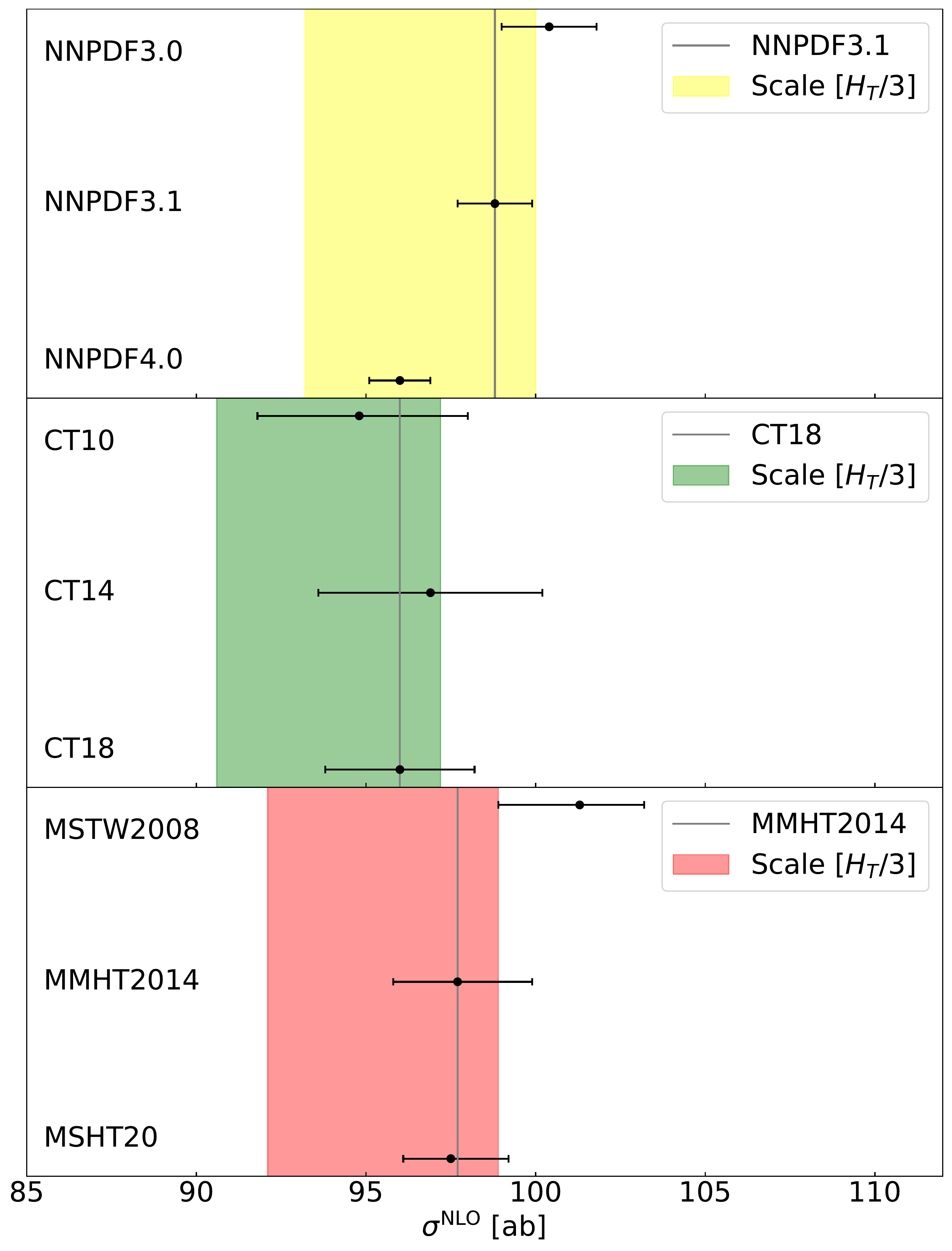}
\end{center}
\caption{\label{fig:pdf-individual} \it  NLO integrated fiducial
  cross sections for $pp \to e^+\nu_e\,
  \mu^-\bar{\nu}_\mu\, b\bar{b}\, \tau^+\tau^-+X$ at the LHC with
  $\sqrt{s}= 13$  TeV. Theoretical results are provided for
  $\mu_R=\mu_F= \mu_0=H_T/3$ as well as for various PDF sets. Scale
  uncertainties are reported for the NNPDF3.1 PDF set (yellow band),
  for CT18 (green band) as well as for MMHT2014 (red band).}
\end{figure}
% =============================================

We begin our presentation of the results of our analysis with a
discussion of the integrated fiducial cross section at the central
value of the scale $\mu_R=\mu_F=\mu_0=m_t+m_Z/2$. We first note that
at the central scale the $gg$ channel dominates the LO $pp$ cross section
by about $65\%$ followed by the $q\bar{q}$ channels with about
$35\%$. With the input parameters and cuts specified in Section
\ref{setup}, we arrive at the following predictions
\begin{equation}
  \begin{split}
\sigma^{\rm LO}_{pp\to e^+ \nu_e \,\mu^- \bar{\nu}_\mu \, b\bar{b} \,
  \tau^+ \tau^- }({\tt NNPDF3.1_{-}lo_{-}as_{-}0118},\mu_0=m_t+m_Z/2) 
  &=
76.98^{\,+24.30\,(32\%)}_{\,-17.17\,(22\%)} \,{\rm ab} \,,\\[0.2cm]
\sigma^{\rm NLO}_{pp\to e^+ \nu_e \,\mu^- \bar{\nu}_\mu \, b\bar{b} \,
  \tau^+ \tau^- }({\tt NNPDF3.1_{-}nlo_{-}as_{-}0118}, \mu_0=m_t+m_Z/2) &=
97.86^{\,+1.08\,(1\%)}_{\,-6.16\,(6\%)} \, {\rm ab}\,.
   \end{split} 
\end{equation} 
The ${\cal K}$-factor, defined as the ratio of NLO to LO cross
section, reads ${\cal K}=1.27$. In our case both LO and NLO integrated
fiducial cross sections are calculated for LO and NLO PDF sets as
obtained with $\alpha_s(m_Z)=0.118$. Had we used the LO NNPDF3.1 PDF
set with $\alpha_s(m_Z) = 0.130$ or the NLO NNPDF3.1 PDF set also for
our LO cross section we would rather have
\begin{equation}
  \begin{split}
\sigma^{\rm LO}_{pp\to e^+ \nu_e \,\mu^- \bar{\nu}_\mu \, b\bar{b} \,
  \tau^+ \tau^- }({\tt NNPDF3.1_{-}lo_{-}as_{-}0130},
\mu_0=m_t+m_Z/2) &=
86.96^{\,+30.13\,(35\%)}_{\,-20.70\,(24\%)} \,{\rm ab} \,,\\[0.2cm]
\sigma^{\rm LO}_{pp\to e^+ \nu_e \,\mu^- \bar{\nu}_\mu \, b\bar{b} \,
  \tau^+ \tau^- }({\tt NNPDF3.1_{-}nlo_{-}as_{-}0118}, \mu_0=m_t+m_Z/2) 
  &=
78.23^{\,+25.49\,(33\%)}_{\,-17.85\,(23\%)} \, {\rm ab}\,.
   \end{split} 
\end{equation} 
These predictions result in a ${\cal K}$-factor of ${\cal K}=1.13$
and ${\cal K}=1.25$ respectively.  Thus, NLO QCD corrections for the
$t\bar{t}Z$ process are below $30\%$ for the fixed scale setting. The
theoretical uncertainties resulting from scale variation taken as a
maximum of the lower and upper bounds are of the order of $32\%$ at LO
and $6\%$ at NLO. Moving from LO to NLO, the theoretical error is reduced 
by a factor larger than $5$. Similar findings are obtained for the 
kinematical dependent scale setting. For $\mu_R=\mu_F=\mu_0=H_T/3$ we have 
\begin{equation}
\label{results-ht}
  \begin{split}
\sigma^{\rm LO}_{pp\to e^+ \nu_e \,\mu^- \bar{\nu}_\mu \, b\bar{b} \,
  \tau^+ \tau^- }({\tt NNPDF3.1_{-}lo_{-}as_{-}0118},
\mu_0=H_T/3) &= 80.32^{\,+25.51\, (32\%)}_{\,-18.02\,(22\%)}
\,{\rm ab} \,,\\[0.2cm]
\sigma^{\rm NLO}_{pp\to e^+ \nu_e \,\mu^- \bar{\nu}_\mu \, b\bar{b} \,
  \tau^+ \tau^- }({\tt NNPDF3.1_{-}nlo_{-}as_{-}0118}, \mu_0=H_T/3) 
  &= 98.88^{\,
  +1.22\, (1\%)}_{\,-5.68\,(6\%)}
\, {\rm ab}\,.
   \end{split} 
\end{equation} 
In this case the ${\cal K}$-factor is slightly smaller, i.e. we have 
${\cal K}=1.23$. Had we used the LO NNPDF3.1 PDF set with 
$\alpha_s(m_Z) = 0.130$ we would rather obtain ${\cal K}=1.09$ and 
finally with the NLO NNPDF3.1 PDF we would have ${\cal K}=1.21$. For 
the $H_T$ based scale choice NLO QCD corrections are reduced, i.e. 
they are below $25\%$. The size of theoretical uncertainties remain 
the same.

Subsequently, we turn our attention to a graphical representation of
the scale dependence for the integrated fiducial LO and NLO cross
sections. In Figure \ref{fig:scale} we present the behaviour of
$\sigma^{\rm LO}$ and $\sigma^{\rm NLO}$ upon varying the central
value of $\mu_R$ and $\mu_F$ by a factor of $\xi\in
\left\{0.125,\dots,8\right\}$.  Both $\mu_0 = m_t+m_Z/2$ and
$\mu_0 = H_T /3$ are shown, allowing us to compare the two scale
settings. Also given in Figure \ref{fig:scale} is the variation of
$\mu_R$ with the fixed value of $\mu_F$ and the variation of $\mu_F$
with fixed $\mu_R$ for each case of $\mu_0$ separately. We note that,
as far as the integrated cross sections are concerned, both scale
settings are similar and might be used in phenomenological
applications. Furthermore, in the range employed for the scale
variation, i.e. when $\xi\in \left\{0.5,\dots,2\right\}$, the scale
uncertainty is driven by the changes in $\mu_R$. Consequently, varying
$\mu_R$ and $\mu_F$ simultaneously up and down by a factor of $2$ around
$\mu_0$ would produce  uncertainty bands very similar to those obtained
using the full prescription of  Eq. \eqref{variation}.

In the next step, while employing the dynamical scale
$\mu_R=\mu_F=\mu_0=H_T/3$, we examine the second main source of 
theoretical uncertainties at NLO in QCD, that comes 
from the choice of PDF. We use the corresponding prescription from each PDF
fitting  group to provide the $68\%$ confidence level (C.L.) PDF
uncertainties. For the default NNPDF3.1 PDF set they are at the $1\%$ 
level. Comparable results have been obtained for NNPDF3.0 and
NNPDF4.0. For the CT family (CT10, CT14 and CT18), on the other hand, 
we have received slightly higher PDF errors, $\delta_{\rm PDF} \in
(2\%-3\%)$. For MSTW2008, MMHT2014 and MSHT20 as well as for ABMP16
the PDF uncertainties are consistently of the order of $\delta_{\rm
PDF} =2\%$. Thus, the NNPDF PDF sets yield the smallest PDF
uncertainties.  Next, we compare the NLO integrated fiducial cross
section obtained with the help of these PDFs to the result generated
with the default setup. The largest difference with respect to our 
central predictions can be noticed for the ABMP16 PDF set.
This result is smaller than the NNPDF3.1 prediction by
$5\%$. For the CT18 (MMHT2014) PDF set we see differences at the 
level of  $3\%$ $(1\%)$, so comparable to the individual estimates 
of PDF systematics. By and large, the scale uncertainties are still 
the dominant source of theoretical error 
($\delta_{\rm scale} =6\%$ at NLO).

A graphical representation of these findings is given in Figure
\ref{fig:pdf-all}, where theoretical results are provided for
$\mu_R=\mu_F= \mu_0=H_T/3$. Scale uncertainties are reported for the
NNPDF3.1 PDF set and are shown as a blue band. At the bottom of the
plot we display the results for the three main global fitting groups:
CT18, MMHT2014 and NNPDF3.1, which are commonly used to provide
theoretical predictions for the LHC Run II. In this way we can see the
relative difference between the various sets and the overall
agreement. As a bonus of our study, in Figure \ref{fig:pdf-individual}
we show the NLO integrated fiducial cross sections for three main
families of PDFs separately. We display NNPDF (NNPDF3.0, NNPDF 3.1 and
NNPDF4.0), CTEQ-TEA (CT10, CT14 and CT18) as well as MSHT (MSTW2008,
MMHT14 and MSHT20). In this way we can track the relative changes from
the older versions to the newest ones within the given set. In the
case of NNPDF sets, scale uncertainties are reported for the NNPDF3.1
PDF set (yellow band), for CTEQ-TEA PDFs they are provided for CT18
(green band) and finally for MSHT scale uncertainties are depicted for
MMHT2014 (red band). For all three cases we obtain $\delta_{\rm
scale}= {}^{\,+1\%}_{\,-6\%}$. While the CT and MSHT PDF families show
perfect agreement within the errors, the NNPDF4.0 result shows a small
tension with the prediction obtained with the NNPDF3.1 PDF set. In other 
words, the evolution from NNPDF3.0 to NNPDF4.0 does not converge equally 
well as observed for the other PDF families. In all three cases, however,  
as expected the errors are reduced  since the newer PDF sets include
more data in their respective fits. On the other hand, when comparing 
the most recent PDF sets, i.e. NNPDF4.0, CT18 and MSHT20, we observe 
that the obtained results are in perfect agreement within the  
corresponding errors.
%
% ============================================
\begin{table}[t]
  \begin{center}
\begin{tabular}{cccccccc}
\hline \hline &&&&&&&\vspace{-0.3cm}\\
\textsc{Scale} & $p_{T,\,b}$ & $\sigma^{\mathrm{LO}}$ [ab] &
 $\delta_{\mathrm{scale}}$
& $\sigma^{\mathrm{NLO}}$ [ab] & $\delta_{\mathrm{scale}}$
& $\delta_{\mathrm{PDF}}$ &  ${\cal K}
=\sigma^{\mathrm{NLO}}/\sigma^{\mathrm{LO}}$ \\
 &&&&&&&\vspace{-0.3cm}\\
\hline \hline &&&&&&&\vspace{-0.3cm}\\
  $\mu_0=m_t+m_Z/2$ & 25 & 76.98 &
${}^{\,+32 \%}_{\,-22 \%}$
& 97.86 & ${}^{\,+1 \%}_{\,-6 \%}$
& ${}^{\,+1 \%}_{\,-1 \%}$ & 1.27 \\[0.2cm]
& 30 & 71.36  & ${}^{\,+32 \%}_{ \,-22 \% }$
& 89.53  & ${}^{\,+1 \% }_{ \,-6 \%}$
& ${}^{\,+1 \%}_{\,-1 \% }$ & 1.25 \\[0.2cm]
& 35 & 65.19 & ${}^{\,+32 \%}_{\,-22 \%}$
& 80.96 & ${}^{\,+1 \%}_{\,-6 \%}$
& ${}^{\,+1 \% }_{ \,-1 \% }$ & 1.24 \\[0.2cm]
 & 40 & 58.84 & ${}^{\,+32 \%}_{\,-22 \%}$
 & 72.42 & ${}^{\,+1 \% }_{ \,-6 \% }$
 & ${}^{\,+1 \% }_{ \,-1 \% }$ & 1.23 \\[0.2cm]
  \hline \hline &&&&&&&\vspace{-0.3cm}\\
    $\mu_0=H_T/3$ & 25 & 80.32
 & ${}^{\,+32 \%}_{\,-22  \%}$
 & 98.88 & ${}^{\,+1 \%}_{\,-6 \%}$
 & ${}^{\, +1 \% }_{ \, -1 \% }$ & 1.23 \\[0.2cm]
 & 30 & 74.22 & ${}^{\, +32 \% }_{ \, -22 \% }$
 & 90.27 & ${}^{\, +1 \% }_{ \, -5 \% }$
  & ${}^{\, +1 \% }_{ \, -1 \% }$ & 1.22 \\[0.2cm]
& 35 & 67.52 & ${}^{\,+32 \% }_{\,-22 \%}$
& 81.51 & ${}^{\,+1 \%}_{\,-5 \%}$
 & ${}^{\,+1 \% }_{\, -1 \% }$ & 1.21 \\[0.2cm]
 & 40 & 60.61 & ${}^{\,+32 \% }_{\, -22 \%}$
 & 72.82 & ${}^{\,+1 \%}_{\,-5 \%}$
 & ${}^{\,+1 \% }_{\,-1 \% }$ & 1.20 \\[0.2cm]
   \hline \hline &&&&&&&\vspace{-0.3cm}\\
\end{tabular}
\end{center}
\caption{\label{tab:2} \it
 LO and NLO integrated fiducial cross sections for the
$pp\to e^+ \nu_e \,\mu^- \bar{\nu}_\mu \, b\bar{b} \, \tau^+ \tau^-+X$
process at the LHC with $\sqrt{s}=13$ TeV. Results are evaluated using
$\mu_R=\mu_F=\mu_0$ where $\mu_0=H_T/3$ and $\mu_0=m_t+m_Z/2$.
The LO and NLO NNPDF3.1 PDF sets are used. We display results for four
different values of the $p_{T,\,b}$ cut. Also given are the theoretical
uncertainties coming from scale variation $(\delta_{\rm scale}) $ and
PDFs $(\delta_{\rm PDF})$. In the last column the ${\cal K}$-factor is
shown. Monte Carlo errors are at the permille or sub-permille level.}
\end{table}
% ============================================
\begin{table}[t]
\begin{center}
\begin{tabular}{ccccccc}
\hline \hline\\[-0.4cm]
$|M_{\tau^+\tau^-} -m_Z|< X$& $\sigma^{\rm LO}$ 
[ab] & $\delta_{\rm scale}$ &
$\sigma^{\rm NLO}$ [ab] & $\delta_{\rm scale}$ & $\delta_{\rm PDF}$
& ${\cal K}=\sigma^{\rm NLO}/\sigma^{\rm LO}$ \\[0.2cm]
\hline\hline\\[-0.4cm]
\multicolumn{7}{c}{$\mu_0=m_t+m_Z/2$} \\ [0.2cm]
\hline \hline \\ [-0.4cm]
  $-$ & $76.98$ & ${}^{\,+32\%}_{\,-22\%}$ & $97.86$
& ${}^{\,+1\%}_{\,-6\%}$
& ${}^{\,+1\%}_{\,-1\%}$ & $1.27$\\[0.2cm]\\ [-0.4cm]
  $25$ GeV & $71.06$ & ${}^{\,+32\%}_{\,-22\%}$
& $90.10$ & ${}^{\,+1\%}_{\,-6\%}$ &
${}^{\,+1\%}_{\,-1\%}$& $1.27$\\[0.2cm]
\\ [-0.4cm]
  $20$ GeV & $70.26$ & ${}^{\,+32\%}_{\,-22\%}$
& $89.07$ & ${}^{\,+1\%}_{\,-6\%}$ &${}^{\,+1\%}_{\,-1\%}$
& $1.27$\\[0.2cm]
\\ [-0.4cm]
  $15$ GeV & $69.10$ & ${}^{\,+32\%}_{\,-22\%}$
& $87.57$ & ${}^{\,+1\%}_{\,-6\%}$ &${}^{\,+1\%}_{\,-1\%}$
& $1.27$\\[0.2cm]
\\ [-0.4cm]
  $10$ GeV & $66.99$ & ${}^{\,+32\%}_{\,-22\%}$ & $84.90$
& ${}^{\,+1\%}_{\,-6\%}$ & ${}^{\,+1\%}_{\,-1\%}$
& $1.27$\\[0.2cm]
\hline\hline\\[-0.4cm]
\multicolumn{7}{c}{$\mu_0=H_T/3$} \\[0.2cm]
\hline \hline \\[-0.4cm]
  $-$ & $80.32$ & ${}^{\,+32\%}_{\,-22\%}$ & $98.88$
& ${}^{\,+1\%}_{\,-6\%}$
& ${}^{\,+1\%}_{\,-1\%}$ & $1.23$\\[0.2cm]
 \\ [-0.4cm]
  $25$ GeV & $74.06$ & ${}^{\,+32\%}_{\,-22\%}$ & $91.00$
 & ${}^{\,+1\%}_{\,-6\%}$ & ${}^{\,+1\%}_{\,-1\%}$
 & $1.23$\\[0.2cm]
 \\ [-0.4cm]
  $20$ GeV & $73.22$ & ${}^{\,+32\%}_{\,-22\%}$
& $89.96$ & ${}^{\,+1\%}_{\,-6\%}$ &
  ${}^{\,+1\%}_{\,-1\%}$
  & $1.23$\\[0.2cm]
\\ [-0.4cm]
  $15$ GeV & $72.00$ & ${}^{\,+32\%}_{\,-22\%}$
 & $88.44$ & ${}^{\,+1\%}_{\,-6\%}$ &
  ${}^{\,+1\%}_{\,-1\%}$
 & $1.23$\\[0.2cm]
 \\ [-0.4cm]
  $10$ GeV & $69.81$ & ${}^{\,+32\%}_{\,-22\%}$ & $85.74$
 & ${}^{\,+1\%}_{\,-6\%}$ & ${}^{\,+1\%}_{\,-1\%}$
& $1.23$\\[0.2cm]
\hline \hline
\end{tabular}
\end{center}
\caption{ \label{tab:3}  \it 
LO and NLO integrated fiducial cross sections for the $pp\to
e^+\nu_e\, \mu^-\bar{\nu}_\mu\, \tau^+ \tau^- \, b\bar{b} +X$ process
at the LHC with $\sqrt{s}=13$ TeV. Results are evaluated using
$\mu_R=\mu_F=\mu_0$ with $\mu_0=m_t+m_Z/2$ and
$\mu_0=H_T/3$. The LO
and NLO NNPDF3.1 PDF sets are used. We display results for four
different values of the $M_{\tau^+\tau^-}$ cut, where
$|M_{\tau^+\tau^-}-m_Z|< X$ with $X\in (25, 20,15,10)$
GeV. Additionally, the LO and NLO predictions without this requirement
are given. The theoretical uncertainties coming from scale variation
$(\delta_{scale})$ and PDFs $(\delta_{\rm PDF})$ are also provided. In
the last column the ${\cal K}$-factor is shown. Monte Carlo errors are
at the permille or sub-permille level.}
\end{table}
% =============================================
\begin{table}[t]
\begin{center}
  \begin{tabular}{ccccc}
    \hline \hline &&&& \vspace{-0.3cm}\\
    \textsc{Scale}
& \textsc{Order} & PDF & $\sigma$ [ab] & ${\cal K}=\sigma^{\rm
NLO}/\sigma^{\rm LO}$ \\ &&&&\vspace{-0.3cm}\\
      \hline \hline
                  &&&&\vspace{-0.3cm}\\
 $\mu_0=m_t+m_Z/2$
 & LO & ${\tt NNPDF3.1_{-}{lo}_{-}{as}_{-}{0130}}$ & $29.74^{\,
     +35\%}_{\,-24\%}$ & $1.06$\\[0.2cm]
    &LO & ${\tt NNPDF3.1_{-}{lo}_{-}{as}_{-}{0118}}$ &
      $26.34^{\,+32\%}_{\,-23\%}$& $1.19$\\[0.2cm]
    &LO& ${\tt NNPDF3.1_{-}{nlo}_{-}{as}_{-}{0118}}$ &
     $26.84^{\,+33\%}_{\,-23\%}$ & $1.17$\\[0.2cm]
    &NLO & ${\tt NNPDF3.1_{-}{nlo}_{-}{as}_{-}{0118}}$ &
      $31.44^{\,+1\%}_{\,-5\%}$            & $-$\\
                  &&&&\vspace{-0.3cm}\\ \hline \hline
                &  &&&\vspace{-0.3cm}\\
  $\mu_0=H_T/3$ & LO &${\tt NNPDF3.1_{-}{lo}_{-}{as}_{-}{0130}}$
     & $30.15^{\,+35\%}_{\,-24\%}$& $1.05$\\[0.2cm]
   & LO &${\tt NNPDF3.1_{-}{lo}_{-}{as}_{-}{0118}}$
    & $26.66^{\, +32\%}_{\,-23\%}$ & $1.18$\\[0.2cm]
   & LO&${\tt NNPDF3.1_{-}{nlo}_{-}{as}_{-}{0118}}$
    & $27.14^{\,+33\%}_{\,-23\%}$ & $1.16$\\[0.2cm]
   & NLO &${\tt NNPDF3.1_{-}{nlo}_{-}{as}_{-}{0118}}$
     &  $31.55^{\,+1\%}_{\,-5\%}$ & $-$\\
&&&&\vspace{-0.3cm}\\ \hline \hline   
\end{tabular}
\end{center}
\caption{\label{tab:1} \it  LO and NLO integrated fiducial cross
sections for the $pp\to e^+ \nu_e \,\mu^- \bar{\nu}_\mu \, b\bar{b} \,
\tau^+ \tau^-+X$ process at the LHC with $\sqrt{s}=13$ TeV. Results are
evaluated for the more exclusive setup, which considers the cuts of
Eq. \eqref{exclusive} on top of the standard kinematical cuts, using
$\mu_R=\mu_F=\mu_0$ where $\mu_0=m_t+m_Z/2$ and $\mu_0=H_T/3$. Three PDF
sets are employed for LO predictions with different values of
$\alpha_s(m_Z)$. In the last column the ${\cal K}$-factor, defined as
$\sigma^{\rm NLO}/\sigma^{\rm LO}$, is shown. Monte Carlo errors are
at the permille or sub-permille level.}
\end{table}
% ============================================

A stability test of LO and NLO fiducial cross sections with respect to
the $b$-jet transverse momentum cut is shown in Table \ref{tab:2}. The
cut is varied in steps of $5$ GeV within the following range:
$p_{T,\,b} \in (25 - 40)$ GeV. We also show theoretical uncertainties
as estimated from the scale variation and from PDFs as well as the
${\cal K}$-factor. We  notice that NLO QCD corrections are almost
constant in size for the $p_{T,\,b}$ cut within the  tested 
range. In addition, NLO QCD predictions show a very stable behaviour
with respect to both sources of  theoretical uncertainties.

In Table \ref{tab:3}, on the other hand, we give LO and NLO integrated
fiducial cross sections for four different values of the cut on the
invariant mass of the $\tau^+\tau^-$ pair. We require
$M_{\tau^+\tau^-}$ to be in the following $Z$ boson mass window,
$|M_{\tau^+\tau^-} -m_Z|< X$, where $X\in \left\{25,20,15,10\right\}$ 
GeV. The latter set covers the values frequently used in different
experimental analyses for the process. This requirement favours 
configurations where the $\tau$ lepton pair originates from a decaying 
$Z$ boson  with possible interferences induced by a very off-shell 
photon in the same mass window.  We compare these
theoretical predictions to the result without this condition. We 
notice  that already for the smallest window, i.e. for
$|M_{\tau^+\tau^-} -m_Z| <$ $10$ GeV, we can recover $87\%$ of the
full $pp$ cross section for the $pp\to
e^+\nu_e\, \mu^-\bar{\nu}_\mu\, \tau^+ \tau^- \, b\bar{b} +X$ process.
 For the enlarged window of $25$ GeV this number is instead 
$92\%$.  Moreover, the NLO QCD corrections are extremely stable with 
respect to this cut, regardless of the window size applied around the 
nominal $Z$ boson mass. We shall use this cut at a later time, when 
we compare our state-of-the-art theoretical predictions with those 
calculated in the NWA. 

Finally, we also provide theoretical predictions for a slightly
modified setup. Specifically, the following more exclusive set of
selection cuts is additionally imposed on the final state charged
leptons and $b$-jets:
\begin{equation}
    \label{exclusive}
  p_{T,\,b} >40 ~{\rm GeV}, \quad \quad \quad \quad \quad
  p_{T,\,\ell} >30 ~{\rm GeV}, \quad \quad \quad \quad \quad
  \Delta R_{b\,\ell} > 0.4\,.
\end{equation}  
 These more exclusive selection cuts are motivated by the
various experimental analysis carried out by ATLAS and CMS
collaborations for the $pp\to t\bar{t}Z$ process, see e.g. Refs.
\cite{CMS:2017ugv,CMS:2019too,ATLAS:2019fwo,ATLAS:2021fzm}.
The LO and NLO integrated fiducial cross sections for both scale
settings calculated with these modified cuts are presented in Table
\ref{tab:1}. When comparing to the default setup we notice the reduced
size of NLO QCD corrections. Specifically, they are now below
$20\%$. With the {\tt{NNPDF3.1-lo-as-0130}} LO PDF set, however, they
are as small as $5\%-6\%$. On the other hand, theoretical
uncertainties due to the scale variation remain the same.  Similarly
to the default setup also here such differences in the $ {\cal
K}$-factor are driven by changes in the LO cross section and in
particular, by changes in the value of $\alpha_s (m_Z) $ used as an
input parameter in the corresponding LO NNPDF3.1 PDF set. We also
note, that employing LO and NLO NNPDF3.1 PDF sets with the same value
of $\alpha_s(m_Z)$ for the LO cross section results in very similar
predictions for the process at hand.

To summarise this part, for the integrated fiducial cross sections,
where effects of the phase-space regions close to the particle
threshold for $t\bar{t}Z$ production dominate, both scale choices that
we employed describe the $pp \to e^+ \nu_e\,\mu^- \bar{\nu}_\mu \,
b\bar{b} \, \tau^+ \tau^-+X$ process very well. They agree within
their respective theoretical errors and have similar theoretical
systematics.  Thus, a choice between a fixed and dynamical scale
setting does not play a crucial role. On the other hand, for the
differential cross sections the off-shell effects of the top quarks
and massive gauge bosons as well as the single- and non-resonant
contributions become more important, making the usage of the dynamical
scale setting a necessity. We will come back to this point in Section
\ref{differential}.

% =============================================
%
\section{Differential fiducial cross sections}
\label{differential}
%
% =============================================

%=============================================
\begin{figure}[t]
  \begin{center}
    \includegraphics[width=0.49\textwidth]{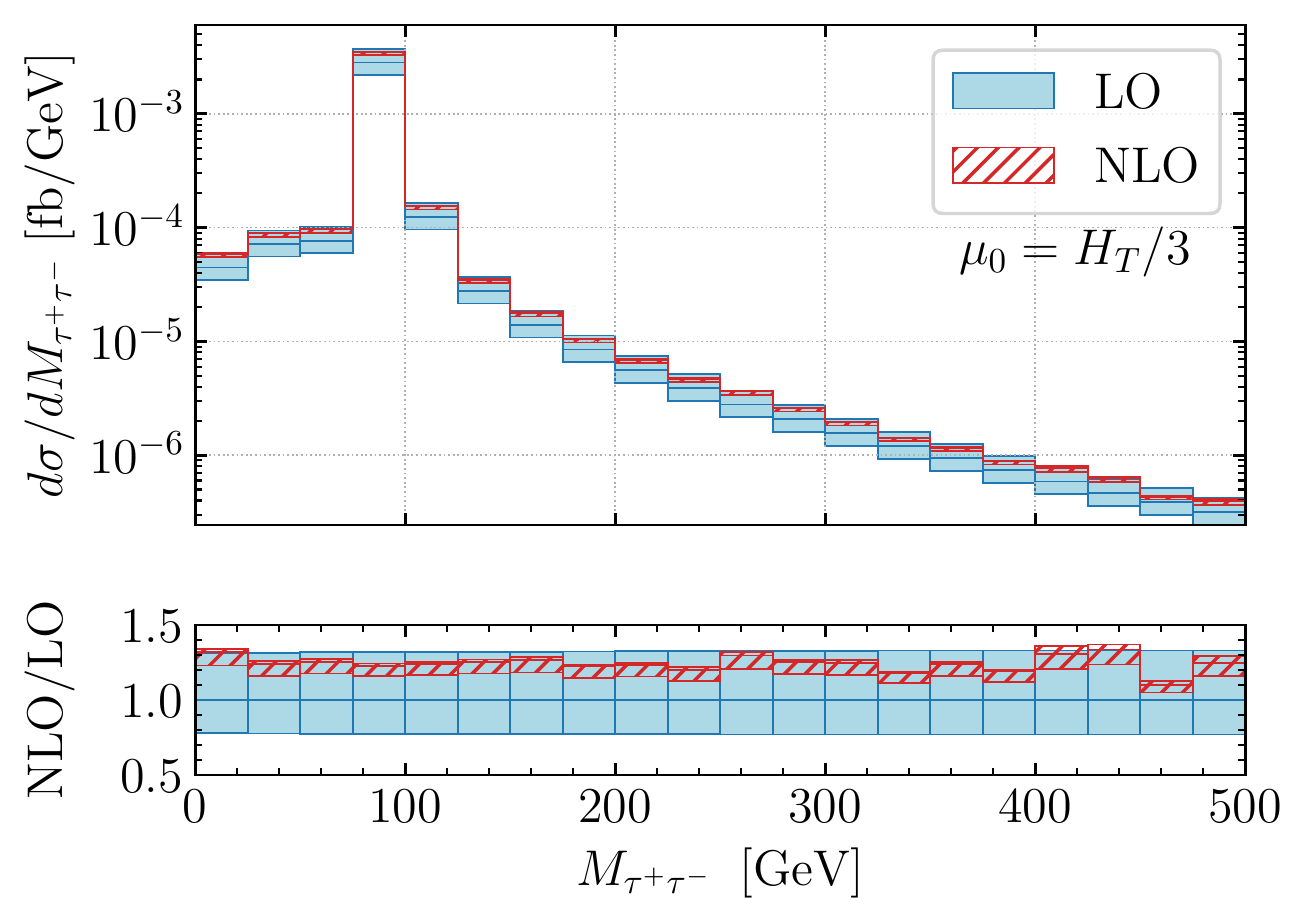}
    \includegraphics[width=0.49\textwidth]{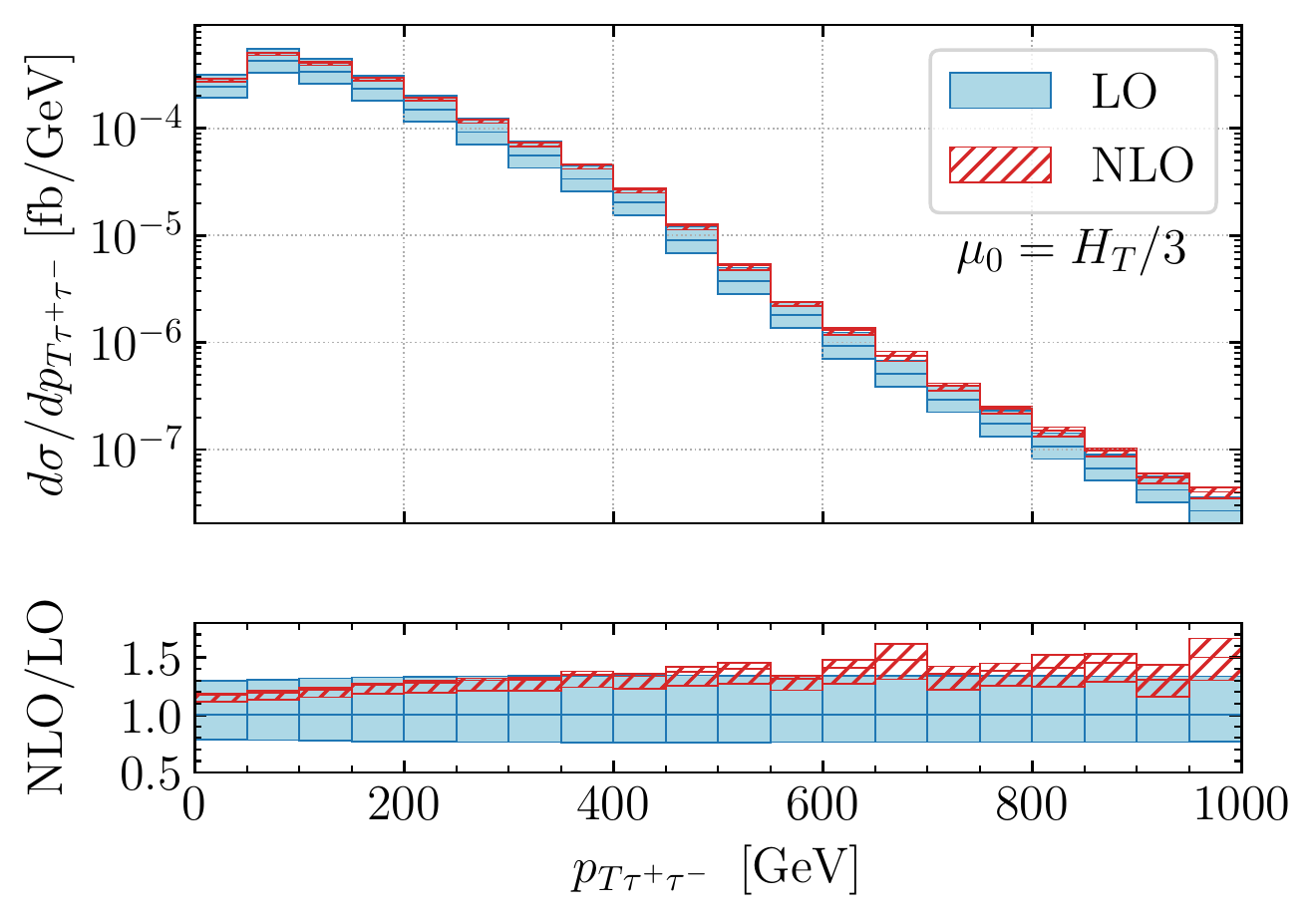}
    \includegraphics[width=0.49\textwidth]{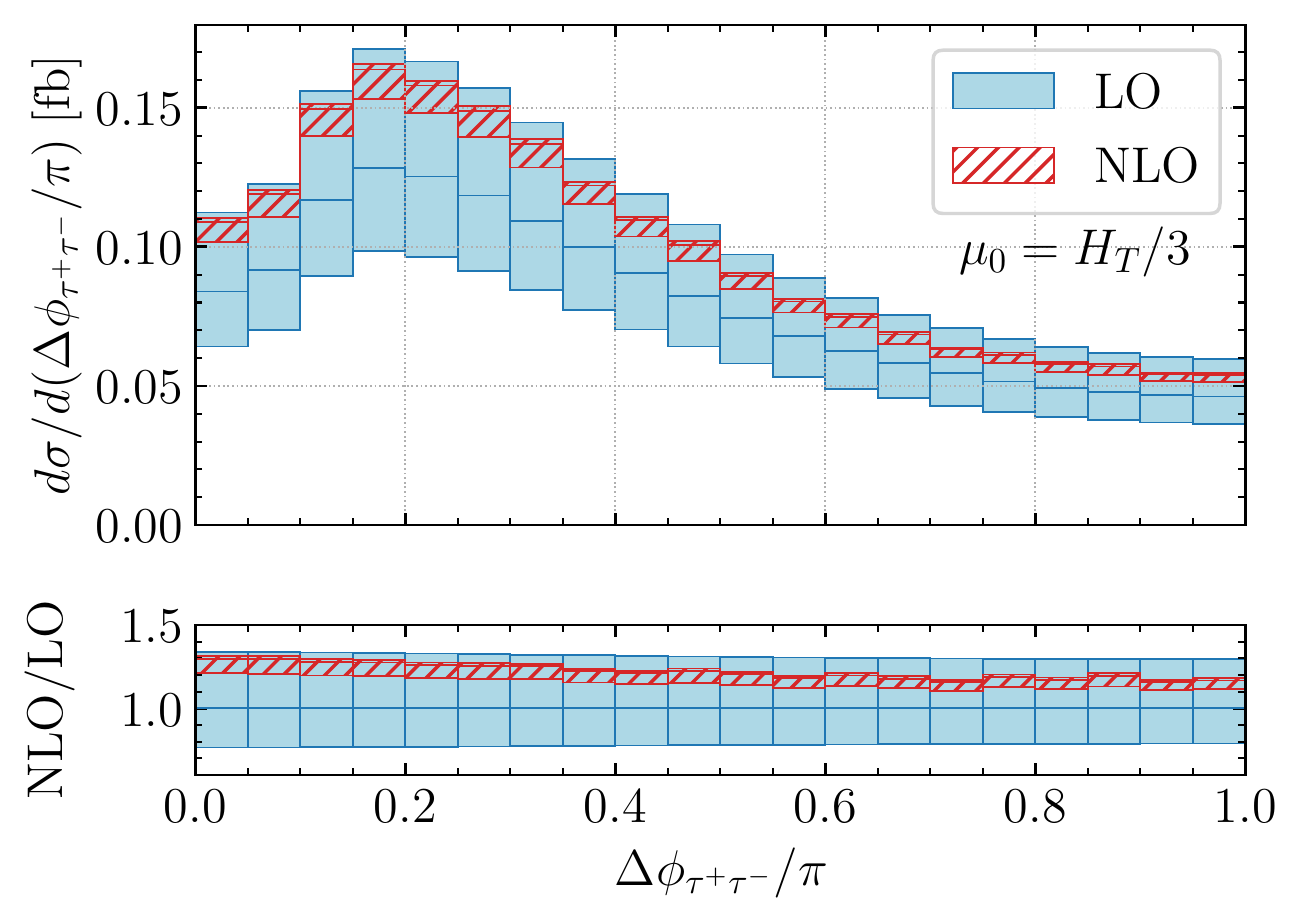}
    \includegraphics[width=0.49\textwidth]{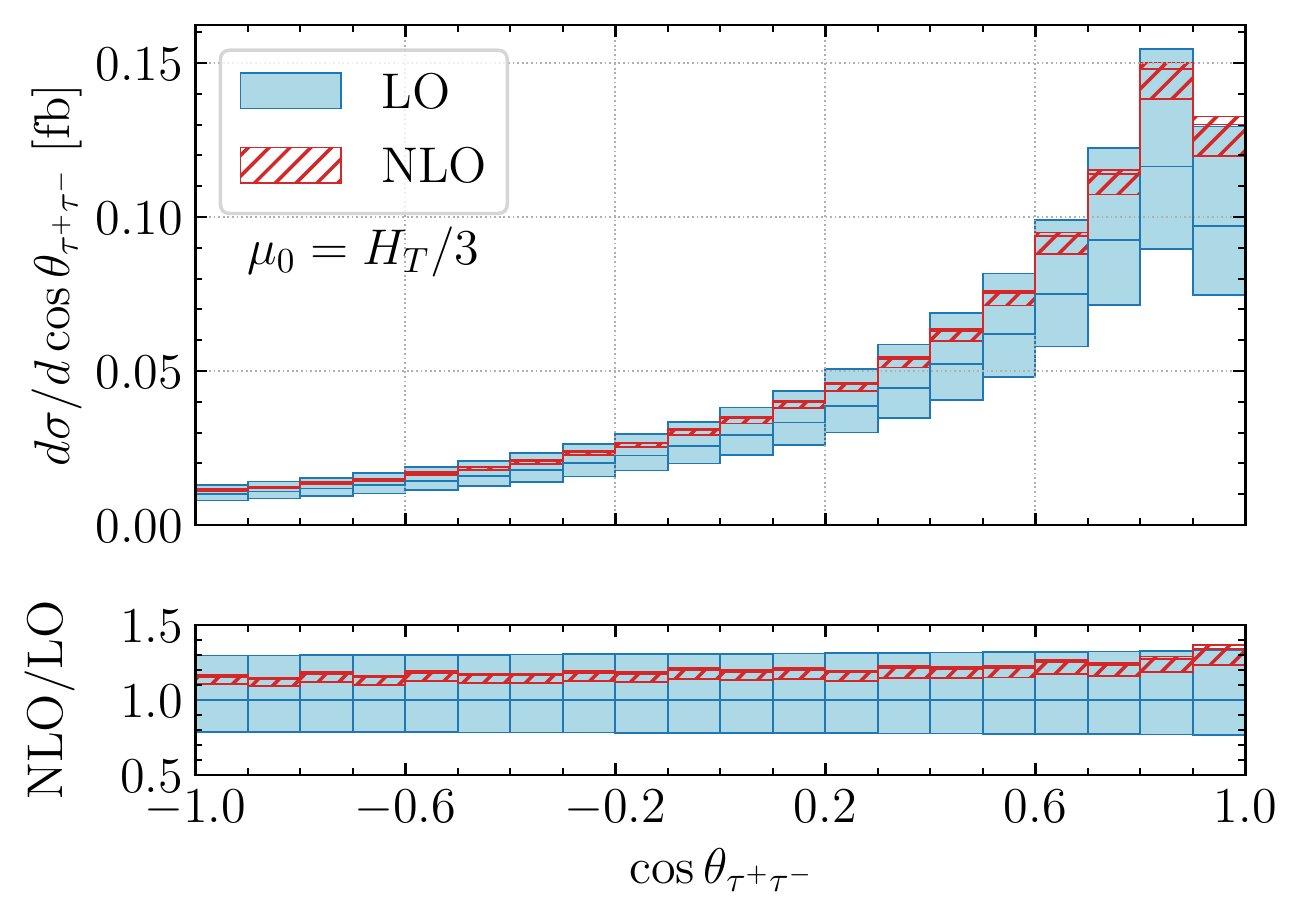}
    \includegraphics[width=0.49\textwidth]{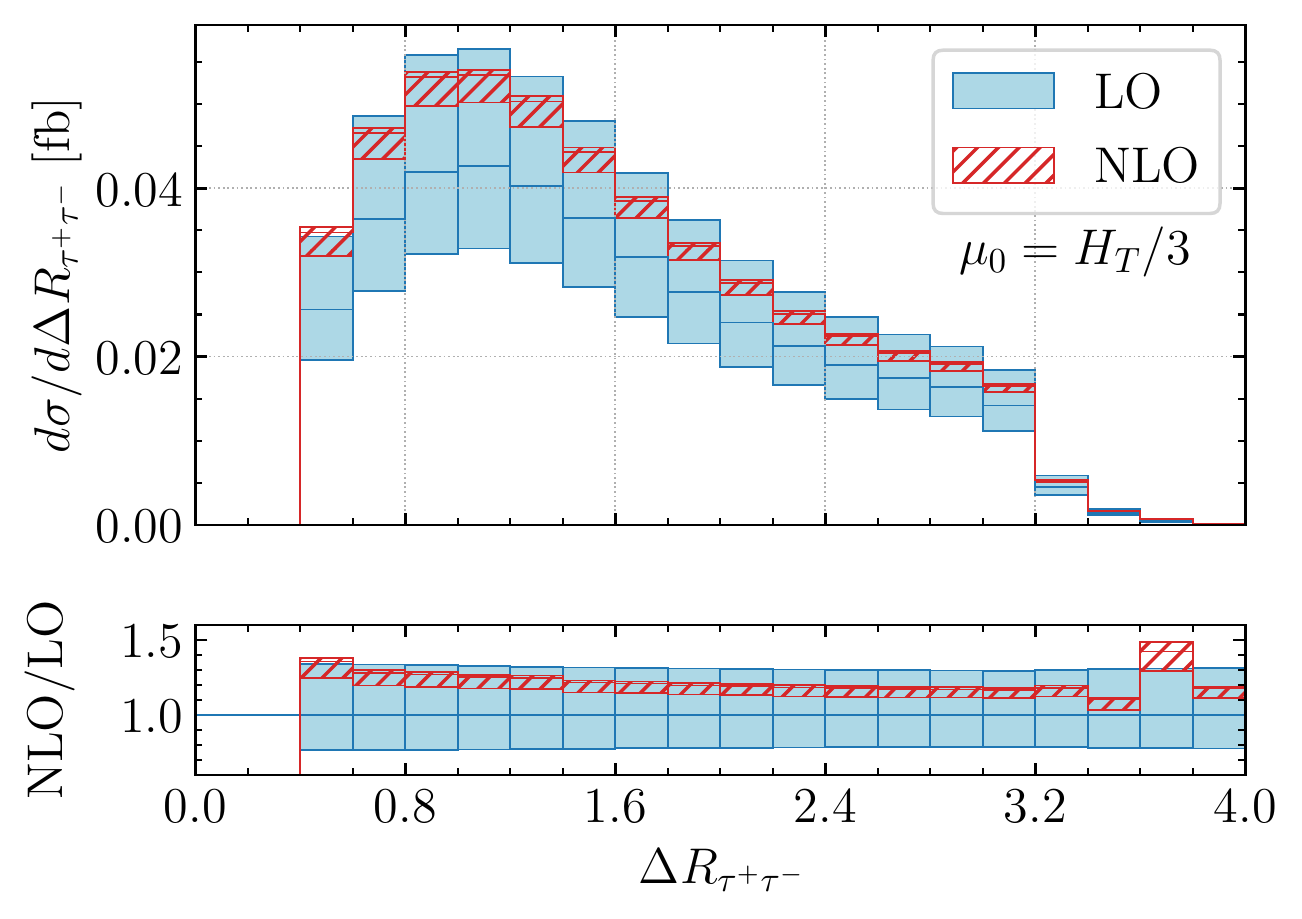}
     \includegraphics[width=0.49\textwidth]{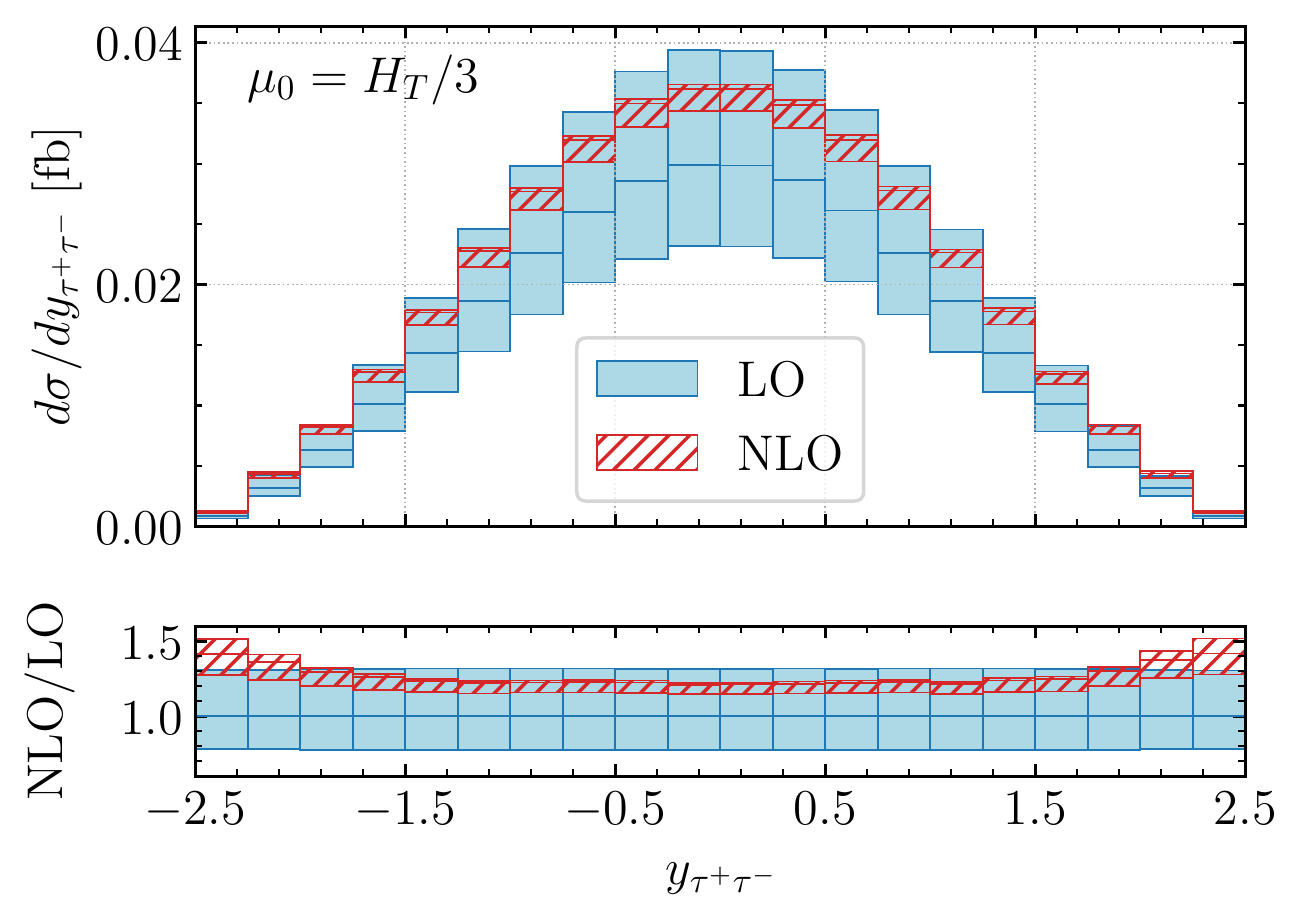}
\end{center}
\caption{\label{fig:diff1} \it Differential cross-section 
  distributions for $pp \to
  e^+ \nu_e \, \mu^- \bar{\nu}_\mu\, b\bar{b}\, \tau^+ \tau^- +X$ at
  the LHC with $\sqrt{s}=13$ TeV as a function of $M_{\tau^+\tau^-}$,
  $p_{T\, \tau^+\tau^-}$, $\Delta \phi_{\tau^+\tau^-}$, $\cos
  \theta_{\tau^+\tau^-}$, $\Delta R_{\tau^+\tau^-}$ and
  $y_{\tau^+\tau^-}$. The blue curve corresponds to the LO and the
  red curve to the NLO result. Also shown are the corresponding
  uncertainty bands resulting from scale variations. The lower panels
  display the differential ${\cal K}$-factor together with the
  uncertainty band and the relative scale uncertainties of the LO
  cross section. The scale choice is  $\mu_R=\mu_F=\mu_0=H_T/3$. The
  cross sections are evaluated with the NNPDF3.1 PDF sets.}
\end{figure}
%=============================================
\begin{figure}[t]
  \begin{center}
     \includegraphics[width=0.49\textwidth]{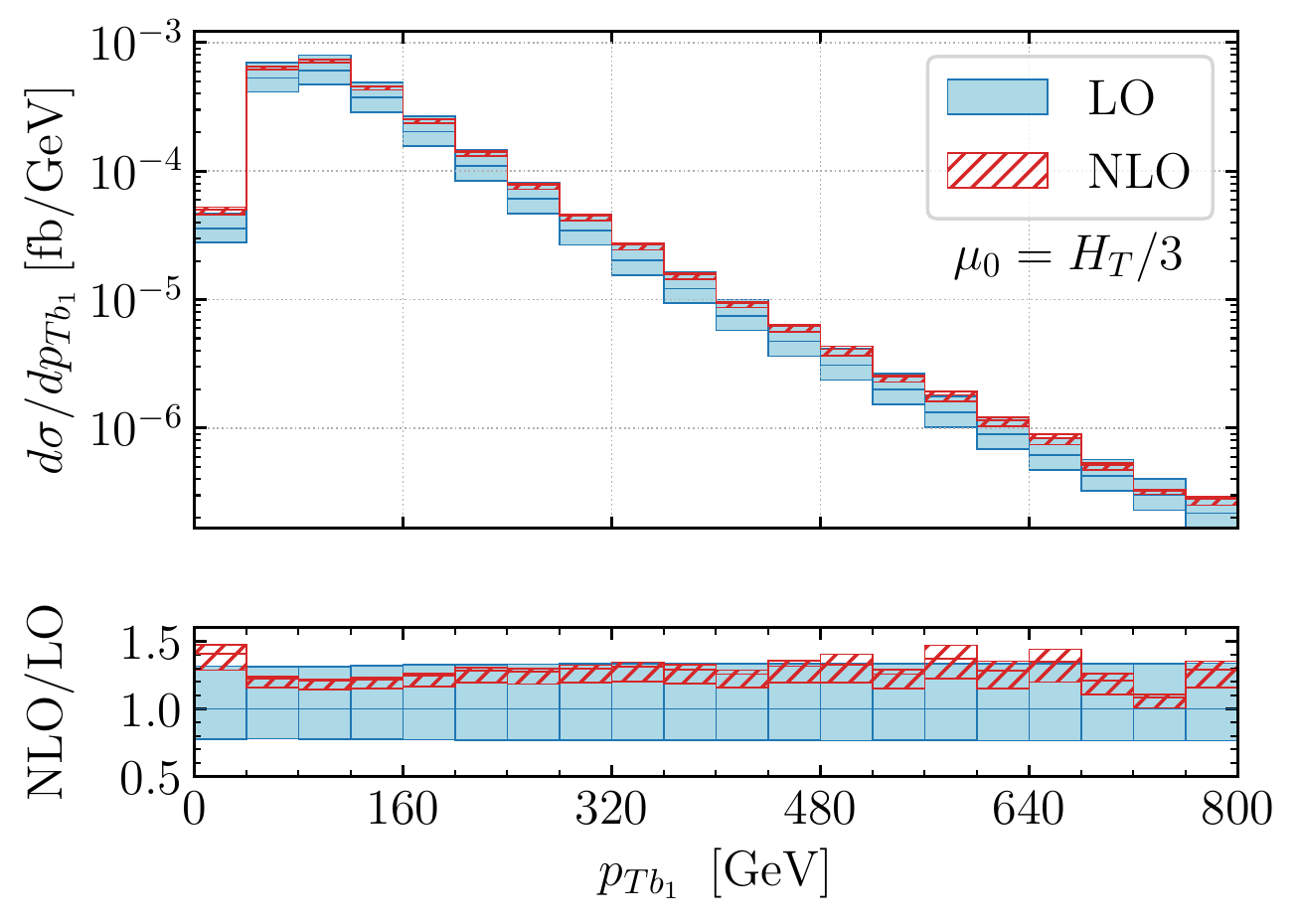}
    \includegraphics[width=0.49\textwidth]{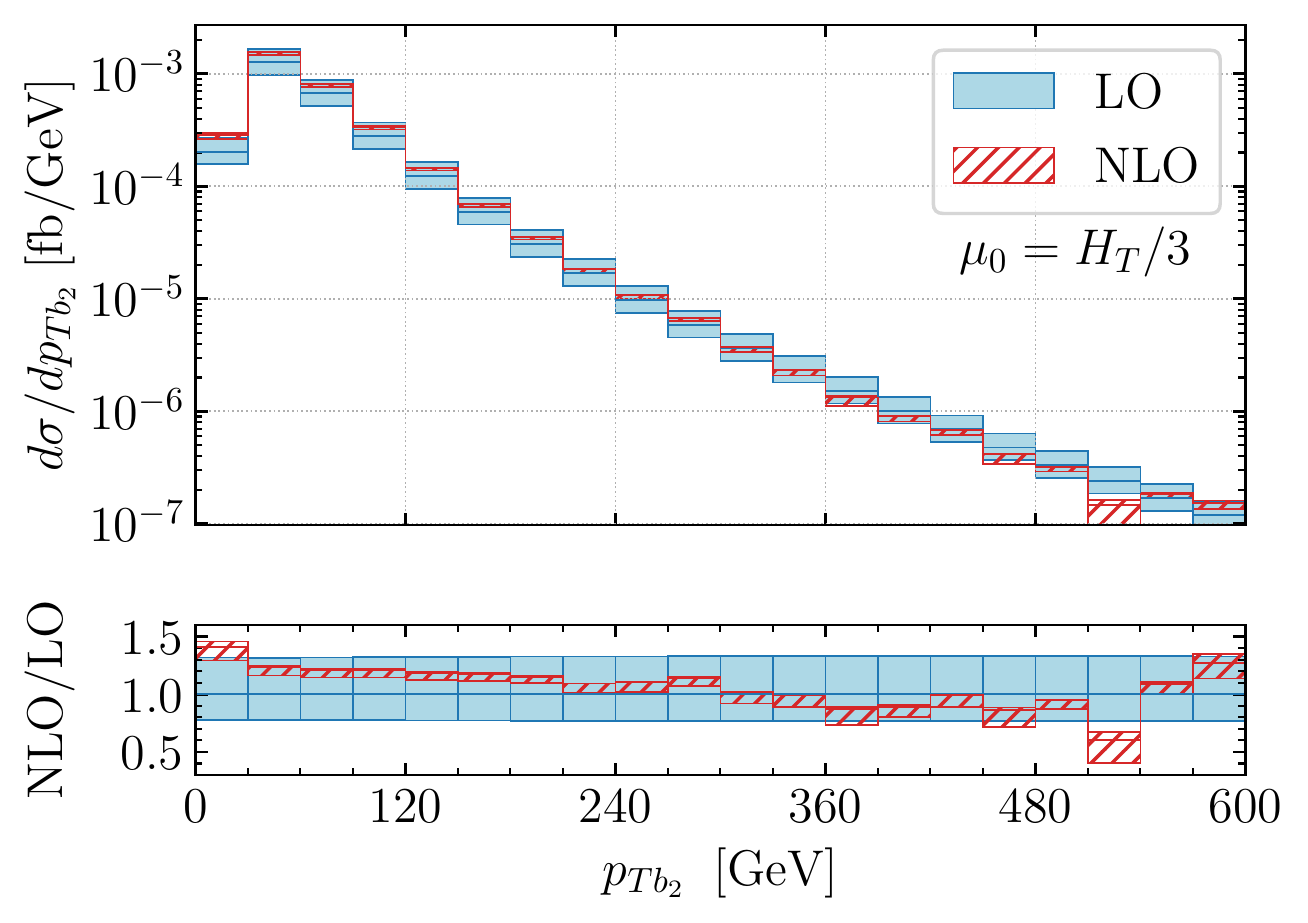}
     \includegraphics[width=0.49\textwidth]{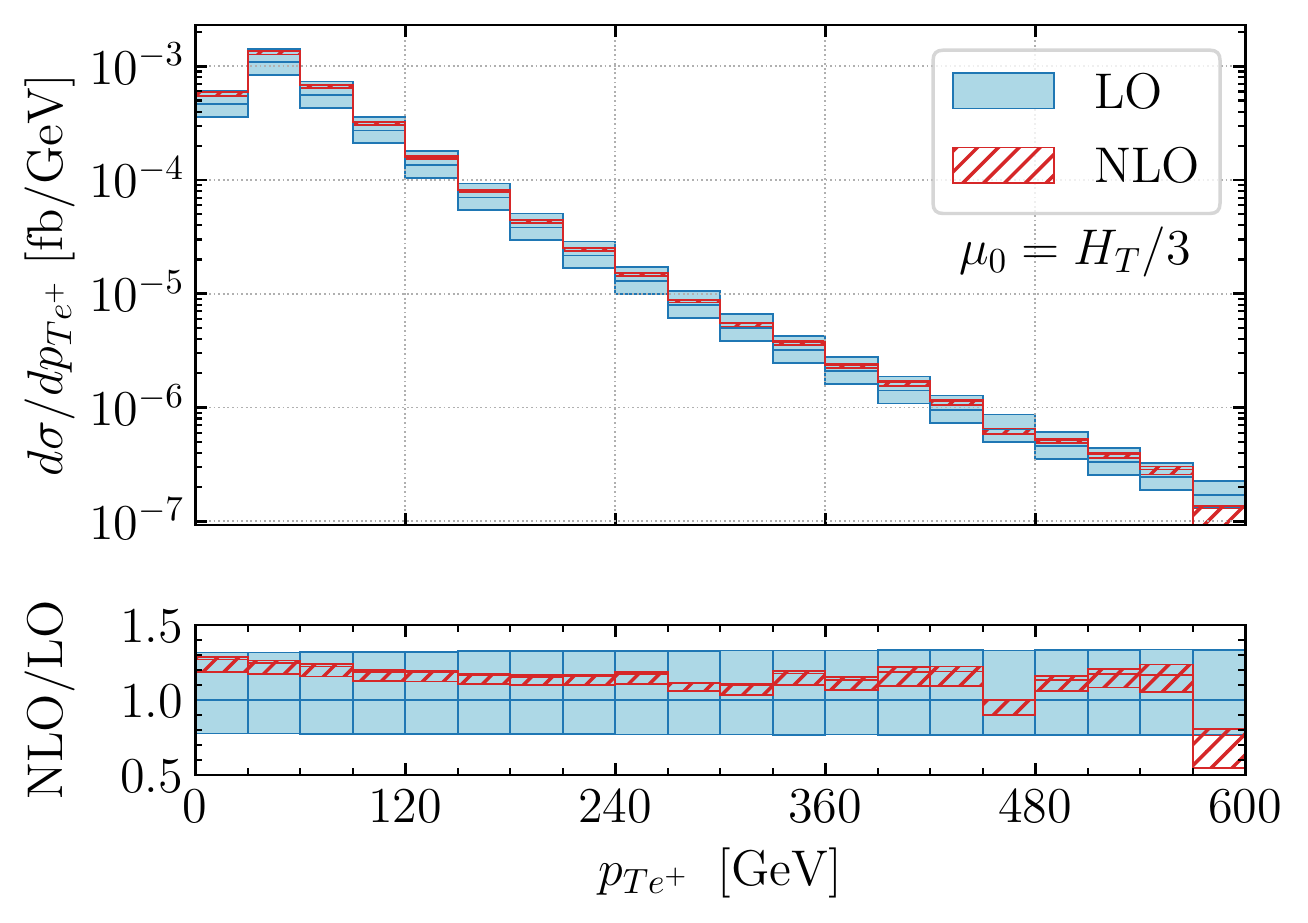}
     \includegraphics[width=0.49\textwidth]{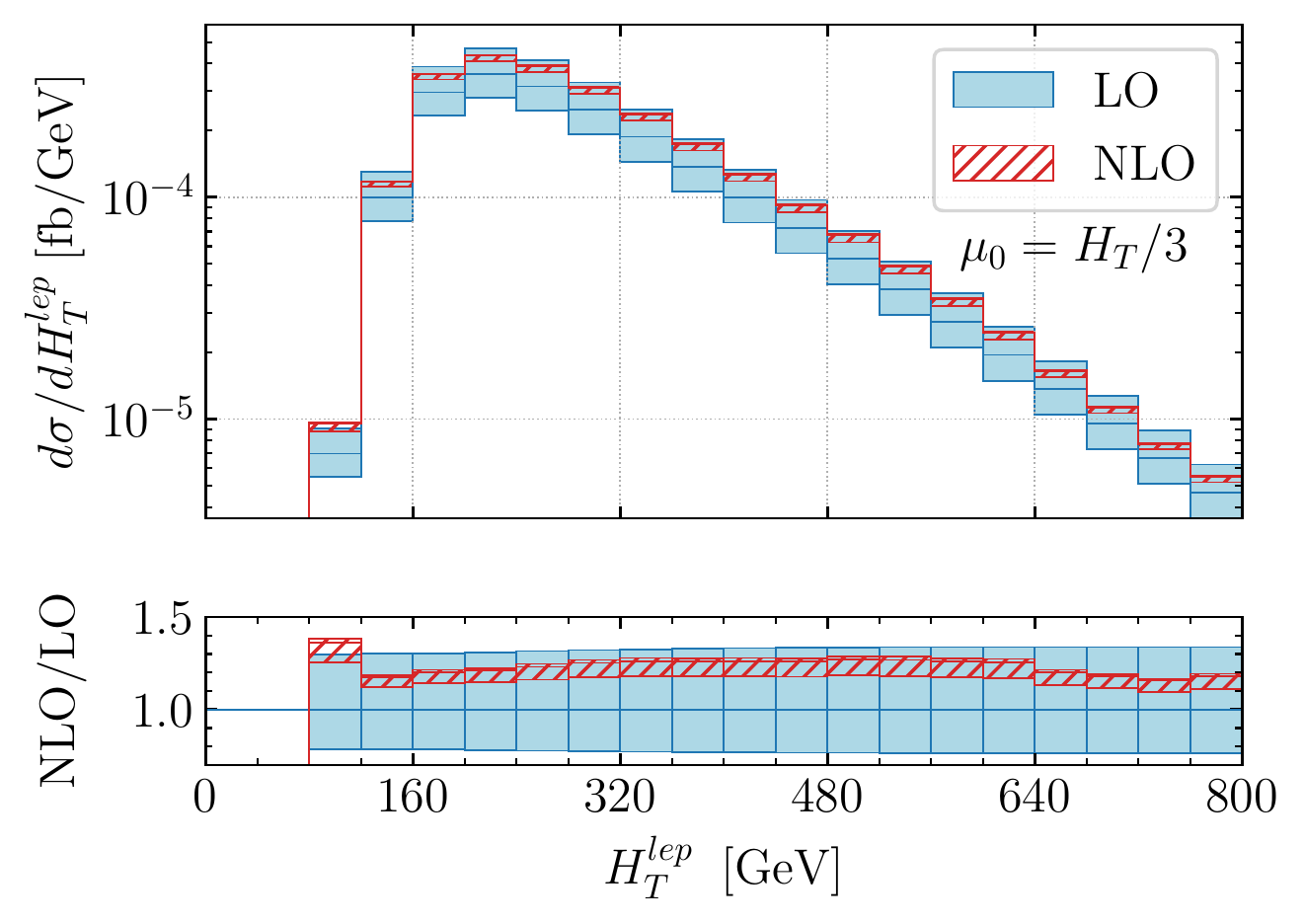}
\end{center}
\caption{\label{fig:diff2} \it
  Differential cross-section  distributions for $pp \to
  e^+ \nu_e \, \mu^- \bar{\nu}_\mu\, b\bar{b}\, \tau^+ \tau^- +X$ at
  the LHC with $\sqrt{s}=13$ TeV as a function of $p_{T\, b_1}$, 
  $p_{T\, b_2}$, $p_{T\, e^+}$ and $H_T^{lep}$. The blue curve corresponds 
  to the LO and the red curve to the NLO result. Also shown are the 
  corresponding uncertainty bands resulting from scale variations. The 
  lower panels display the differential ${\cal K}$-factor together with 
  the uncertainty band and the relative scale uncertainties of the LO
  cross section. The scale choice is  $\mu_R=\mu_F=\mu_0=H_T/3$. The
  cross sections are evaluated with the NNPDF3.1 PDF sets.}
\end{figure}
%=============================================
\begin{figure}[t]
  \begin{center}
    \includegraphics[width=0.49\textwidth]{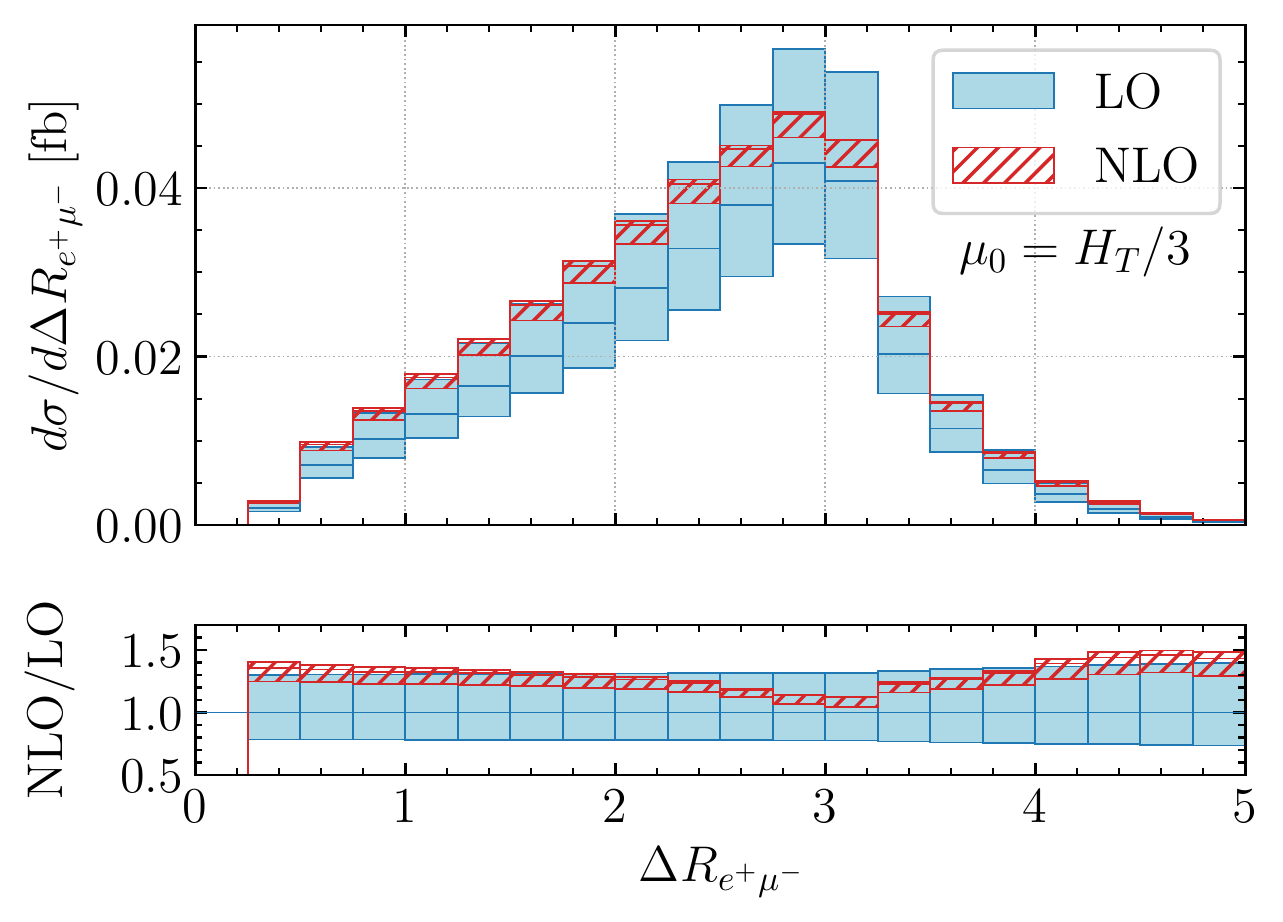}
    \includegraphics[width=0.49\textwidth]{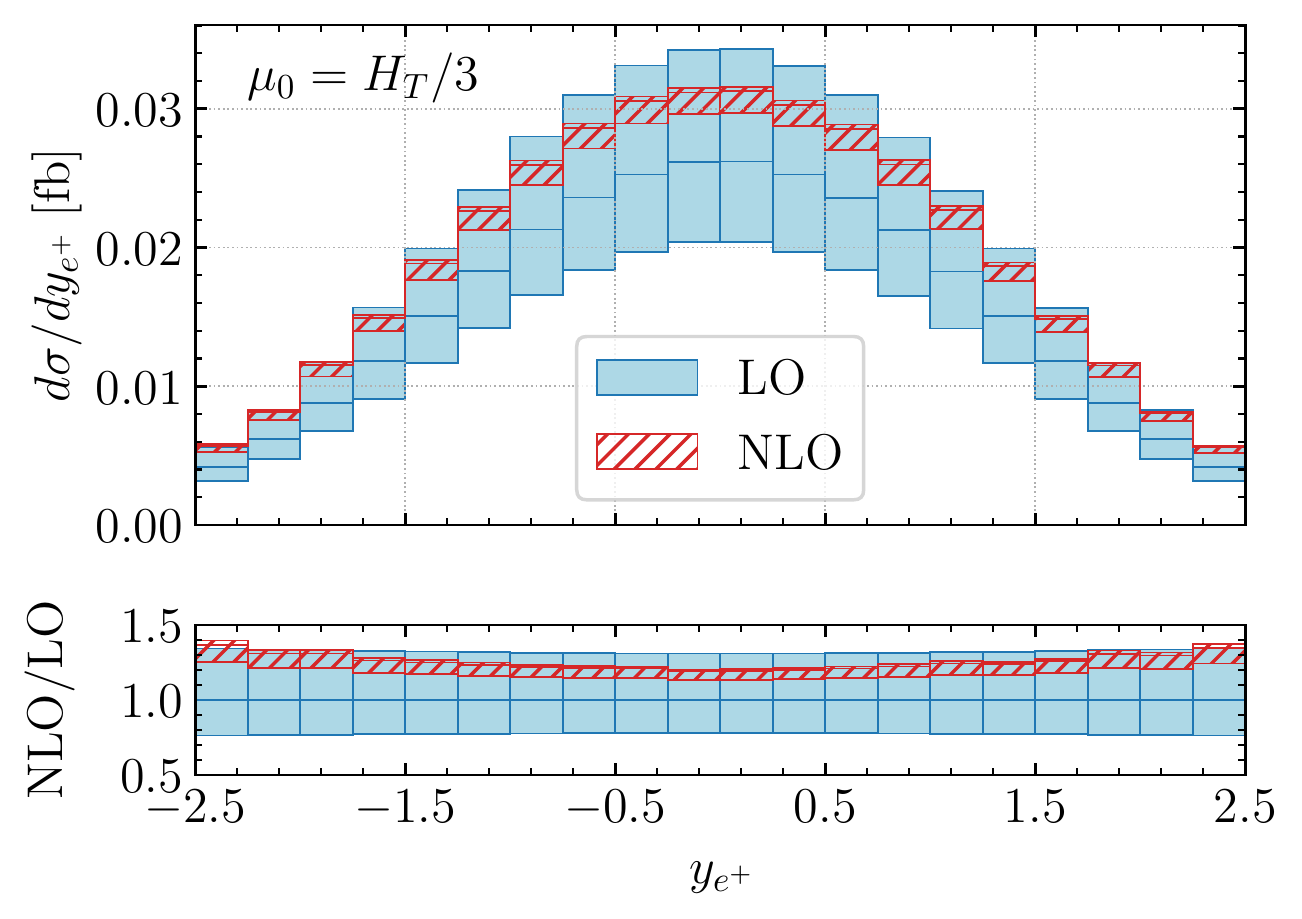}
    \includegraphics[width=0.49\textwidth]{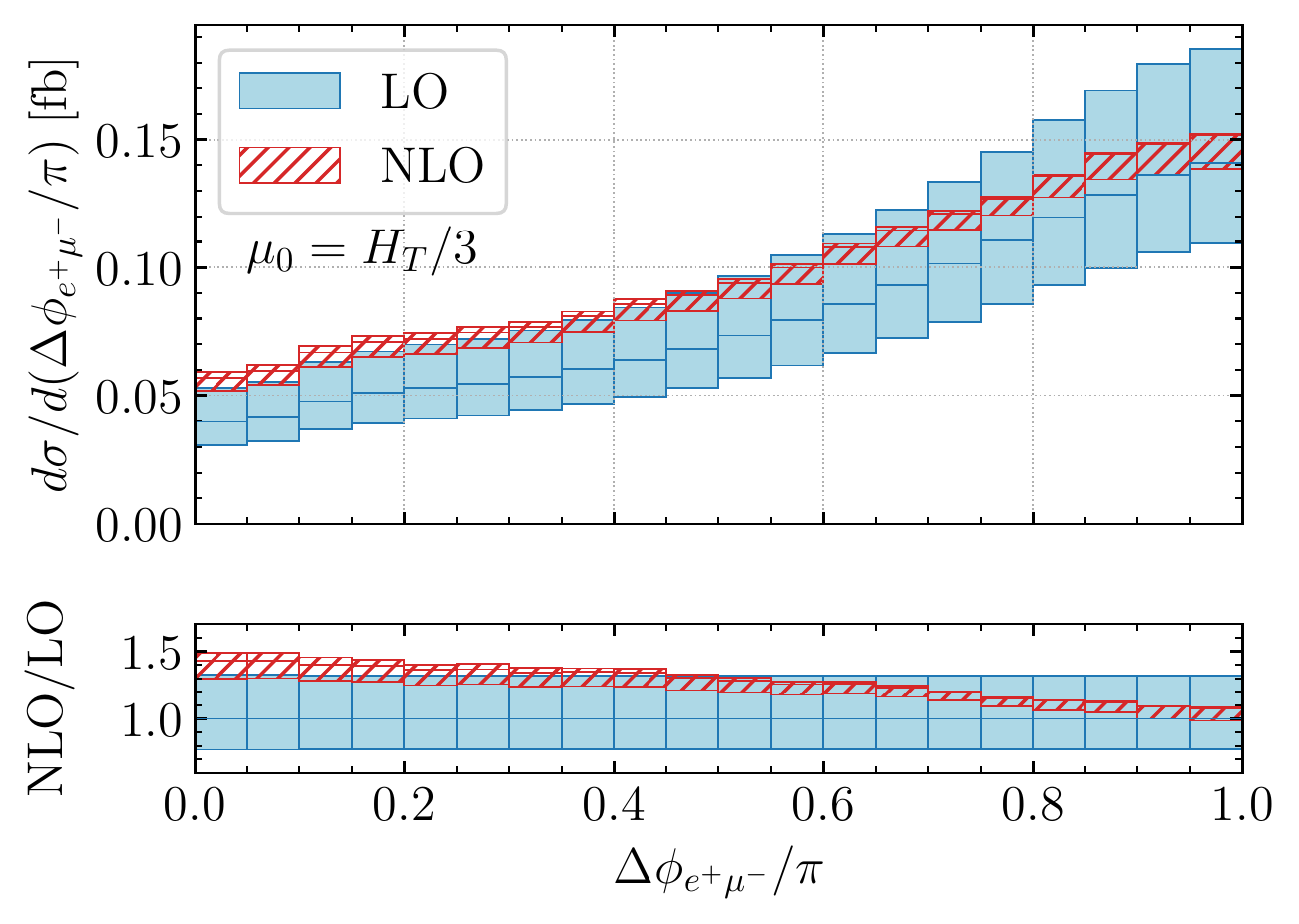}
       \includegraphics[width=0.49\textwidth]{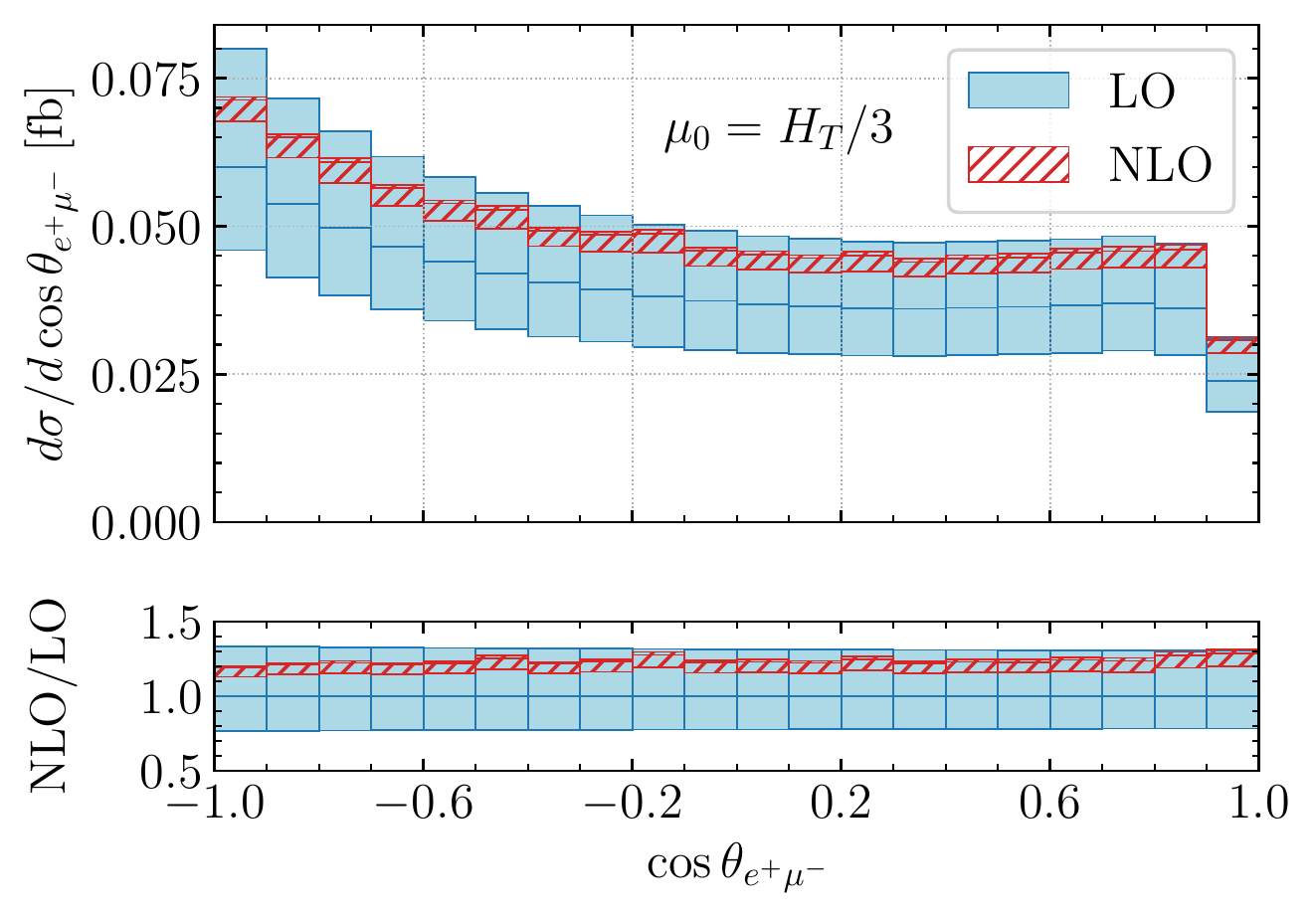}
\end{center}
\caption{\label{fig:diff3} \it 
  Differential cross-section  distributions for $pp \to
  e^+ \nu_e \, \mu^- \bar{\nu}_\mu\, b\bar{b}\, \tau^+ \tau^- +X$ at
  the LHC with $\sqrt{s}=13$ TeV as a function of $\Delta R_{e^+\mu^-}$, 
  $y_{e^+}$, $\cos\theta_{e^+\mu^-}$ and $\Delta\phi_{e^+\mu^-}$.
  The blue curve corresponds to the LO and the red curve to the NLO result. 
  Also shown are the corresponding uncertainty bands resulting from scale 
  variations. The lower panels display the differential ${\cal K}$-factor 
  together with the uncertainty band and the relative scale uncertainties 
  of the LO cross section. The scale choice is  $\mu_R=\mu_F=\mu_0=H_T/3$. 
  The cross sections are evaluated with the NNPDF3.1 PDF sets.}
\end{figure}
%=============================================

Integrated cross sections are mostly influenced by final-state
production relatively close to the threshold as defined by particle
masses. On the other hand, differential cross sections extend
themselves up to energy scales that are much larger than the
threshold, and may show larger shape distortions in such high-energy
regions. Therefore, in the next step we turn our attention to the
differential cross section distributions for the $pp \to e^+ \nu_e \,
\mu^- \bar{\nu}_\mu\, b\bar{b}\, \tau^+ \tau^- +X$
process.   In the following, we examine the size of the
NLO QCD corrections to the relevant observables at the LHC using our
default setup with LO and NLO NNPDF3.1 PDF sets obtained with
$\alpha_s(m_Z) = 0.118$. Had we used the same NLO NNPDF3.1 PDF set
also for the LO predictions our results and conclusions that we
provide below would not change significantly.  For each plot the upper
panels always show absolute LO and NLO predictions together with the
corresponding scale uncertainty bands. The lower panels display the
differential ${\cal K}$-factor together with its scale uncertainty
band as well as the relative scale uncertainties of the LO cross
section. Specifically, we plot
\begin{equation}
{\cal K}^{\rm NLO}(\mu) = 
\frac{d\sigma^{\rm NLO}(\mu)/dX}{d\sigma^{\rm LO}(\mu_0)/dX} \,,
\quad \quad \quad \quad 
{\rm and}
\quad \quad \quad \quad 
{\cal K}^{\rm LO}(\mu) = 
\frac{d\sigma^{\rm LO}(\mu)/dX}{d\sigma^{\rm LO}(\mu_0)/dX}\,,
\end{equation}
where $\mu_0$ is the central value of the scale and $X$ denotes the
observable under consideration.  The error band is determined
similarly to the integrated cross section case, i.e. with the help of
a $7$-point scale variation. We provide our results at the
differential level only for the dynamical scale setting, i.e. for
$\mu_R=\mu_F=\mu_0=H_T/3$. The reason for that is as follows. For the
fixed scale choice that we employed, i.e. for $\mu_0=m_t+m_Z/2$, we
could observe that for some choices of the $\xi$ parameter, where
$\mu_R=\mu_F=\xi \mu_0$, the NLO results become negative. This happens
in the high-energy tails of dimensionful distributions. Furthermore,
there are kinematic regions where the scale variation bands at LO and
NLO do not overlap anymore. In addition, it can even happen that 
the scale variation at NLO actually exceeds the scale variation of the
LO predictions. All these effects have already been
observed in our previous studies for the $t\bar{t}+X$
process, where $X=j,\,\gamma,\, Z(\nu\bar{\nu}),\, W^\pm,\, H$
\cite{Bevilacqua:2016jfk,Bevilacqua:2018woc,Bevilacqua:2019cvp,
Bevilacqua:2020pzy,Stremmer:2021bnk}.  They might be accommodated by a
judicious choice of a dynamical scale setting \footnote{ We note,
however, that for the missing transverse momentum, $p_{T}^{miss}$ (not
shown here), the fixed scale setting performs better. Similar results
have already been observed in the case of the $pp \to e^+ \nu_e \,
\mu^- \bar{\nu}_\mu\, b\bar{b}\, \nu_\tau \bar{\nu}_\tau +X$ process
and discussed in Ref.  \cite{Bevilacqua:2019cvp}.}.

We start with the observables constructed from the two $\tau$ leptons
originating from the $Z/\gamma^*$ boson decay. In Figure \ref{fig:diff1} we
show differential cross sections as a function of the invariant mass of
the $\tau^+\tau^-$ system, $M_{\tau^+\tau^-}$, the transverse momentum
of the $\tau^+\tau^-$ system, $p_{T\,\tau^+\tau^-}$ and the azimuthal
separation between the two $\tau$ leptons in the plane transverse to
the beam, $\Delta \phi_{\tau^+\tau^-}$. The latter is always taken in such 
a way as to ensure $0\le \Delta \phi_{\tau^+\tau^-} \le \pi$, consequently, 
we display $\Delta \phi_{\tau^+\tau^-}/\pi$. Furthermore, we present the 
cosine of the angle between the two $\tau$ leptons,
$\cos\theta_{\tau^+\tau^-}$, where the angle $\theta_{\tau^+\tau^-}$ is 
given by $\hat{p}_{\tau^+} \cdot \hat{p}_{\tau^-}=\cos\theta_{\tau^+\tau^-}$.
Also given  are the  rapidity-azimuthal-angle distance 
between the two $\tau$ leptons, $\Delta R_{\tau^+\tau^-}$ and the rapidity 
of the $\tau^+\tau^-$ system, $y_{\tau^+\tau^-}$. For $M_{\tau^+\tau^-}$, 
in the vicinity of the $Z$ resonance as well as outside of the $Z$ boson
peak, the NLO QCD corrections are similar in size and significant.
Specifically, they are of the order of $10\%-30\%$. The inclusion of
higher-order corrections reduces the scale dependence from $33\%$ to $8\%$
for our dynamical scale setting. For the transverse momentum of the
$\tau^+\tau^-$ pair the NLO corrections vary between $17\%$ and $50\%$
introducing shape distortions at the level of $33\%$. Also the NLO
theoretical uncertainties are slightly larger for this observable up to
$13\%$.  In the case of the following two dimensionless observables, 
$\Delta \phi_{\tau^+\tau^-}/\pi$ and $\cos\theta_{\tau^+\tau^-}$ we 
observe  a similar reduction of the theoretical scale uncertainty 
from around  $30\%$ at LO to $8\%$ at NLO after the inclusion of NLO QCD
effects.  The NLO QCD corrections can go up to about $35\%$, introducing
the overall shape distortions not larger than $20\%$. The situation is
slightly different for $\Delta R_ {\tau^+\tau^-}$, for which we received
larger NLO QCD corrections up to about $40\%$ and shape distortions of the
order of $30\%$.  At last, for $y_{\tau^+\tau^-}$ in the central rapidity
regions higher  order effects are of the order of $20\%$. At forward and
backward rapidity regions, on the other hand, they increased to 
$30\%-40\%$. For the last two observables NLO scale uncertainties are  
below $10\%$.

Similar to the integrated fiducial cross sections we observe a strong 
reduction in the unphysical scale dependence when NLO QCD corrections are
included. Moreover, scale dependence bands for LO and NLO predictions
indicate a well behaved perturbative convergence.   
However, even with our dynamical scale setting
the NLO QCD corrections to the differential cross-section
distributions, are significant. In a few cases shape distortions are
comparable in size to the LO theoretical uncertainties, i.e. they are
of the order of $30\%$. Thus, a suitably chosen global ${\cal
K}$-factor can not be applied to all LO predictions at the same time 
to obtain results that approximate well the full NLO QCD predictions. 
As such, full NLO predictions should be used consistently. Furthermore, 
because of the key role of the top quark interaction with the $Z$ boson in 
many beyond the Standard Model (BSM) scenarios, theoretical predictions for
benchmark observables calculated within the SM framework must be
provided as accurately as possible.  A realistic assessment of
systematic uncertainties for these theoretical predictions is required
as well.  Various differential fiducial cross-section distributions
might be modified by anomalous $t\bar{t}Z$ couplings.  Thus, it is
vital to have the SM predictions under excellent theoretical
control. We note that, from the observables that have  been
presented, $p_{T\, \tau^+\tau^-}$ and $\Delta \phi_{\tau^+\tau^-}$
are particularly interesting as they have already proved to be
good analysers of the $t\bar{t}Z$ coupling, see
e.g. Refs. \cite{Baur:2004uw,Rontsch:2014cca,Rontsch:2015una}.

In the next step we look at  observables related to the two top
quarks. They are especially important from the point of view of the
modelling of top-quark decays. In Figure \ref{fig:diff2} we 
present the following dimensionful observables: the transverse 
momentum of the hardest $b$-jet, $p_{T\,
b_1}$, the transverse momentum of the second hardest $b$-jet, $p_{T\,
b_2}$, the transverse momentum of the positron, $p_{T\, e^+}$ and the
scalar sum of the transverse momenta of the charged leptons,
$H_T^{lep}$.  The last observable is defined as follows
\begin{equation}
H_T^{lep}=p_{T\, e^+} + p_{T\, \mu^-} + p_{T\, \tau^+}+p_{T\, \tau^-}\,.
\end{equation}  
NLO QCD corrections for the $b$-jet related observables, $p_{T\,b_1}$ and 
$p_{T\,b_2}$, go up to $40\%$, whereas for the leptonic observables these
corrections are below $30\%$, excluding the second bin  in $H_T^{lep}$ 
which goes up to about $40\%$.  NLO theoretical uncertainties for all 
these observables are similar and around $10\%$, even though they can be
larger for a few single bins and reach up  to $30\%$.

Finally, in Figure \ref{fig:diff3} angular correlations of the leptons
coming from the top-quark decays are displayed. Specifically, we show
the rapidity-azimuthal-angle distance between the positron and muon,
$\Delta R_{e^+\mu^-}$, the rapidity of the positron, $y_{e^+}$, the
azimuthal angle between the positron and muon in the transverse plane,
$\Delta \phi_{e^+\mu^-}$, and the cosine of the angle between them,
$\cos \theta_{e^+\mu^-}$. These observables are also sensitive to
various BSM models, however, they are mostly used to study spin
correlations in top-quark decays. We can observe that compared to 
similar observables for the $\tau^+\tau^-$ system, the shapes are vastly
different here. This of course is not surprising given that the positron 
and muon are predominantly produced
in  back-to-back configurations. NLO QCD corrections for
dimensionless observables constructed from the two charged leptons
originating from the top quarks are not small. Specifically, for
$\Delta R_{e^+\mu^-}$ they are up to $45\%$, for $y_{e^+}$ in the
central rapidity regions they are of the order of $20\%$ whereas in
the forward and backward regions they increase up to about $40\%$. In the
case of $\Delta \phi_{e^+\mu^-}$ higher-order effects are below $40\%$
and finally for $\cos\theta_{e^+\mu^-}$ NLO QCD corrections are of the
order of $25\%$. For all four differential cross-section distributions
theoretical uncertainties coming from scale variation are reduced from
$30\%-40\%$ at LO to maximally up to $10\%$ at NLO in QCD.
%
%=============================================
\begin{figure}[t]
  \begin{center}
     \includegraphics[width=0.49\textwidth]{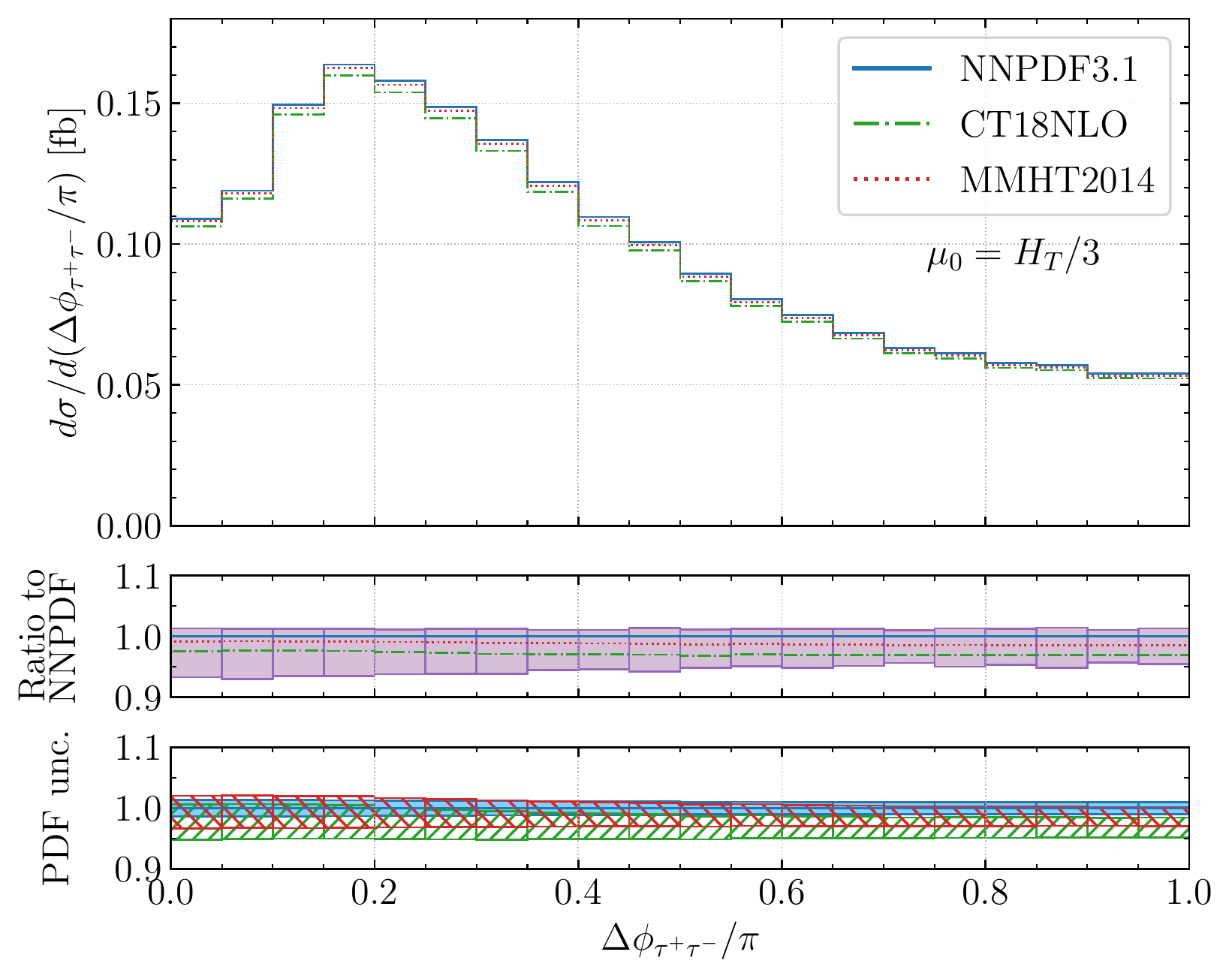}
    \includegraphics[width=0.49\textwidth]{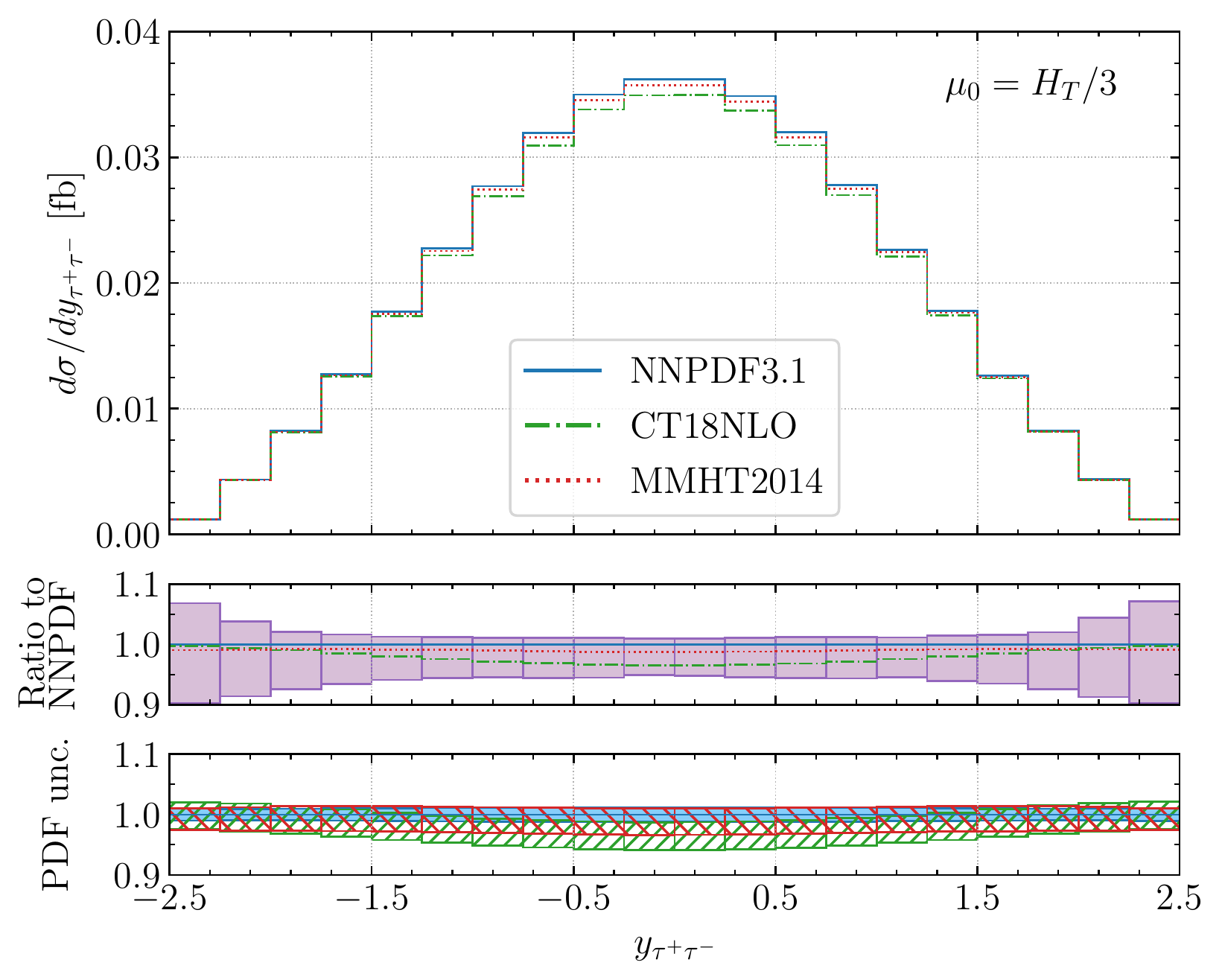}
    \includegraphics[width=0.49\textwidth]{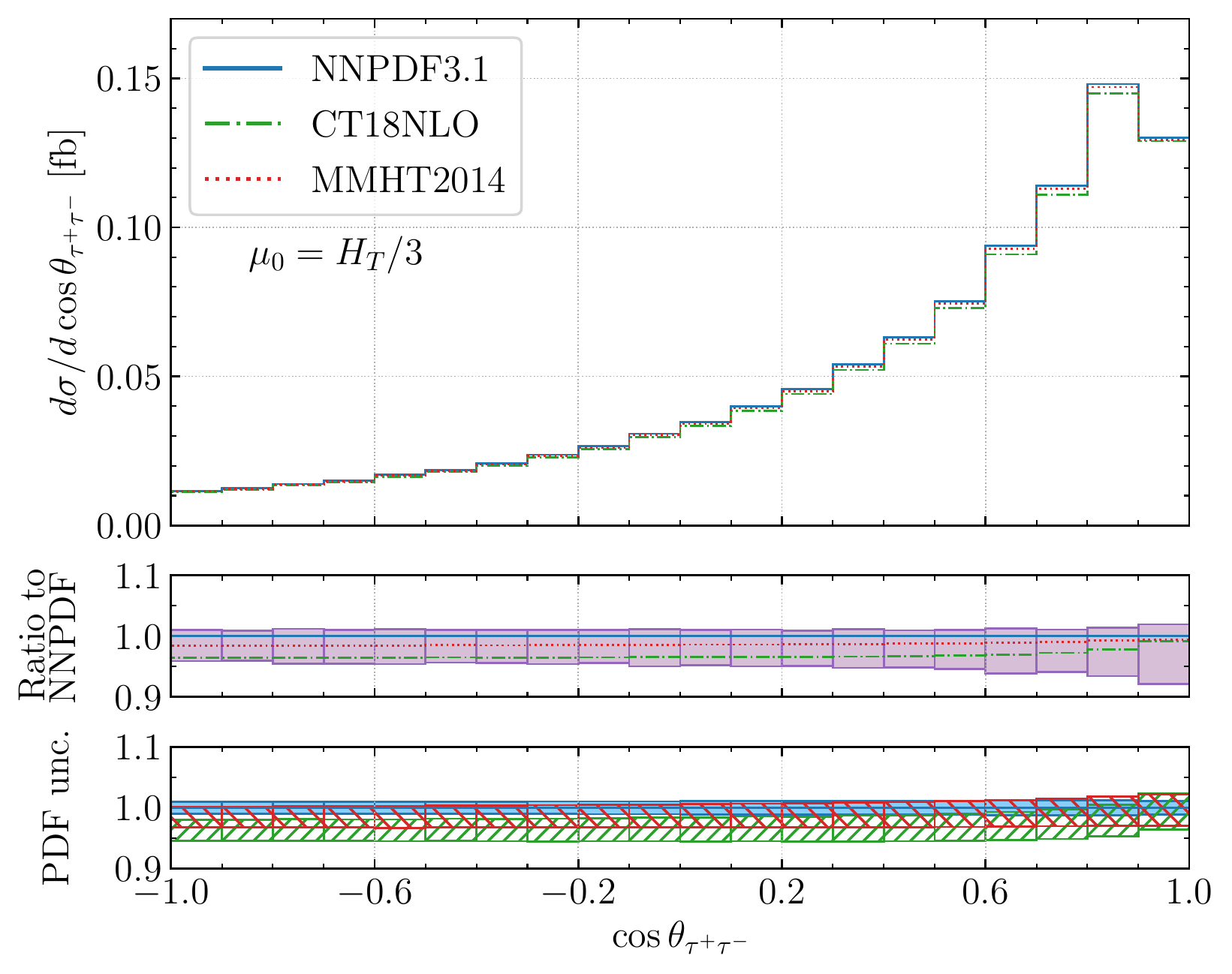}
    \includegraphics[width=0.49\textwidth]{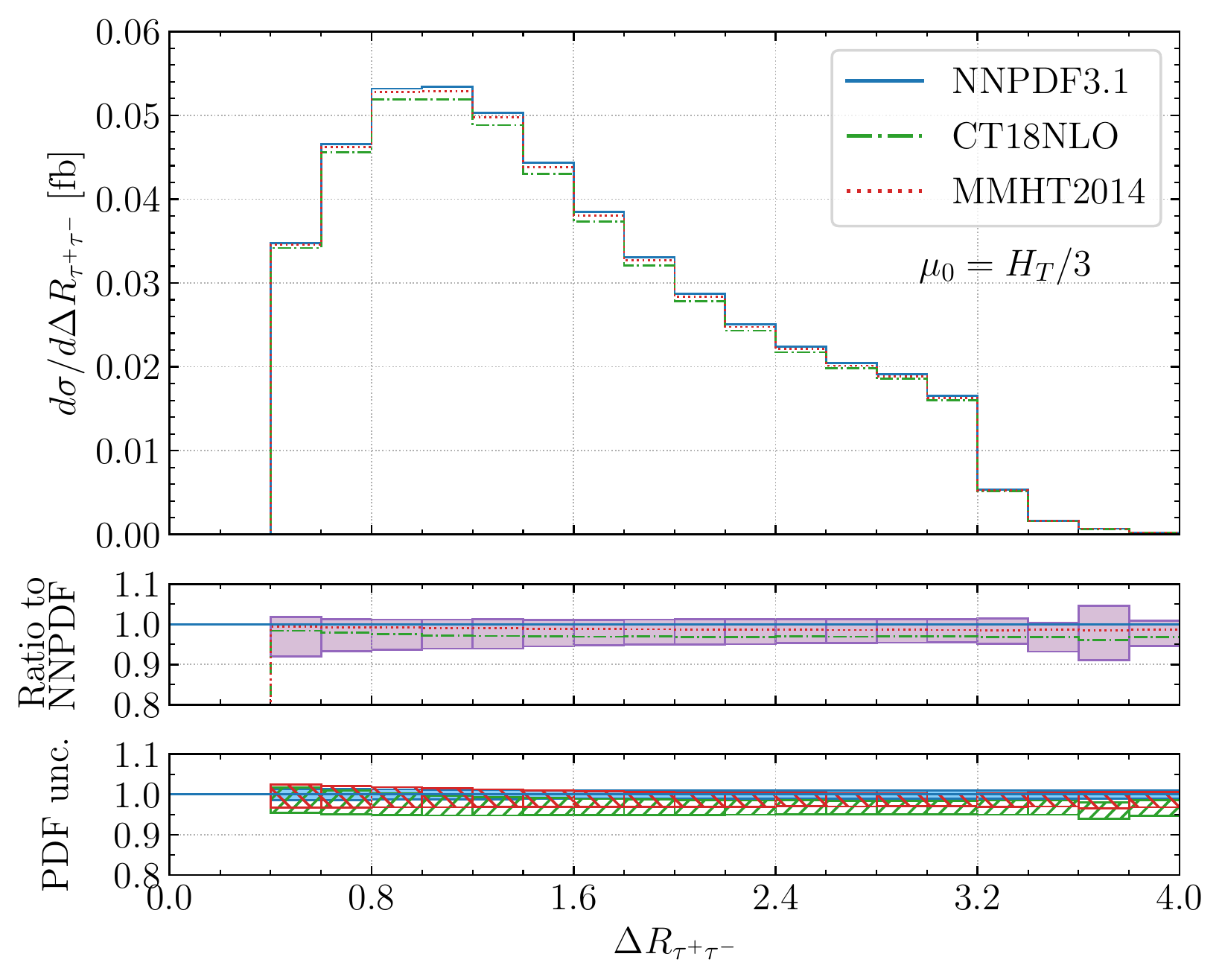}
     \includegraphics[width=0.49\textwidth]{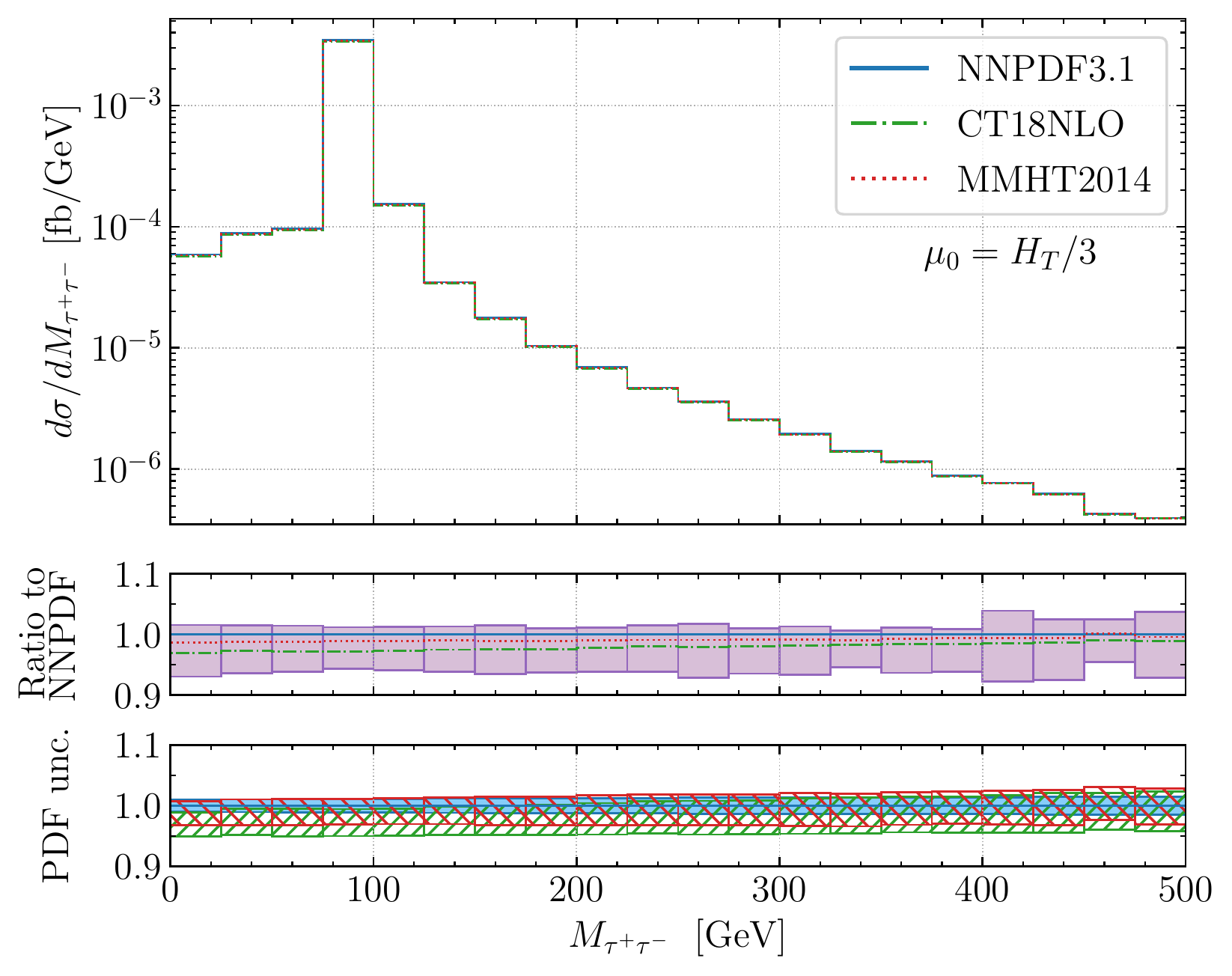}
     \includegraphics[width=0.49\textwidth]{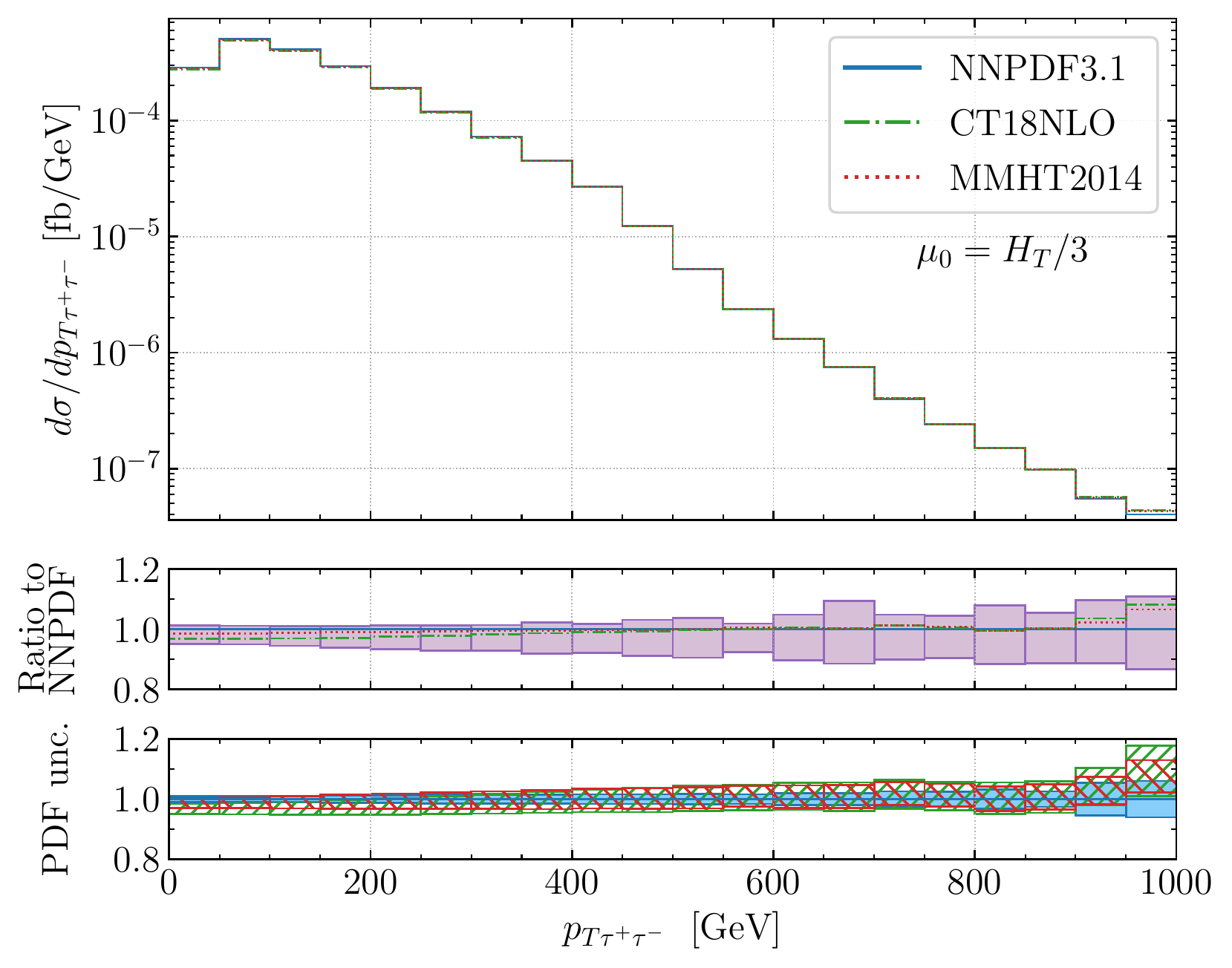}
\end{center}
\caption{\label{fig:diff_pdf} \it Differential cross-section 
distributions for $pp \to e^+\nu_e \, \mu^-\bar{\nu}_{\mu} 
\, b\bar{b}\, \tau^+\tau^- +X$ at the LHC with $\sqrt{s}=13$ TeV 
as a function of $\Delta \phi_{\tau^+\tau^-}$, $y_{\tau^+\tau^-}$, 
$\cos\theta_{\tau^+\tau^-}$, $\Delta R_{\tau^+\tau^-}$, $M_{\tau^+\tau^-}$ 
and $p_{T\, \tau^+\tau^-}$. The upper panel shows the absolute NLO QCD 
predictions for three different PDF sets with $\mu_R=\mu_F=\mu_0=H_T/3$. 
The middle panel displays the ratio to the result with the default NNPDF3.1 
PDF set as well as its scale dependence. The lower panel presents the 
internal PDF uncertainties calculated separately for each PDF set. }
\end{figure}
%=============================================

Having examined the size of scale uncertainties for the differential 
cross section distributions we turn our attention to the PDF uncertainties.  
We have already checked that the latter are smaller than the theoretical 
uncertainties stemming from scale variation at the integrated fiducial level. 
We would like to check whether this is still the case at the differential 
level. We display the six observables that have been shown previously. 
Specifically,  in Figure \ref{fig:diff_pdf} we present differential cross
section distributions as a function of $\Delta \phi_{\tau^+\tau^-}$,
$y_{\tau^+\tau^-}$, $\cos\theta_{\tau^+\tau^-}$, $\Delta R_{\tau^+\tau^-}$,
$M_{\tau^+\tau^-}$ and $p_{T\, \tau^+\tau^-}$. We plot them afresh for the
CT18 and MMHT14 PDF sets. Each plot comprises three panels. The upper panel
displays the absolute NLO prediction for three different PDF sets at the
central scale value, $\mu_R=\mu_F=\mu_0=H_T/3$. The middle panel shows the
NLO QCD scale dependence band normalised to the NLO prediction for $\mu_0$
and NNPDF3.1. Also given is the ratio of NLO QCD predictions generated for
CT18 and MMHT14 to the default NNPDF3.1 PDF set. The lower panel exhibits 
the internal PDF uncertainties for each PDF set separately, which are also 
normalised to  central NLO predictions with NNPDF3.1

We observe that also at the differential  level the NNPDF3.1 PDF 
uncertainties are very small and well below  the corresponding
theoretical uncertainties due to scale dependence. Indeed, the internal
NNPDF3.1 PDF uncertainties are maximally up to  $1.5\%$ for dimensionless
observables and for $M_{\tau^+\tau^-}$. At the same time scale uncertainties
are up to $7\%-10\%$ for these observables. For $p_{T\, \tau^+\tau^-}$, the
last  observable that we have shown,  the NNPDF3.1 PDF uncertainties in the
tails of the distribution do not exceed $5\%-6\%$. However, these are 
phase-space regions where the scale uncertainties can be as large as $13\%$.
When analysing  the internal PDF uncertainties  for CT18 and MMHT14 we 
notice that they behave similarly and their PDF uncertainties are almost 
a factor of $2$ larger than those for NNPDF3.1. This is true for MMHT14 
for all observables except for $p_{T\, \tau^+\tau^-}$ where the PDF
uncertainties  are not bigger than $6\%$ and thus comparable to those from
NNPDF3.1. In all cases, the selected PDF sets are still within the 
theoretical uncertainties due to scale dependence for all observables. 
Looking at the theoretical predictions for MMHT14 and CT18, we note that 
the relative difference to the NNPDF3.1 result is of the order of $1.5\%$ 
and $4\% $ respectively. Therefore, the PDF uncertainties are of a similar 
size as the difference between the results obtained using various PDF sets.

% =============================================
%
\section{Off-shell vs on-shell modelling of top quarks
  and gauge bosons}
\label{comparison}
% =============================================
%

In order to examine the size of the non-factorisable corrections 
for the $pp \to e^+\nu_e \, \mu^- \bar{\nu}_\mu\, b\bar{b}\, 
\tau^+\tau^- +X$ process within our setup we compare the NLO QCD results 
with full off-shell effects included with the calculations in the NWA.
The latter results are also generated with the help of the 
\textsc{Helac-NLO} program \cite{Bevilacqua:2019quz}, which recently has 
been extended to provide theoretical predictions also in this approximation.
The NWA results at NLO in QCD are divided in two categories: the
full NWA and the NWA with LO top-quark decays (hereafter referred to as 
${\rm NWA}_{\rm full}$ and ${\rm NWA}_{\rm LOdec}$ respectively). 
The full NWA comprises NLO QCD corrections to both $t\bar{t}Z$ production 
and the subsequent top-quark decays preserving at the same time the
$t\bar{t}$ spin correlations at the same accuracy. The ${\rm NWA}_{\rm
LOdec}$ case contains the results with NLO QCD corrections to the 
production stage only, whereas top-quark decays are calculated at LO. 
In this case $t\bar{t}$ spin correlations are only available with LO
accuracy. For consistency these findings are calculated with
$\Gamma^{\rm LO}_{t,\, {\rm NWA}}$. The contribution from the $t \to WbZ$
decay are neglected  due to the tiny available phase space and the size of
the $t\to WbZ$ branching ratio.  With the input parameters and cuts 
specified in Section \ref{setup}, our findings can be summarised as follows
\begin{equation}
\sigma^{\rm LO}_{\rm NWA} ({\tt NNPDF3.1{}_{-}lo{}_{-}as{}_{-}0118}, 
\mu_0=H_T/3) = 72.69^{\,+22.97 \,(32\%)}_{\,-16.24 \, (22\%)} \, 
{\rm ab}\,,
\end{equation}
\begin{equation}
\sigma^{\rm NLO}_{\rm NWA_{\rm full}} ({\tt
NNPDF3.1{}_{-}nlo{}_{-}as{}_{-}0118}, \mu_0=H_T/3)
=88.75^{\, -2.16\, ( 2\%)}_{\,-4.87 \,(5\%)} \,\rm{ab}\,.
\end{equation}
In the case where NLO QCD corrections to top-quark decays are omitted we 
obtain instead
\begin{equation}
\sigma^{\rm NLO}_{\rm NWA_{\rm LOdec}} ({\tt
NNPDF3.1{}_{-}nlo{}_{-}as{}_{-}0118}, \mu_0=H_T/3)
= 96.74^{\, +7.21\,( \,\,\,7\%)}_{\, -9.98 \,(10\%)} \, \rm{ab}\,.
\end{equation}
We can immediately make the following conclusions based on our NWA results. 
First, we can see that the ${\rm NWA}_{\rm full}$ prediction has the same 
theoretical uncertainties due to the scale dependence as the full off-shell
result (see Eq. \eqref{results-ht}). At LO they are at the level of $32\%$ 
(to be compared to $32\%$ for the full-off shell case), while at NLO in QCD
they are reduced to $5\%$ ($6\%$ for the full off-shell prediction). 
The size of NLO QCD corrections  is also similar. Indeed, we find ${\cal
K}=\sigma^{\rm NLO}_{\rm NWA_{\rm full}}/\sigma^{\rm LO}_{\rm NWA} =1.22$ 
for our dynamical scale setting (${\cal K}=\sigma^{\rm NLO}_{\rm
off-shell}/\sigma^{\rm LO}_{\rm off-shell}=1.23$). When the higher-order
corrections in top-quark decays are omitted, however, the scale dependence
increases and theoretical uncertainties are at the level of $10\%$. 
Therefore, they are almost a  factor of $2$  larger than  the theoretical
uncertainty estimate for the ${\rm NWA}_{\rm full}$ case. In addition, 
the NLO QCD corrections are larger as well for this case as we have 
${\cal K}=\sigma^{\rm NLO}_{\rm NWA_{\rm LO decay}}/\sigma^{\rm LO}_{\rm
NWA}=1.33$. Having  both results  $\sigma^{\rm NLO}_{\rm NWA_{\rm full}}$ 
and $\sigma^{\rm NLO}_{\rm NWA_{\rm LO decay}}$ we can furthermore assess  
that higher-order  QCD corrections to top-quark decays are negative 
and of the order of $9\%$. We observe that ${\rm NWA}_{\rm LOdec}$ is very
close to the full off-shell prediction, and the two results agree well 
within the estimated uncertainties. We stress that this agreement is  
accidental and is an interplay of two effects. Indeed, 
when NLO QCD corrections to top-quark decays are included 
(${\rm NWA}_{\rm full}$), the cross section receives negative 
corrections. On the other  hand, when off-shell contributions  are 
included, the correction to the cross section is positive.
%
%=============================================
\begin{table}[t!]
\begin{center}
  \begin{tabular}{ccc}
    \hline \hline\vspace{-0.3cm}\\ 
    \textsc{Modelling}  & $\sigma^{\rm NLO}_{i}$ [ab] 
&  $\sigma^{\rm NLO}_{i}/\sigma^{\rm NLO}_{\rm NWA_{\rm full}} -1$ 
  \\
   &&\vspace{-0.3cm}\\
    \hline \hline \vspace{-0.4cm}\\ 
  Off-shell     & $98.88$ & $+11.4\, \%$ \\ [0.1cm]
 Off-shell $M^{25\, {\rm GeV}}_{\tau^+\tau^-}$  & $91.00$ & $+2.5\, \%$ 
 \\[0.1cm] 
 Off-shell $M^{20\, {\rm GeV}}_{\tau^+\tau^-}$  & $89.96$ & $+1.4\, \%$ 
  \\ [0.1cm]
  Off-shell $M^{15\, {\rm GeV}}_{\tau^+\tau^-}$ & $88.44$ & $-0.3\, \%$ 
  \\[0.1cm] 
 Off-shell $M^{10\, {\rm GeV}}_{\tau^+\tau^-}$  & $85.74$ & $-3.4\, \%$ 
\\[0.1cm]
 ${\rm NWA}_{\rm full}$ & $88.75$ & $-$  \\[0.1cm]
 ${\rm NWA}_{\rm LOdec}$ &  $96.74$  & $+9.0\, \%$ \\[0.1cm]
    \hline \hline
 \end{tabular}
\end{center}
\caption{\label{tab:nlo-nwa} \it 
NLO integrated fiducial cross sections for the for $pp \to e^+\nu_e \, 
\mu^-\bar{\nu}_{\mu} \, b\bar{b}\, \tau^+\tau^- +X$ at the LHC with
$\sqrt{s}=13$ TeV.  The full off-shell prediction and  various NWA results
are presented. Also given are the full off-shell findings with the 
additional  $|M_{\tau^+\tau^-}-m_Z|<X$ cut, where  $X\in
\left\{25,20,15,10\right\}$ GeV. The latter is denoted as
$M_{\tau^+\tau^-}^{X}$. The NNPDF3.1 PDF sets are dynamical scale setting,
$\mu_R=\mu_F=\mu_0=H_T/3$, are used. }
\end{table}
%=============================================

With all the theoretical predictions available, we can check the 
importance of  the full off-shell effects for the process, 
first at the integrated fiducial cross-section level, 
and in the next step differentially 
for various  observables. As we have already mentioned  the full 
off-shell predictions at NLO QCD accuracy consist of multiple effects.
Firstly, they comprise the NLO QCD corrections not only to the production 
and decay stages of top quarks separately, but also radiative interference
effects between the production and decay stage which represent genuine
non-factorisable effects. Secondly, all heavy unstable particles are
described by Breit-Wigner propagators in the matrix element calculations.
Thirdly, they incorporate single- and non-resonant $t$ and $W^\pm/Z$
contributions.  Finally, photon-induced contributions and $Z/\gamma^*$
interference effects are included as well.

We start with the results for NLO integrated fiducial cross
sections that are given in Table \ref{tab:nlo-nwa}. For the default set of
cuts we observe large effects. Specifically, the difference between the 
full off-shell result and  ${\rm NWA}_{\rm full}$ is of the order of $11\%$,
which is well above the NWA uncertainty determined by ${\cal O}(\Gamma/m)$
for sufficiently inclusive processes \cite{Fadin:1993kt}. The latter
uncertainty for the $pp\to t\bar{t}Z$ process with our rather inclusive 
setup is given by ${\cal O}(\Gamma/m) \approx 2.7\%$.  When the additional
$|M_{\tau^+\tau^-}-m_Z|<X$ cut, where $X\in \left\{25,20,15,10\right\}$ 
GeV, is employed we can notice that full off-shell effects are 
substantially reduced and fall into the range $0.3\%-3.4\%$
depending on the $X$ value. Thus, this larger $11\%$ effect is mostly 
driven by the $Z/\gamma^*$ interference effects and  photon-induced
contributions. This is  further confirmed by a  previous study for 
the $pp\to t\bar{t}Z$ process with $Z\to \nu \bar{\nu}$. Specifically, 
in Ref.  \cite{Hermann:2021xvs} the final state $pp \to e^+\nu_e\, \mu^-
\bar{\nu}_\mu \, b\bar{b}\,  \nu_{\tau} \bar{\nu}_\tau +X $ was considered
with a fairly inclusive cut selection. For such a setup full off-shell
effects  of the order of  $3\%-4\%$ have been found, see Table 2 in 
Ref.  \cite{Hermann:2021xvs}.

\begin{figure}[t!]
  \begin{center}
    \includegraphics[width=0.49\textwidth]{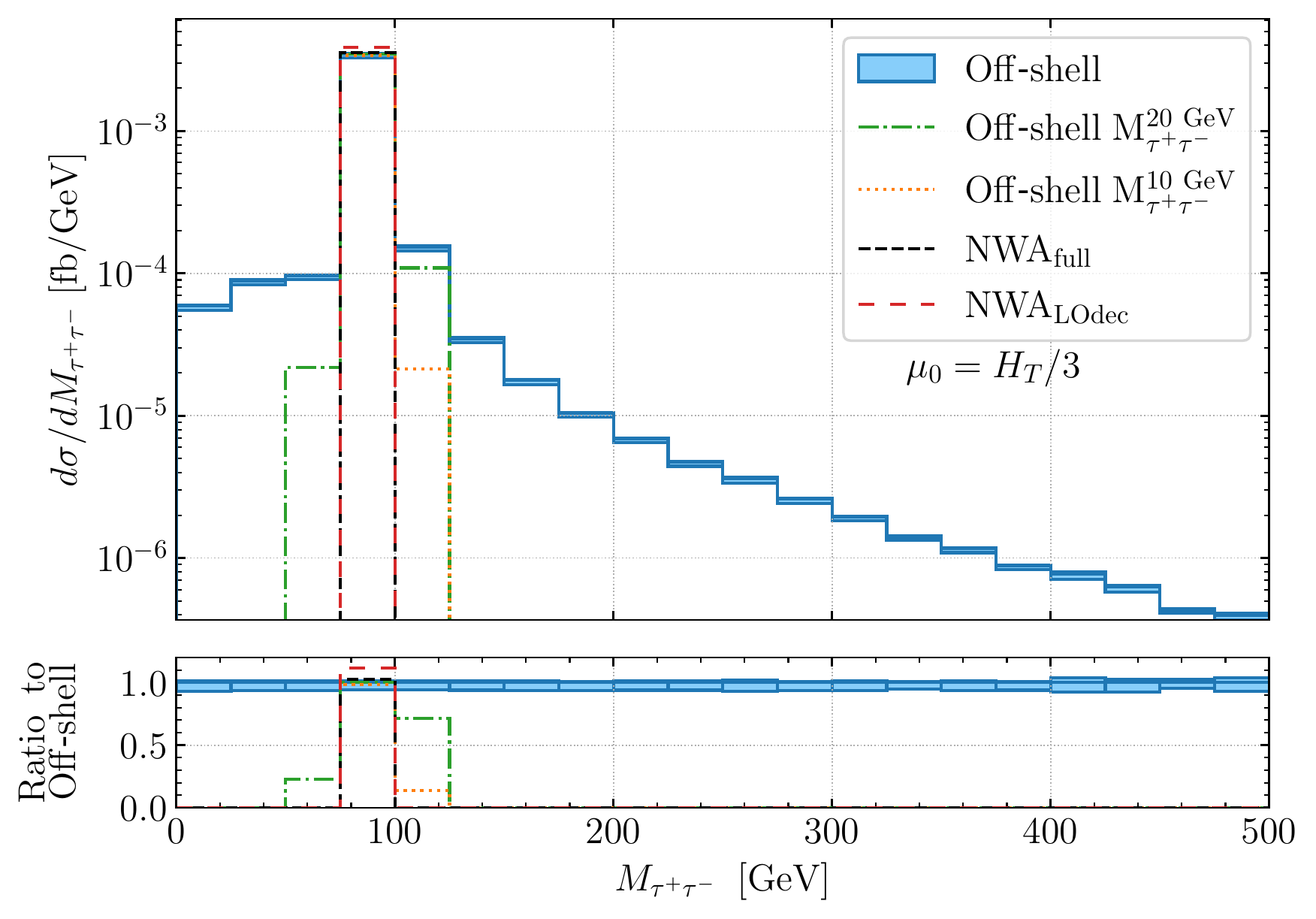}
\end{center}
\caption{\label{fig:diff_mtautau} \it Differential cross-section 
distribution for $pp \to e^+\nu_e \, \mu^-\bar{\nu}_{\mu} \, b\bar{b}\,
\tau^+\tau^- +X$ at the LHC with $\sqrt{s}=13$ TeV as a function of 
$M_{\tau^+\tau^-}$. The upper panel shows the NLO QCD prediction for the 
full off-shell case as well as for ${\rm NWA}_{\rm full}$ and 
${\rm NWA}_{\rm LOdec}$.  Also given is the full off-shell result with the
additional $|M_{\tau^+\tau^-} -m_Z|< 20 \, (10)$ GeV cut.
Furthermore, we provide theoretical uncertainties as obtained from the 
scale dependence for the full off-shell case. The lower panel displays the
ratios to the full off-shell result. The scale choice is
$\mu_R=\mu_F=\mu_0=H_T/3$. The cross sections are evaluated with the 
NNPDF3.1  PDF set. }
\end{figure}

To visualise this better we display in Figure \ref{fig:diff_mtautau} 
the NLO integrated fiducial cross section as a function of 
$M_{\tau^+\tau^-}$. The upper panel shows the absolute prediction at NLO 
for various approaches for the modelling of top-quark production and 
decays. Specifically, we  show the full off-shell case without and with 
the $|M_{\tau^+\tau^-}-m_Z|< 20\, (10)$ GeV cut. Also reported are
the  ${\rm NWA}_{\rm full}$  and ${\rm NWA}_{\rm LOdec}$ results. We
additionally provide theoretical uncertainties  as obtained from the 
scale dependence for the  full off-shell case. The lower panel displays 
the ratios to the full off-shell prediction. We can observe that in the
full off-shell case large tails from the Breit-Wigner $Z$ propagator and 
the $t\bar{t}Z/t\bar{t}\gamma^*$ interference are present. We can 
additionally notice substantial contributions from the $t\bar{t}\gamma^*$
process for $8 \, {\rm GeV} \lesssim M_{\tau^+\tau^-} \lesssim m_Z$ 
\footnote{ We note that the separation between the $\tau^\pm$ leptons, 
implied by the $\Delta R_{\tau^+\tau^-}$ cut, together with the requirement 
of having both $\tau$ leptons with $p_{T\,\tau} > 20$ GeV sets an effective
lower limit on the invariant mass $M_{\tau^+\tau^-}$. Indeed, we can write 
$(M_{\tau^+\tau^-})_{min} = (p_{T, \tau})_{min} \sqrt{2(1-\cos(\Delta
R_{\tau^+\tau^-}))}\approx 8$ GeV.}. On the other hand, when the
invariant mass of the $\tau^+\tau^-$ system is set within  a range of 
$\pm\, 10$ GeV  $(\pm\, 20 \, {\rm GeV})$ around the nominal $Z$ boson 
mass we are more sensitive to on-shell-like $Z$ boson decays. In these
phase-space  regions the photon-induced contribution as  well as 
$Z/\gamma^*$  interference effects are highly suppressed.
%
%============================================= 
%============================================= 
\begin{figure}[t]
  \begin{center}
     \includegraphics[width=0.49\textwidth]{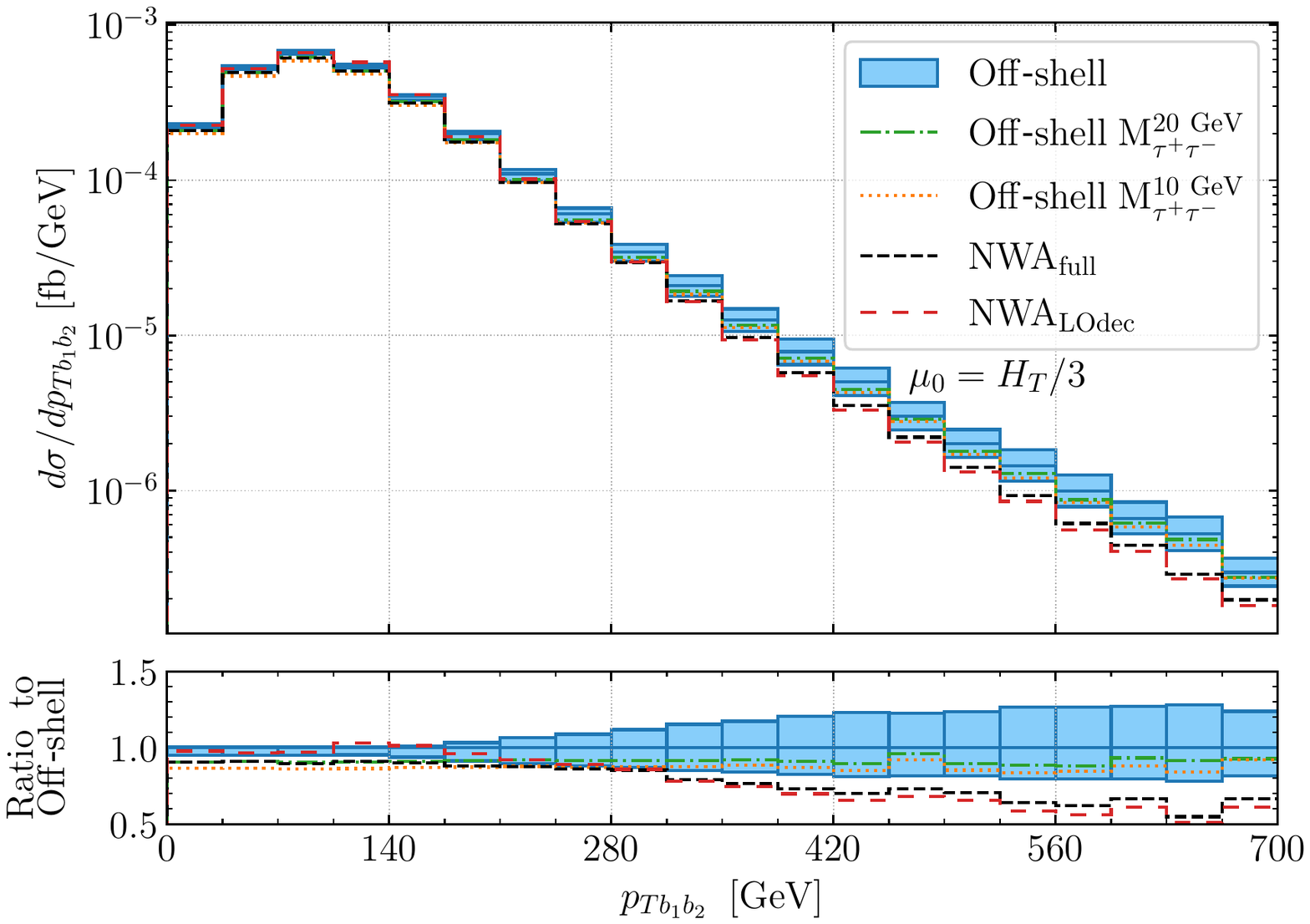}
    \includegraphics[width=0.49\textwidth]{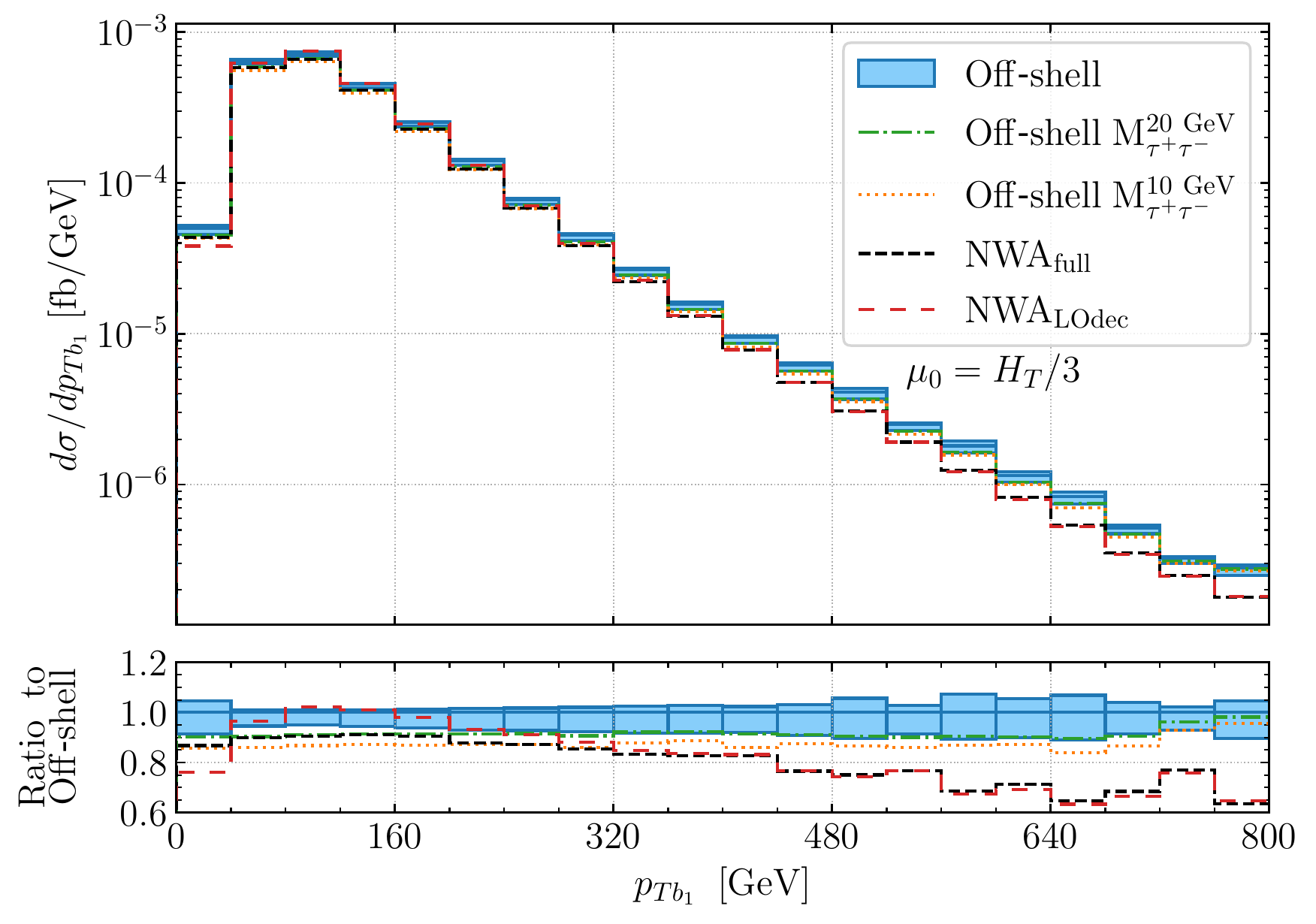}
    \includegraphics[width=0.49\textwidth]{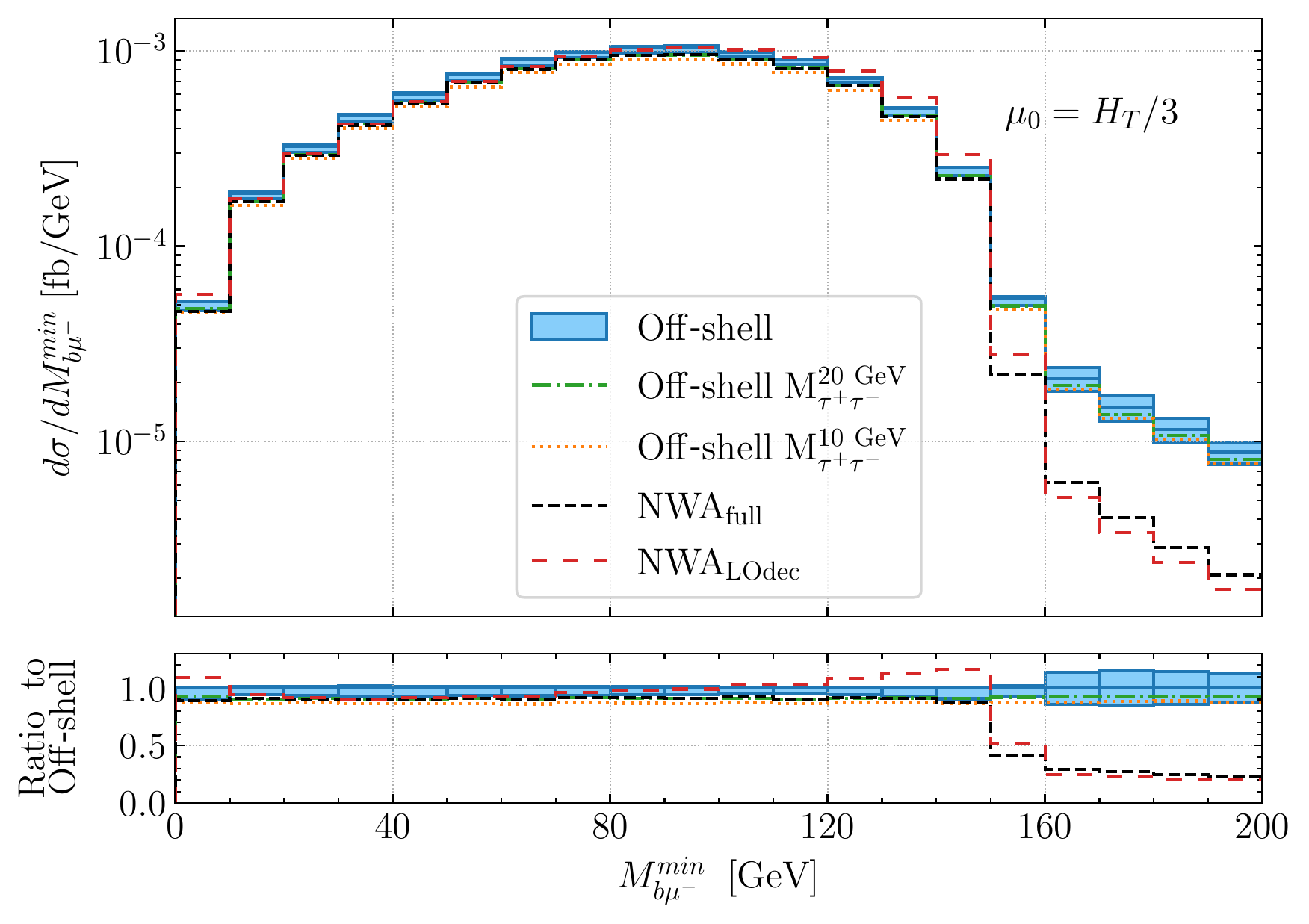}
    \includegraphics[width=0.49\textwidth]{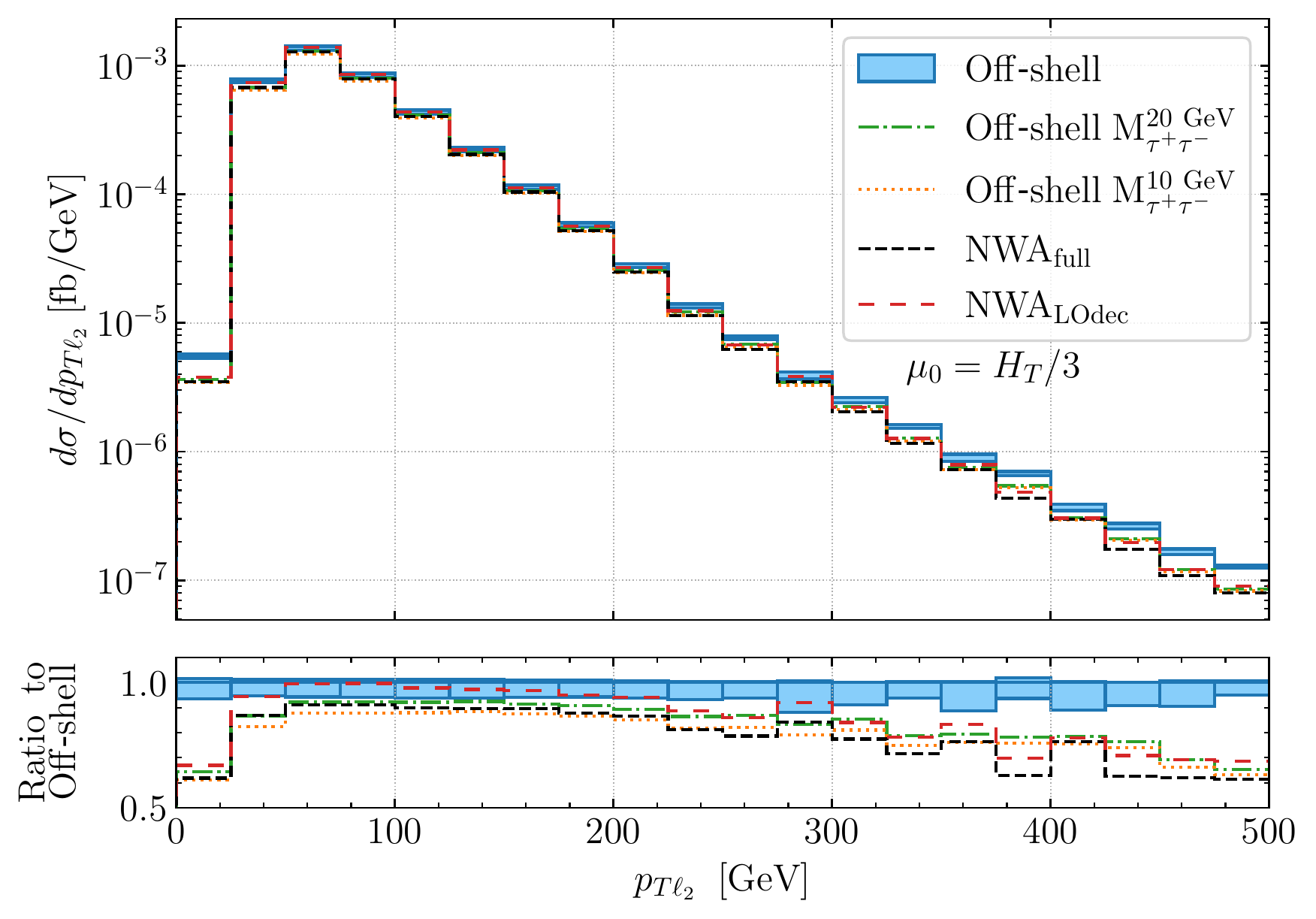}
     \includegraphics[width=0.49\textwidth]{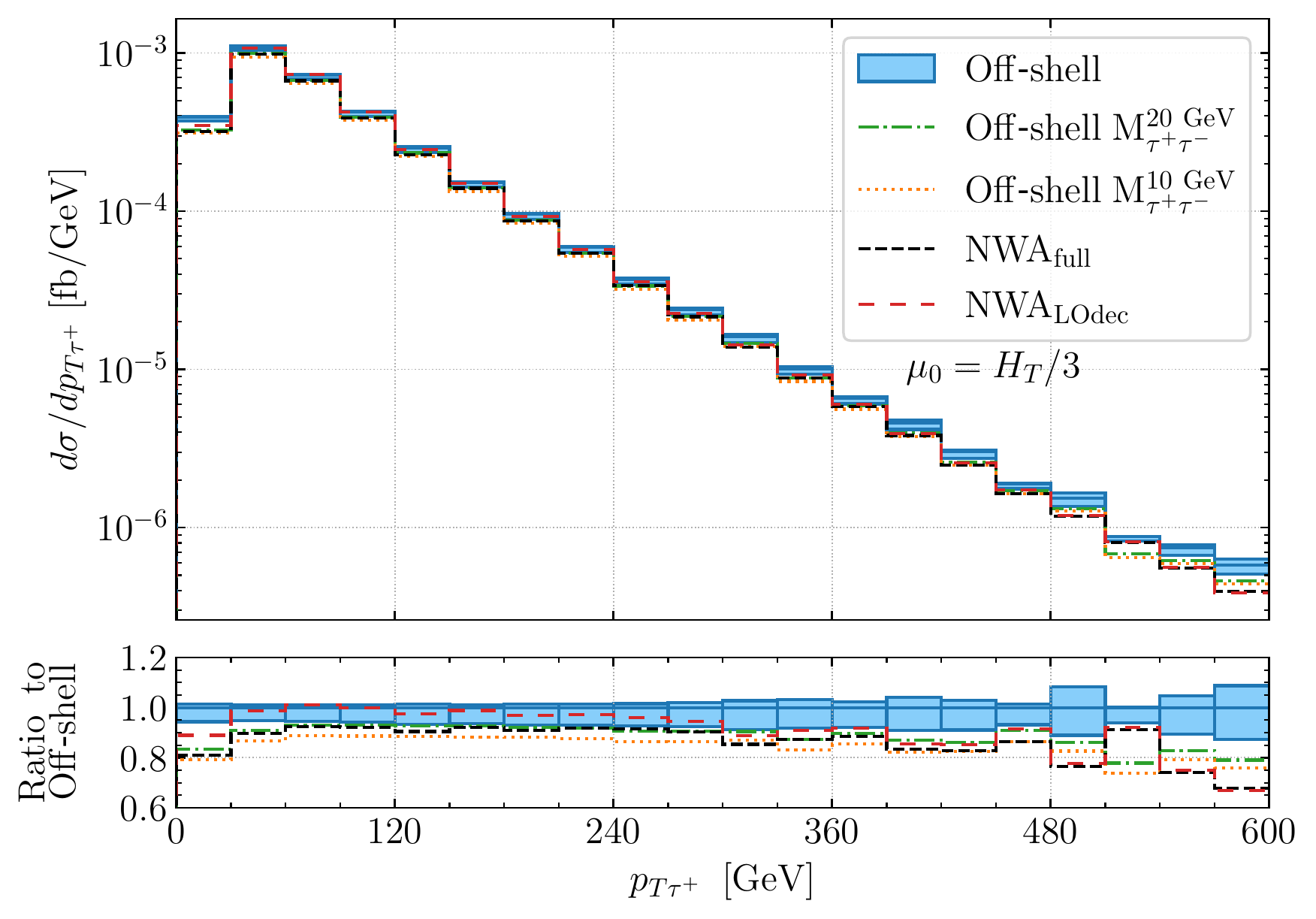}
     \includegraphics[width=0.49\textwidth]{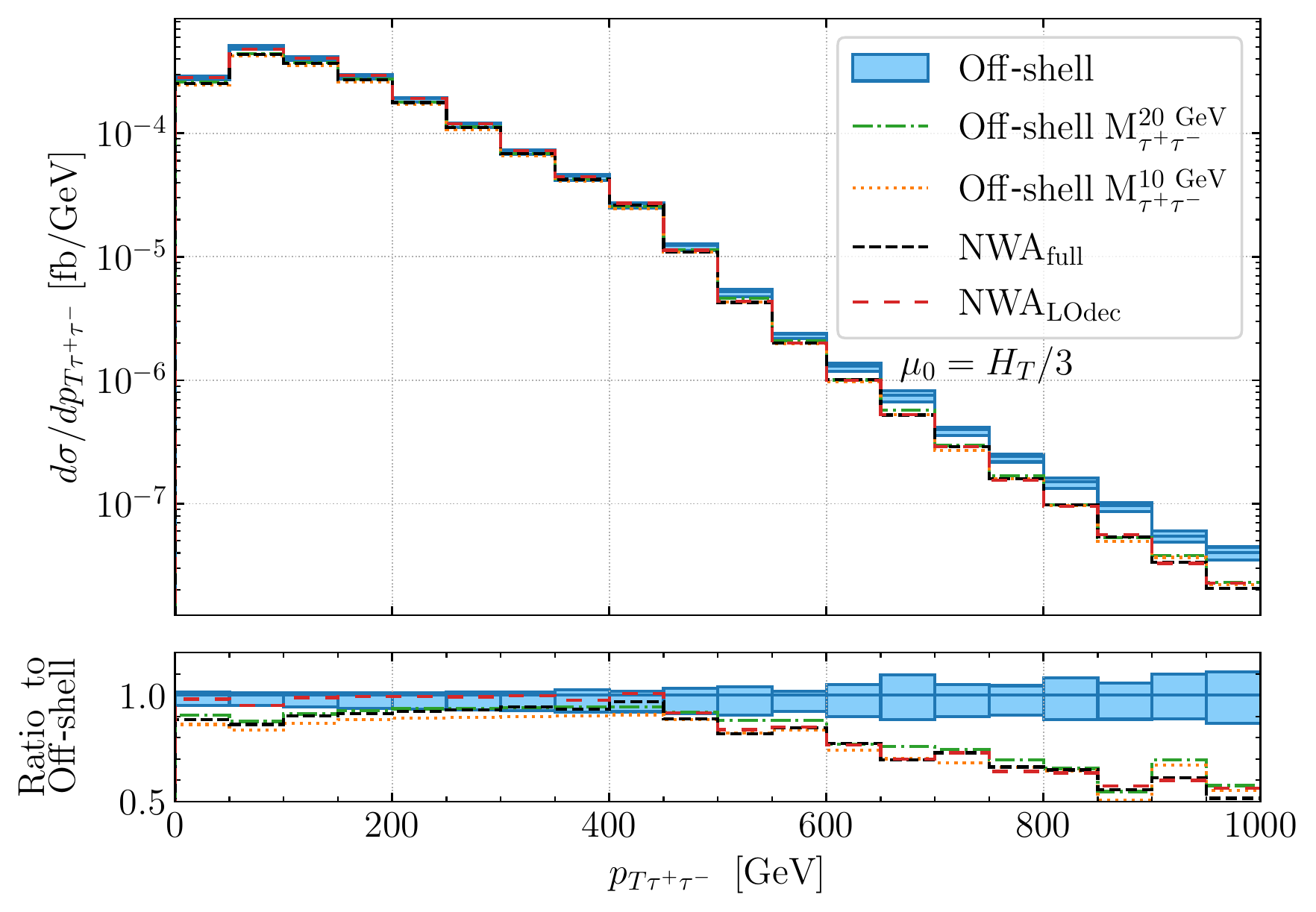}
\end{center}
\caption{\label{fig:diff_nwa_dimension} \it 
Differential cross-section  distributions for 
$pp \to e^+\nu_e \, \mu^-\bar{\nu}_{\mu} \, b\bar{b}\, \tau^+\tau^- +X$ 
at the LHC with $\sqrt{s}=13$ TeV as a function of $p_{T\, b\bar{b}}$, 
$p_{T\, b_1}$, $M^{\rm min}_{b\mu^-}$, $p_{T\, \ell_2}$, $p_{T\, \tau^+}$ 
and  $p_{T\, \tau^+\tau^-}$. The upper panel shows the NLO QCD prediction 
for the full off-shell case as well as for ${\rm NWA}_{\rm full}$ and 
${\rm NWA}_{\rm LOdec}$. Also given is the full off-shell result with 
the additional $|M_{\tau^+\tau^-} -m_Z|< 20\, (10)$ GeV cut.
Furthermore, we  provide theoretical uncertainties as obtained from 
the scale dependence for the full off-shell case. The lower panel displays
the ratios to the full off-shell result. The scale choice is
$\mu_R=\mu_F=\mu_0=H_T/3$. The cross sections are evaluated with the 
NNPDF3.1  PDF set. }
\end{figure}
%============================================= 
\begin{figure}[t]
  \begin{center}
     \includegraphics[width=0.49\textwidth]{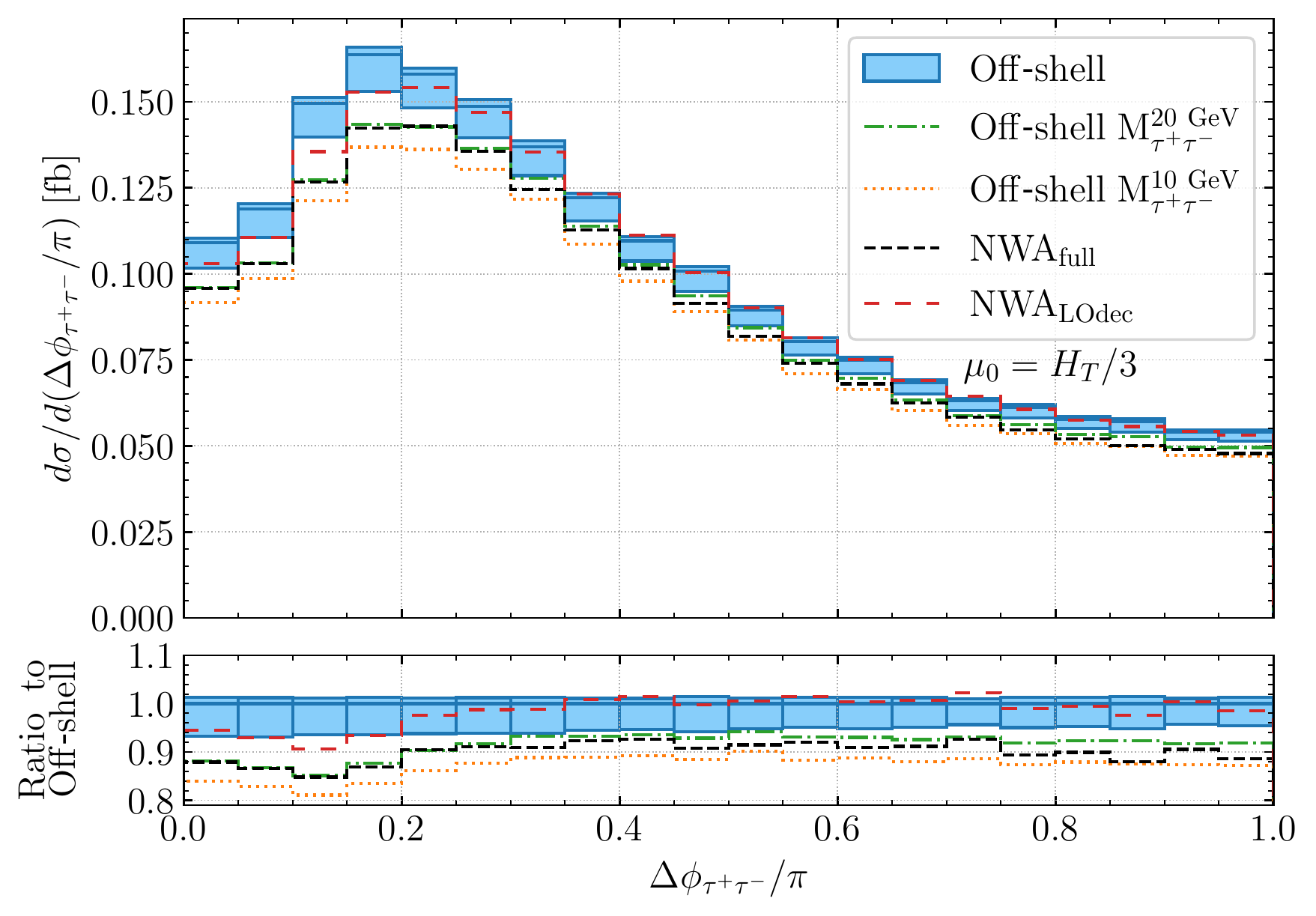}
    \includegraphics[width=0.49\textwidth]{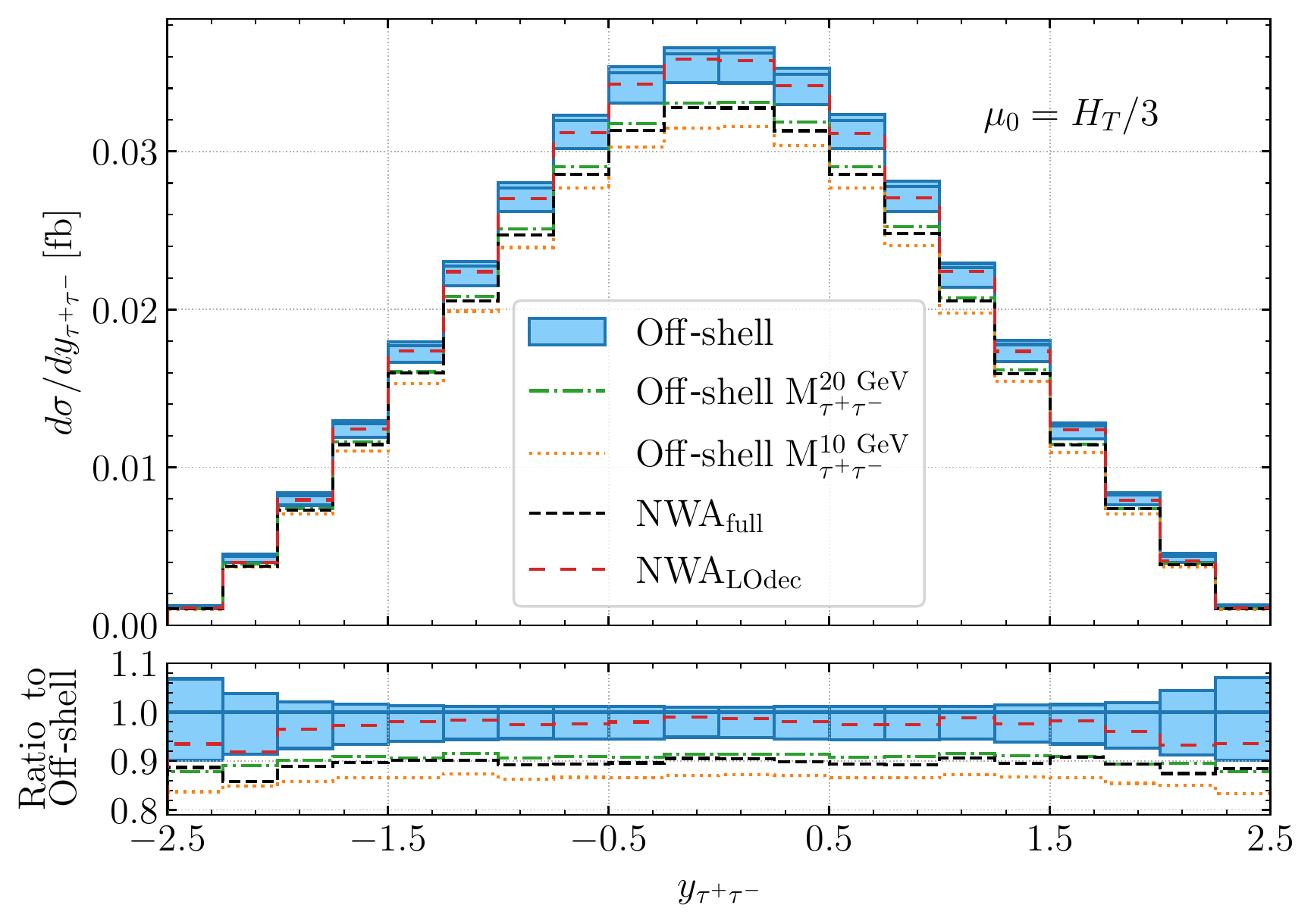}
    \includegraphics[width=0.49\textwidth]{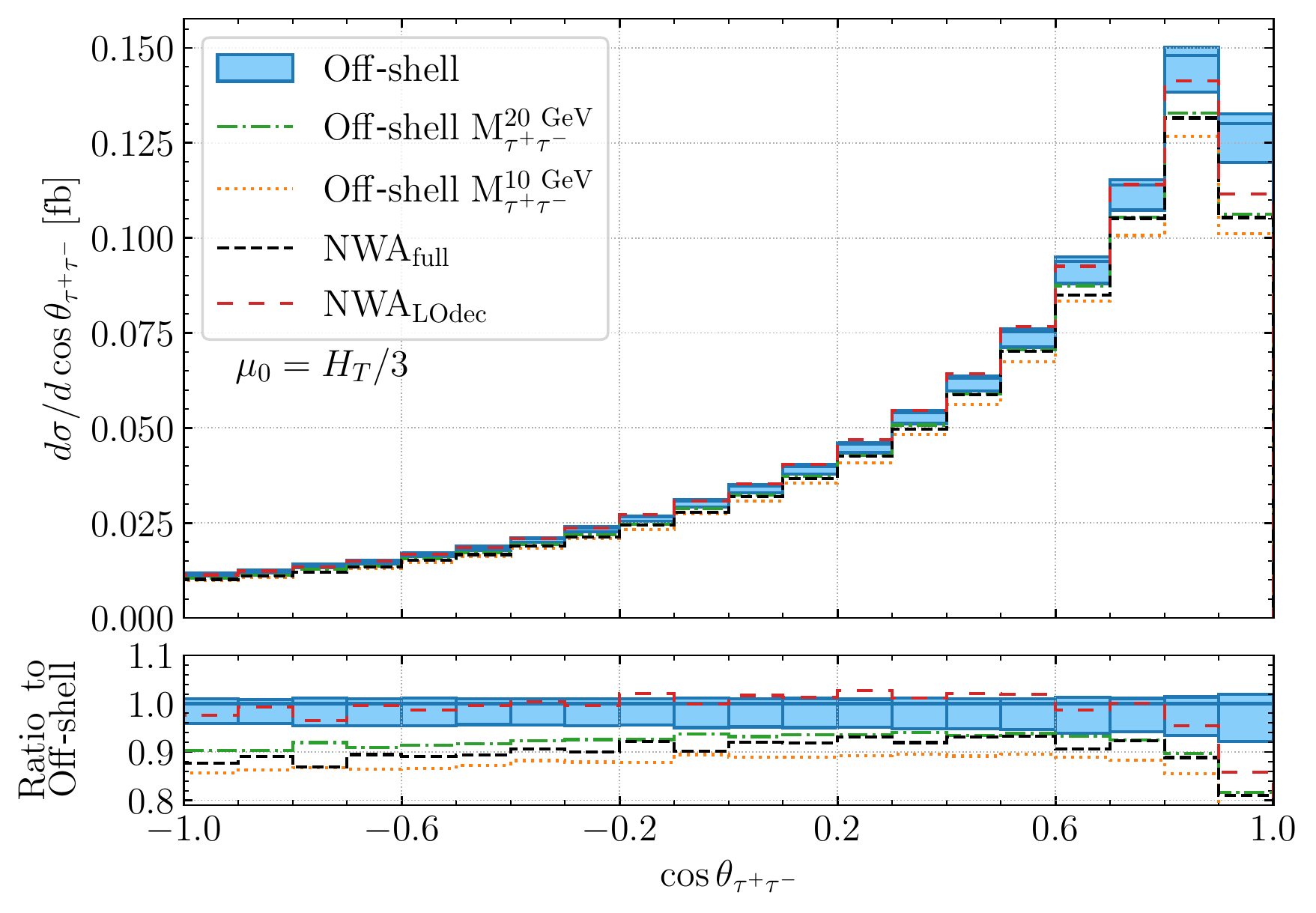}
    \includegraphics[width=0.49\textwidth]{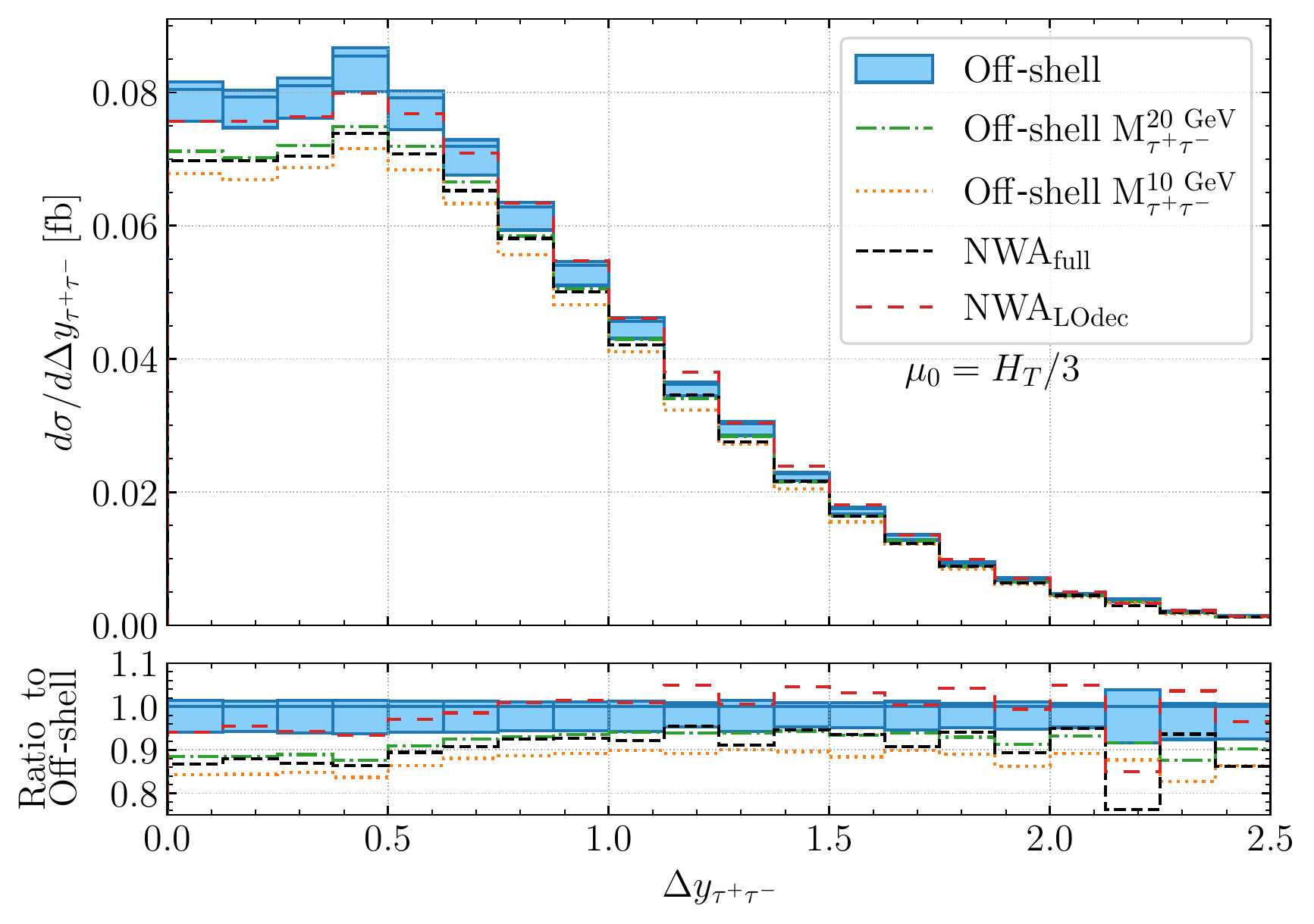}
    \includegraphics[width=0.49\textwidth]{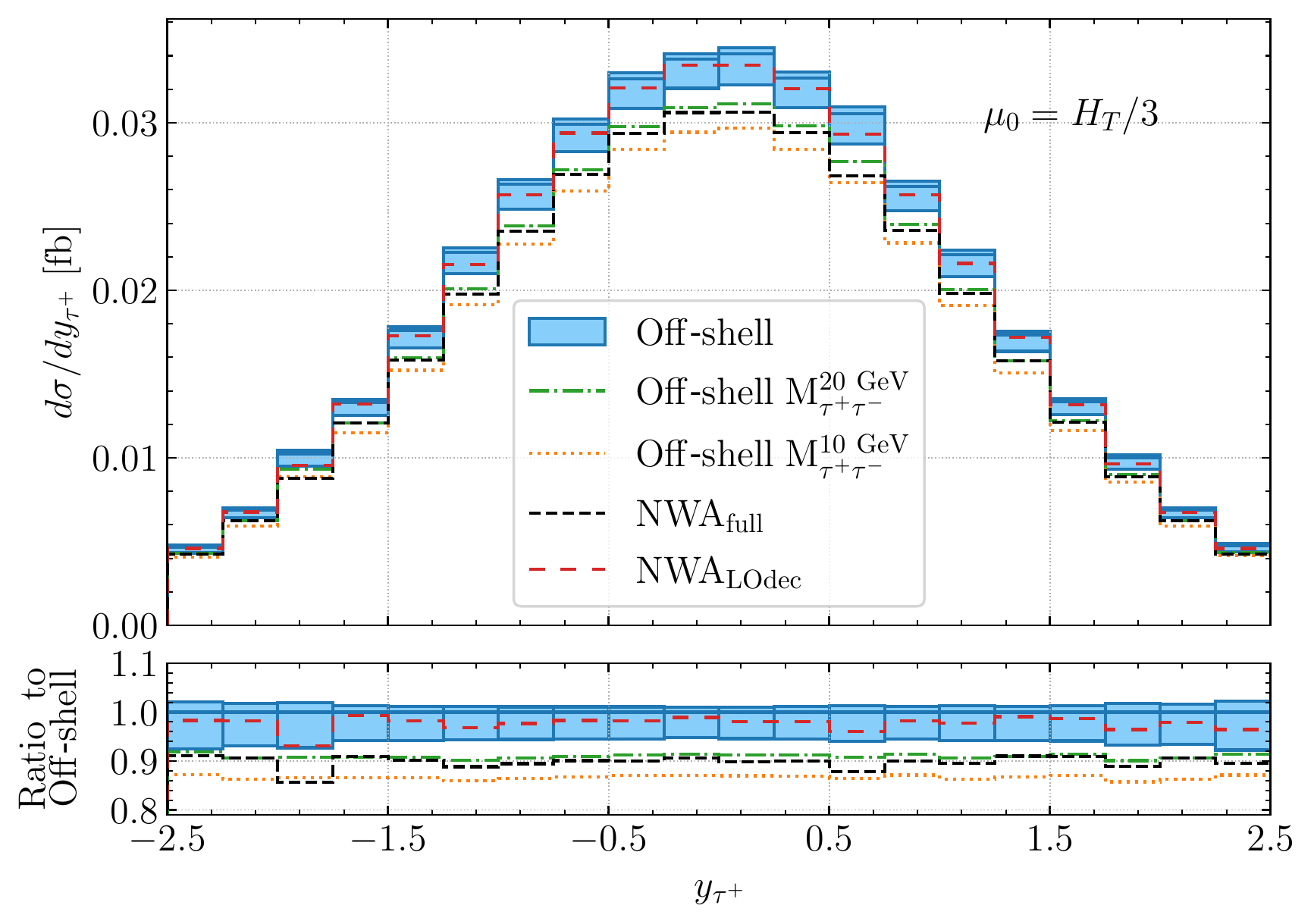} 
    \includegraphics[width=0.49\textwidth]{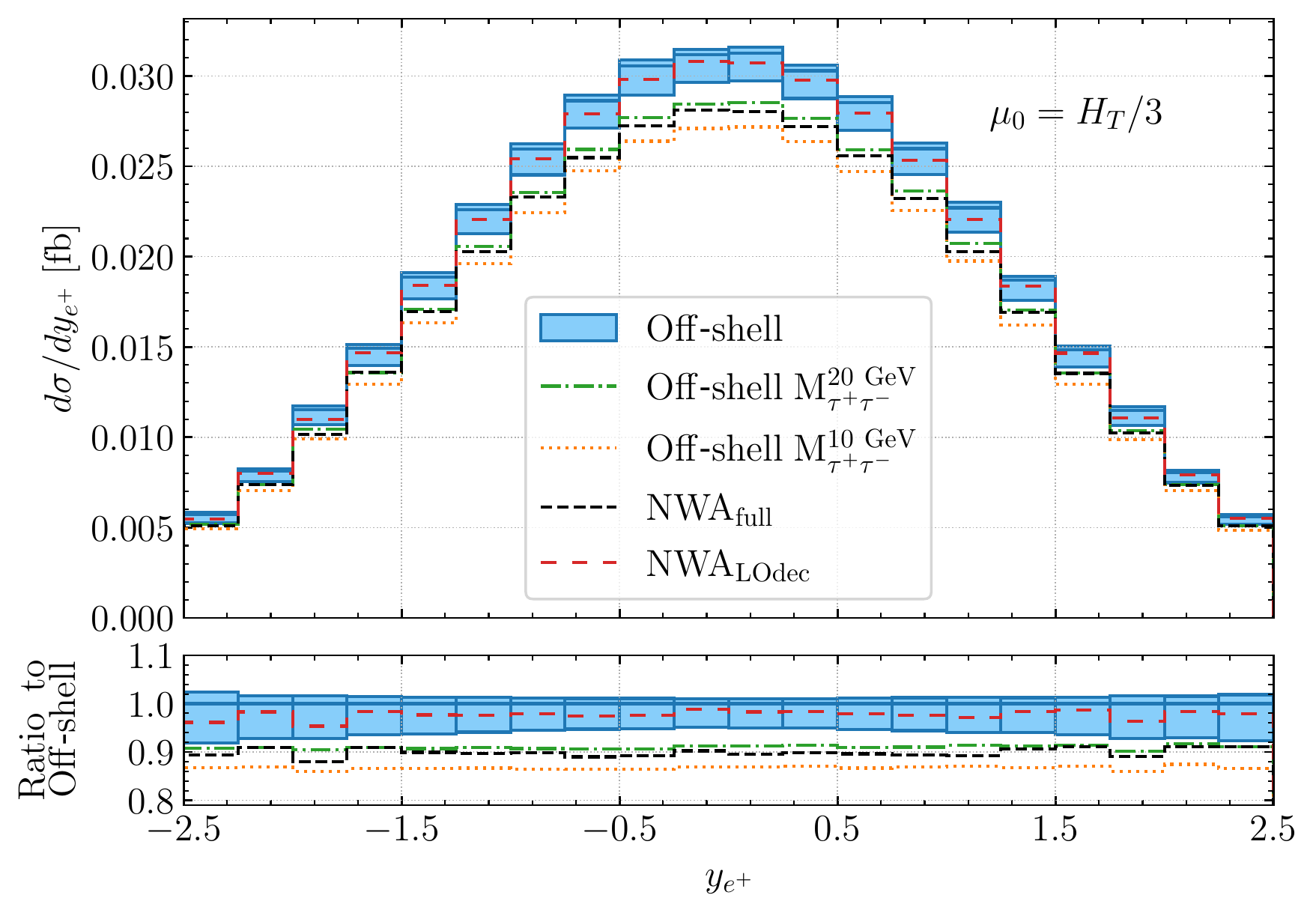}
\end{center}
\caption{\label{fig:diff_nwa_angular} \it 
Differential cross-section  distributions for 
$pp \to e^+\nu_e \, \mu^-\bar{\nu}_{\mu} \, b\bar{b}\, \tau^+\tau^- +X$ 
at the LHC with $\sqrt{s}=13$ TeV as a function of $\Delta
\phi_{\tau^+\tau^-}$, $y_{\tau^+\tau^-}$, $\cos\theta_{\tau^+\tau^-}$,
$\Delta y_{\tau^+\tau^-}$, $y_{\tau^+}$ and  $y_{e^+}$. The upper panel 
shows the NLO QCD prediction for the  full off-shell case as well as for
${\rm NWA}_{\rm full}$ and ${\rm NWA}_{\rm LOdec}$.  Also given is the full
off-shell result with the additional $|M_{\tau^+\tau^-} -m_Z|< 20 \, (10)$ 
GeV cut. Furthermore, we  provide theoretical
uncertainties as obtained from the scale dependence for the full off-shell
case. The lower panel displays the ratios to the full off-shell result. 
The scale choice is $\mu_R=\mu_F=\mu_0=H_T/3$. The cross sections are
evaluated with the  NNPDF3.1  PDF set.}
\end{figure}
%=============================================

In the next step we analyse further various dimensionful observables. 
Specifically,  in Figure \ref{fig:diff_nwa_dimension} we show differential
cross-section distributions as a  function of $p_{T\, b_1 b_2}$, $p_{T\,
b_1}$, $M^{\rm min}_{b\mu^-}$, $p_{T\, \ell_2}$, $p_{T\, \tau^+}$ and 
$p_{T\, \tau^+\tau^-}$. We focus on observables where the full off-shell
effects are most pronounced. The latter are nevertheless visible for the
majority of dimensionful differential cross sections. Very generally 
speaking the full off-shell effects are noticeable  in the following two
phase-space regions: in high $p_T$ tails and in the vicinity of 
kinematical endpoints. In the following we shall provide examples for 
the both cases. The structure of each plot is the same as in Figure
\ref{fig:diff_mtautau}.  Some of the observables displayed in Figure
\ref{fig:diff_nwa_dimension}  have already been introduced. However,
presented for the first time is the transverse momentum of the $b\bar{b}$
system, $p_{Tb_1b_2}$, the transverse momentum of the positively charged 
tau lepton, $p_{T\tau^+}$, and the transverse momentum of the second hardest charged lepton, $p_{T\ell_2}$, where $\ell$ stands for
$\ell=e^+,\mu^-,\tau^+,\tau^-$. Finally, also given is the minimum  
invariant mass of the $b$-jet and the muon, $M^{\rm min}_{b\mu^-}$, 
which is defined as 
\begin{equation}
M^{\rm min}_{b\mu^-}=\min \left\{M_{b_1\mu^-},M_{b_2 \mu^-}\right\}\,.
\end{equation}
In the case when the top quarks  and $W$ gauge bosons are on-shell and the
masses of all final decay products are neglected the $M^{\rm min}_{b\mu^-}$
observable is bounded from above by $\sqrt{m^2_t - m_W^2} \approx 153$ GeV.
For the off-shell prediction this kinematic limit is smeared, nevertheless
there is a sharp fall of the cross section for $M^{\rm min}_{b\mu^-}  
\gtrsim 153$ GeV. Even for the ${\rm NWA}_{\rm full}$ and
${\rm NWA}_{\rm LOdec}$ cases at NLO in QCD this endpoint is smeared by
real emission contributions. 

For the transverse momentum distribution of the $b\bar{b}$ system we can
detect substantial effects. Already for $p_{T\, b_1 b_2} \approx 400$ GeV
the difference between the full off-shell result and the ${\rm NWA}_{\rm
full}$ prediction is about $30\%$. At the end of the plotted 
spectrum the full off-shell effects are even larger up to $40\%-45\%$. 
In spite of the fact that theoretical uncertainties due to the scale 
dependence are particularly large for this observable, ${\rm NWA}_{\rm
full}$ and ${\rm NWA}_{\rm LOdec}$ are still outside of the uncertainty
bands.  The latter are at most at the $20\%-30\%$ level in the
aforementioned phase-space regions.  If the additional $|M_{\tau^+\tau^-}
-m_Z|< 20 \, (10)$ GeV cut  is imposed the full off-shell effects are
reduced to $15\%-20\%$ for  $p_{T\, b_1 b_2} \approx 400$ GeV, whereas 
they are up to $30\%-35\%$  towards the  end of the $p_{T\, b_1 b_2}$
spectrum. We can monitor the impact of higher-order corrections to 
top-quark decays by analysing the difference between ${\rm NWA}_{\rm full}$
and ${\rm NWA}_{\rm LOdec}$ predictions. We can see that NLO  QCD
corrections to top-quark decays are up to $10\%-13\%$, however, the shape  
distortions introduced by these higher order effects  are larger 
and increase up to $23\%$.

For the transverse momentum of the hardest $b$-jet, $p_{T\, b_1}$, we can
observe quite similar effects. The off-shell effects are up to 
$20\%-35\%$ in $p_T$ tails. Moreover, in the whole plotted range the
${\rm NWA}_{\rm full}$ curve is outside of the uncertainty bands of the
full off-shell result.  NLO QCD corrections to top-quark decays affect the 
distribution only for  moderate $p_{T\,b_1}$ values, i.e. up to about 
$300$ GeV, and amount to $12\%-13\%$. They provide, however, the shape  
distortions up to $25\%$. The situation does not change significantly 
when the invariant mass of the two $\tau$ leptons is restricted due 
to the extra cut.

We continue with differential cross section distributions as a function of
the top-quark decay products and analyse in the next step the 
$M^{\rm min}_{b\mu^-}$ observable.  For the phase-space regions up to the
kinematical endpoint, the relative difference between the full off-shell
prediction and the ${\rm NWA}_{\rm full}$ result is moderate and rather
constant of the order of $10\%$. For $M^{\rm min}_{b\mu^-}  \approx 153$ 
GeV off-shell effects increase to about $60\%$ whereas for $M^{\rm
min}_{b\mu^-}  \approx 200$  GeV they are as high as $80\%$. For 
$M^{\rm min}_{b\mu^-}  \gtrsim 153$ GeV  the ${\rm NWA}_{\rm full}$
prediction simply fails to adequately describe this observable. This is 
not surprising since this kinematic range is dominated by a single-resonant
top-quark contribution, which is missing from the ${\rm NWA}_{\rm full}$ 
result. As for the NLO QCD corrections to top-quark decays we also observe  
a different behaviour in the two distinct phase-space regions, i.e. below 
and above the kinematic edge. For $M^{\rm min}_{b\mu^-}  < 153$ GeV 
higher-order effects are negative and in the range $1\%-30\%$. After 
reaching the kinematical limit of $M^{\rm min}_{b\mu^-}  \approx 153$ 
GeV  they become positive and constant, of the order of $15\%$. 

Next in line is the transverse momentum of the second hardest charged
lepton, $p_{T\, \ell_2}$. In this case full off-shell effects are in the 
range of $10\%-40\%$. On the other hand, the scale dependence is only up to
$10\%$ in the whole plotted range. Higher-order effects in top-quark 
decay are negative and fairly constant at the level of $10\%$. As expected 
the additional $M_{\tau^+\tau^-}$ cut has a sizable impact on the 
$p_{T\, \ell_2}$ spectrum. This is due to the fact that $p_{T\, \ell_2}$ 
comprises all charged leptons, thus, also $\tau^+$ and $\tau^-$ that are 
directly affected by this cut. Moreover, among the four charged leptons, 
the decay products of the $Z$ boson have harder spectra than the leptons 
from the decays of top quarks. We can observe that the full off-shell 
prediction with the $|M_{\tau^+\tau^-} -m_Z|< 10 \,(20)$ GeV cut closely 
follows the  ${\rm NWA}_{\rm full}$ result. For the transverse momentum 
of the $\tau^+$, $p_{T\, \tau^+}$, we observe qualitatively  similar 
effects as for the  $p_{T\, \ell_2}$ observable. 

The last observable plotted in Figure \ref{fig:diff_nwa_dimension}
is the transverse momentum  of the $\tau^+\tau^-$
system.  Already at $p_{T\, \tau^+\tau^-} \approx 600$ GeV full off-shell
effects are at the level of $20\%$, and at the same time theoretical
uncertainties are two times  smaller. As we approach the end of the 
$p_{T\, \tau^+\tau^-}$ spectrum the off-shell effects increase and 
are as high as $40\%-50\%$.  Theoretical uncertainties though are 
only slightly higher there, i.e. they  are at most $15\%$. Although 
NLO QCD corrections to top-quark decays are far  from constant in the 
plotted range they reach maximally $10\%$. By employing the 
$|M_{\tau^+\tau^-} -m_Z|< 10 \,(20)$ GeV cut we essentially reproduce 
the on-shell prediction as given by the ${\rm NWA}_{\rm full}$ curve. 

Finally, in Figure \ref{fig:diff_nwa_angular} we display a few examples of
dimensionless observables. In detail, we show differential cross-section 
distribution as a  function of $\Delta \phi_{\tau^+\tau^-}$,
$y_{\tau^+\tau^-}$, $\cos\theta_{\tau^+\tau^-}$ and 
$\Delta y_{\tau^+\tau^-}$, where  $\Delta y_{\tau^+\tau^-}=
|y_{\tau^+} -y_{\tau^-}|$. Also given are rapidity distributions for
$\tau^+$, $y_{\tau^+}$, and for the positron, $y_{e^+}$.  Dimensionless 
observables receive contributions from all phase-space regions, most 
notably from those that are sensitive to the threshold for the $t\bar{t}Z$
production. Thus, the double-resonant top-quark and resonant $Z$ gauge 
boson contributions have a dominant impact on these
observables. Naively we would expect that the NWA should be good enough to 
properly describe the dimensionless observables. Yet, we can clearly observe
that ${\rm NWA}_{\rm full}$ fails to do this job for all the cases that we
have presented. Indeed, the ${\rm NWA}_{\rm full}$ curves are always outside
of the uncertainty  bands of the full off-shell results. These effects are
even enhanced in the phase-space regions where the  bulk of the cross  
section is located. Thus, photon-induced contributions  and the $Z/\gamma^*$
interference effects are very important also for angular distributions. We 
can further observe that the  ${\rm NWA}_{\rm LOdec}$ seems to correctly
predict the dimensionless spectra in all phase-space regions.  As already
argued earlier in this Section, this agreement is accidental and cannot be
understood as ${\rm NWA}_{\rm LOdec}$ giving better predictions than those
provided  by  ${\rm NWA}_{\rm full}$.

More specifically, for the $\Delta \phi_{\tau^+\tau^-}/\pi$ observable full
off-shell effects  are up to $15\%$ for the small values of $\Delta
\phi_{\tau^+\tau^-}/\pi$, whereas for $\Delta \phi_{\tau^+\tau^-} /\pi 
\approx 1$  they decrease to about $10\%$. Theoretical uncertainties 
due to the scale dependence are a factor  of two smaller. Furthermore, we
obtain negative NLO QCD corrections to top-quark decays of the order of 
$10\%.$

For all rapidity distributions, i.e. for $y_{\tau^+\tau^-}$, $y_{\tau^+}$ 
and $y_{e^+}$, full off-shell effects and higher-order corrections to
top-quark decays are alike. Both effects are rather constant
and of the order of $10\%$.  What makes them different is the sign of the
effect. Also scale uncertainties are the same in all three cases. The latter 
are well below $10\%$. We add at this point that, also for the 
$\Delta y_{\tau^+\tau^-}=|y_{\tau^+} -y_{\tau^-}|$ observable  we observe qualitatively similar effects.

At last, in the case of $\cos \theta_{\tau^+\tau^-}$ full off-shell effects
are of the order of $10\%$ in all but the last bin. In the latter the
difference between ${\rm NWA}_{\rm full}$ and the full off-shell prediction
increases up to $20\%$. Theoretical uncertainties  range respectively from
about $5\%$ up to maximally $8\%$. Also here, NLO QCD corrections to 
top-quark decays  are negative and below $10\%$.

We conclude that for the observables that we have presented, which have been
obtained with a rather inclusive set of cuts, the importance of a full
off-shell calculation is rather clear. Not only high $p_T$ tails and
kinematical edges of dimensionful  differential cross section distributions
are poorly described by the NWA results but also angular distributions 
are affected. Thus even the latter would strongly benefit  from the full
off-shell predictions. Having such theoretical predictions at our disposal
we can  also analyse the impact of the additional 
$|M_{\tau^+\tau^-} -m_Z|< 10 \,(20)$ GeV cut on various observables.  
Namely, to understand the effect of enlarging or reducing 
the window of the  invariant  mass of the  $\tau^+\tau^-$ system around 
the nominal value of the mass  of the $Z$ gauge boson.

% =============================================
%
\section{Summary}
\label{summary}
%
% =============================================
%

In this paper we calculated NLO QCD corrections 
to $pp\to e^+\nu_e \, \mu^-\bar{\nu}_{\mu} \, b\bar{b}\, 
\tau^+\tau^- +X$ at the LHC with $\sqrt{s}=13$ TeV. 
In the computation off-shell top quarks and massive gauge 
bosons  have been described by Breit-Wigner propagators, furthermore
double-, single- and non-resonant top-quark contributions along with
resonant and non-resonant $Z$ as well as photon-induced contributions 
have been consistently incorporated already at the matrix element level. 
The total cross section and its scale dependence have been evaluated for 
two different scale choices, i.e.  for the fixed scale
$\mu_R=\mu_F=\mu_0=m_t+m_Z/2$ and for the dynamical one
$\mu_R=\mu_F=\mu_0=H_T/3$. The impact of the NLO QCD corrections on the
integrated cross sections is moderate, of the order of $27\%$ for 
$\mu_0=m_t+m_Z/2$ and $23\%$ for $\mu_0 = H_T/3$. As to the theoretical
uncertainty of our calculation, the contribution related to unknown
higher-order corrections, as obtained by studying the scale dependence of
our NLO predictions, is of the order of $6\%$. We have also analyzed the
theoretical error arising from PDFs, being able to quantify it at the 
level of $1\%$ for the default NNPDF3.1 PDF set. For other PDF sets, 
that we have examined, the PDF uncertainties are of the order of $2\%-3\%$.
Thus, they are well below the uncertainty associated with the 
scale dependence. 

For differential cross section distributions we have provided only
predictions for the dynamical scale setting. The reason why we have not
provided our results at the differential level for $\mu_0=m_t+m_Z/2$ is
summarised in the following. Firstly, for the fixed scale choice we have
noticed that for some choices of the scale the NLO results  
become negative. This happened in the high-energy tails of dimensionful
distributions. Secondly, there are kinematic regions where the scale
variation bands at LO and NLO do not overlap anymore. Thirdly, it has 
even happened that the scale variation at NLO has actually exceeded the
scale variation of the LO predictions. All these effects have already 
been observed in our previous studies for $t\bar{t}$ plus an additional
object(s) and to a large extent have been accommodated by a  judicious 
dynamical scale choice. In the case at hand for the dynamical
scale setting, i.e. for $\mu_0=H_T/3$, NLO QCD effects are up to
$20\%-50\%$. The theoretical uncertainties due to scale variations are, 
on the other hand, not higher than  $10\%$. The latter are  the dominant
source  of the theoretical systematics since the NNPDF3.1 PDF 
uncertainties are very  small and well below the corresponding 
theoretical uncertainties due to  scale dependence.

We have additionally provided theoretical predictions for the $t\bar{t}Z$
process in  the tetra-lepton decay channel using the full NWA and the NWA
with LO top-quark decays. We have noticed that the ${\rm NWA}_{\rm full}$
prediction has the same theoretical uncertainties due to scale dependence 
as the full off-shell result. However, when higher-order corrections in
top-quark decays are omitted, the scale dependence increases. For example,
for the  integrated fiducial cross section these theoretical uncertainties
are at the level of $10\%$. Therefore, they are almost a factor of $2$ larger
than the theoretical uncertainty estimate for the ${\rm NWA}_{\rm full}$
case.  Furthermore, we have studied the impact of NLO QCD corrections to
top-quark decays. These corrections are  of the order of $10\%$  for the
integrated fiducial cross section. For various differential distributions 
the differences between the ${\rm NWA}_{\rm full}$ and  the ${\rm NWA}_{\rm
LOdec}$ case are of the same order. On top of that, having both theoretical
predictions, full off-shell  and ${\rm NWA}_{\rm full}$ we have
examined  the impact of full off-shell effects at the integrated and
differential fiducial level. For the integrated cross sections these
effects are consistent with the expected accuracy  of the NWA only when
the $|M_{\tau^+\tau^-} -m_Z|< X$ GeV cut, with $X\in\left\{25,20,15,10\right\}$ GeV, 
has  been employed. Specifically, with this cut full off-shell effects are
in the range $0.3\%-3.4\%$. Without  this additional cut the
full off-shell effects are as large as  $11\%$. These large effects 
mostly originate from the photon-induced  contributions and $Z/\gamma^*$
interference effects. At the differential level, large non-factorisable
corrections even up to $40\%-80\%$ have been obtained for dimensionful
differential cross sections. Poorly described  by the ${\rm NWA}_{\rm
full}$ are also angular distributions. Thus, even  the  latter would
strongly benefit from the full off-shell calculation. 

In summary, the non-factorisable NLO QCD corrections  impact
significantly the $t\bar{t}Z$ cross section in various phase-space
regions. For this reason they should be included in  future 
comparisons between theoretical predictions and experimental data. In
addition, given the importance of the $t\bar{t}Z$ process as background 
to Higgs boson production in association with a top-quark pair more
detailed phenomenological studies in  the tetra-lepton decay channel 
are required.  We postpone such work for the future.

\acknowledgments{
  
The work of J.N. and M.W.  was supported by the Deutsche
Forschungsgemeinschaft (DFG) under grant 396021762 $-$
TRR 257: {\it P3H - Particle Physics Phenomenology after the Higgs
Discovery} and  by the DFG under grant 400140256 - GRK
2497: {\it The physics of the heaviest particles at the Large Hardon
  Collider}.

Support by a grant of the Bundesministerium f\"ur Bildung
und Forschung (BMBF) is additionally acknowledged.

The research of G.B. is supported by grant K 125105 of the National
Research, Development and Innovation Office in Hungary.

The work of M.K. is supported in part by the U.S. Department of Energy
under grant DE-SC0010102.

H.B.H. has received funding from the European Research Council (ERC) under
the European Union's Horizon 2020 Research and Innovation Programme (grant
agreement no. 683211). Furthermore, the work of H.B.H has been partially
supported by STFC consolidated HEP theory grant ST/T000694/1.

Simulations were performed with computing resources granted by RWTH
Aachen University under projects {\tt rwth0414} and {\tt rwth0847}.
}

%\bibliography{References} 

%\bibliographystyle{JHEP}

%\printbibliography[heading=bibintoc, title={References}]

\providecommand{\href}[2]{#2}\begingroup\raggedright\endgroup

\end{document}